\documentclass[12pt]{spieman}

\usepackage{graphicx}
\usepackage{array}
\usepackage{tabularx}
\usepackage{xltabular}
\usepackage{multirow}
\usepackage{pdflscape}
\usepackage{amsmath,amssymb}
\usepackage{url}
\usepackage{titlesec}
\titleformat{\paragraph}[block]
  {\normalfont\normalsize}
  {\normalfont\small\theparagraph}
  {0.8em}
  {}
\titleformat{\subparagraph}[block]
  {\normalfont\normalsize}
  {\normalfont\footnotesize\thesubparagraph}
  {0.8em}
  {}
\titlespacing*{\paragraph}{0pt}{2ex plus .5ex minus .2ex}{1ex plus .2ex}
\titlespacing*{\subparagraph}{0pt}{1.5ex plus .5ex minus .2ex}{0.8ex plus .2ex}

\title{The Habitable Worlds Observatory Technology Development Plan}

\author[a]{Matthew Bolcar\thanks{*}}
\author[b]{Feng Zhao}
\author[a]{Paul Scowen}
\author[a]{Lee Feinberg}
\author[a]{Breann Sitarski}
\author[a]{Alice Liu}
\author[a]{J. Scott Smith}
\author[a]{Josh Abel}

\affil[a]{NASA Goddard Space Flight Center, 8800 Greenbelt Rd., Greenbelt, MD, USA}
\affil[b]{Jet Propulsion Laboratory, 4800 Oak Grove Dr., Pasadena, CA, USA}

\begin{document}
\maketitle
\begin{abstract}
The Habitable Worlds Observatory (HWO) is NASA’s next large space telescope, selected by the 2020 Decadal Survey in Astronomy and Astrophysics to search for and characterize habitable exoplanets while enabling a broad range of transformative astrophysics. In August 2024, the HWO Technology Maturation Project Office (TMPO) was formed to begin exploring the HWO science, technology, and mission architectures toward a Mission Concept Review (MCR) at the end of the decade. A primary deliverable of this effort is this technology development plan that identifies critical technologies that enable the mission, defines a process for assessing the readiness of those technologies, and outlines a strategy for developing those technologies to a Technology Readiness Level (TRL) of 5 before the MCR. This document covers technologies organized along three ``tracks'': Coronagraph System technologies, Ultra-stable Telescope System technologies, and High-sensitivity Ultraviolet and Visible Instrumentation technologies. Additional emerging and enhancing technologies are also discussed.
\end{abstract}
\keywords{technology development, Habitable Worlds Observatory, ultraviolet, coronagraph, detectors} \\
{\noindent \footnotesize\textbf{*} \linkable{matthew.r.bolcar@nasa.gov} }
\section{Introduction}
\subsection{Purpose}
This document presents the approach for assessing and developing technologies needed to enable or enhance the Habitable Worlds Observatory (HWO). As the technology development efforts progress, this document will also be used to track and report on the technology development accomplishments.

The 2020 Decadal Survey in Astronomy and Astrophysics, ``Pathways to Discovery in Astronomy and Astrophysics for the 2020s'' (Astro2020\cite{Astro2020}), recommended NASA establish a ``Great Observatories Mission and Technology Maturation Program'' (GOMAP) that would ``co-develop the science, mission architecture, and technologies'' for a new series of Great Observatories spanning the X-ray to far-infrared wavelengths. The first mission to enter the GOMAP program is the HWO, which will search for biosignatures on $\sim$25 habitable zone planets and perform transformative general astrophysics. HWO is envisioned to have an $\sim$ 6 m segmented aperture and use an internal coronagraph instrument to perform the high-contrast imaging. A suite of imaging and spectroscopic instruments spanning the far-ultraviolet (FUV, $\sim$100 nm) to near infrared (NIR, $\sim$2 \textmu{}m) will also be used for a broad array of general astrophysics investigations and discoveries.

Following the Astro2020 recommendation, NASA has established the HWO Technology Maturation Project Office (TMPO) to co-develop the science, architecture, design, and technologies for a potential HWO mission. This document describes the TMPO technology development plan.

\subsection{Dual use technology and technology spin-offs}
Historically there have been co-investments in technologies needed for astrophysics missions by multiple government agencies. Some technologies also have national security applications.  Key technologies covered in this document do have potential dual uses. Infrared photon counting sensors have a wide variety of potential applications including Light Detection and Ranging (LIDAR) applications. Precise and stable deformable mirrors, Extreme Ultraviolet (EUV) mirrors, metrology systems, and stability control systems are used in microlithography and will enable next generation applications. Some technologies also have national security applications. HWO TMPO will coordinate investments in technology with investments from other government agencies and industry partners to maximize economic benefits.

\section{Mission Description and Organization}
\subsection{Notional Mission Description}
\begin{figure}[htbp]
\centering
\includegraphics[width=0.9\linewidth]{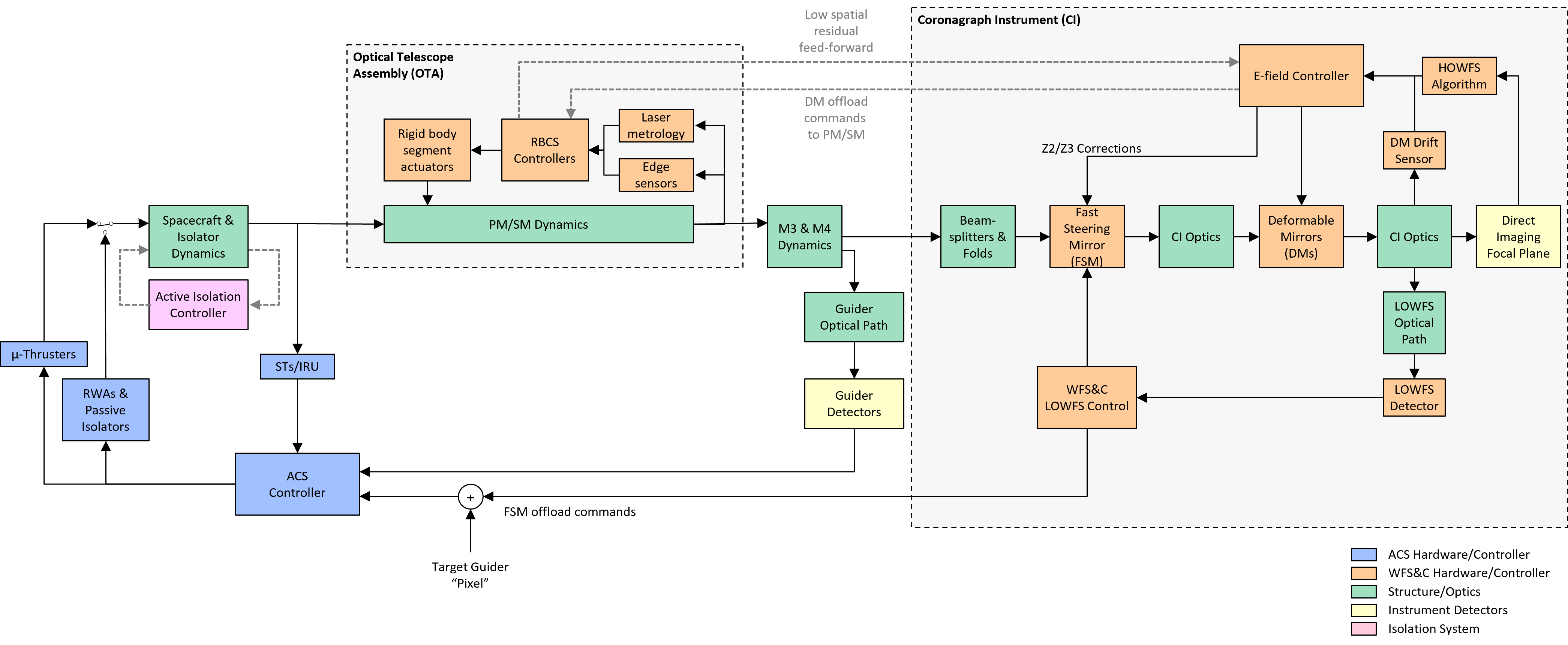}
\caption{Notional control system block diagram.}
\label{fig:1}
\end{figure}

The mission science, architecture, design, and technology are being co-developed by the HWO TMPO, and a more detailed overview of the system architecture and elements can be found in Feinberg et al. (2026)\cite{Feinberg2026}. Generally, HWO will consist of an observatory and a ground system. The observatory is expected to comprise all or some of the following:

\begin{itemize}
\item A segmented aperture optical telescope element (OTE) with a $\sim$6-8.5 m aperture. Varying degrees of deployment are currently being studied.
\item A high-contrast coronagraph instrument (CI), capable of achieving $\sim$1$\times$10{\textsuperscript{-10}} raw contrast over an $\sim$20\% instantaneous bandwidth at $\sim$3 $\lambda$/D inner working angle (IWA). Specific details on the total bandpass of the instrument and its spectroscopic capabilities are still being studied.
\item An ultra-violet multi-object spectrograph (UV MOS) with imaging capabilities at wavelengths as short as 100 nm. Specific long wavelength cutoff and spectroscopic resolution details are still being studied.
\item A high-resolution imager (HRI) with a large ($\sim$6 arcmin{\textsuperscript{2}}) field-of-view over visible and NIR wavelengths. Additional details on instrument capabilities are still being studied.
\item An ultra-violet integral field spectrograph (UV IFS) that will take spatially resolved spectroscopy at wavelengths as short as 100 nm. Specific long wavelength cutoff and spectroscopic resolution details are still being studied.
\item One or more other instruments that may also be used to perform guiding and/or image-based wavefront sensing.
\item A spacecraft bus that will provide typical spacecraft functions (i.e., communication, power delivery, attitude control, command and data handling $[$C\&DH$]$, etc.)
\item A main barrel assembly (MBA) that will surround the OTE to provide protection against stray light and micrometeoroid impacts, as well as potentially support thermal control of the OTE.
\end{itemize}
A key defining challenge associated with HWO is the ability to perform high-contrast coronagraphic imaging and spectroscopy of Earth-like planets around Sun-like stars. Generally, this requires achieving a contrast floor (also referred to as flux-ratio noise, or limiting magnitude) of order 10{\textsuperscript{-11}}, which itself is expected to require post-compensation wavefront stability of a few picometers over durations of minutes or hours. This challenge necessitates a majority of the critical technology elements identified in this plan, including coronagraph masks, deformable mirrors (DMs), sensing and control systems, ultra-stable mirrors and structures, thermal control systems, and disturbance isolation. Figure \ref{fig:1} shows a notional control system block diagram identifying the sensor and actuator interactions.

A second challenge associated with HWO is to enable transformative astrophysics from the FUV to NIR. Generally, this requires better sensitivity over the entire band, but especially in the FUV. This improvement in sensitivity is enabled by advancements in mirror coatings, detectors, and optical component performance.

\subsection{Organization}
Consistent with the challenges defined above, the HWO TMPO Technology Plan is organized along three Tracks

\begin{itemize}
\item Coronagraph System Technologies -- technologies associated with achieving the contrast, contrast stability, and signal-to-noise ratio (SNR) to support exoplanet direct-imaging and spectroscopic functions of the mission.
\item Ultra-stable Telescope System Technologies -- technologies needed to create an optical system capable of achieving the picometer-level wavefront stability required for coronagraphic observations.
\item High-sensitivity Ultraviolet and Visible Instrument Technologies -- technologies associated with improving broadband sensitivity to support an array of transformative astrophysics observations.
\end{itemize}
Each track is further divided into Lanes, where each Lane describes a critical technology need. In some cases, multiple candidate technologies of different maturity may address a particular Lane, offering the opportunity for parallel development and down-selecting to the best performing candidates. Table 1 shows the current set of Lanes for each track.

\begin{xltabular}{\textwidth}{@{} >{\raggedright\arraybackslash}X >{\raggedright\arraybackslash}X >{\raggedright\arraybackslash}X @{}}
\caption{HWO TMPO Technology Tracks and Lanes}\label{tab:1}\\
\hline
\textbf{Track 1: }\textit{Coronagraph System Technologies} & \textbf{Track 2: }\textit{Ultra-stable Telescope System Technologies} & \textbf{Track 3: }\textit{High-sensitivity Ultraviolet and Visible Instrument Technologies} \\
\hline
\endfirsthead
\caption[]{(continued)}\\
\hline
\textbf{Track 1: }\textit{Coronagraph System Technologies} & \textbf{Track 2: }\textit{Ultra-stable Telescope System Technologies} & \textbf{Track 3: }\textit{High-sensitivity Ultraviolet and Visible Instrument Technologies} \\
\hline
\endhead
\hline
\multicolumn{3}{r}{\textit{(continued on next page)}}\\
\hline
\endfoot
\hline
\endlastfoot
Starlight Suppression & Ultra-stable Mirrors & Far-UV Mirror Coatings \\
Contrast Stabilization & Ultra-stable Structures & Near-UV / Visible Detectors \\
Deformable Mirrors & Thermal Control Systems & Far-UV Detectors \\
Low-noise / Noiseless Detectors (visible \& NIR) & Telescope Wavefront Sensing \& Control & Multi-object Selection / Integral Field Units \\
Spectroscopy & Low-Disturbance Systems & UV Gratings \& Filters \\
Post-processing & Deployable Systems &  \\
Near-UV Capability &  &  \\
\end{xltabular}
While this organization implies separation between the technologies, it is important to recognize that each track may have multiple interfaces with the other two. For example, there is a clear interplay between the coronagraph and telescope technologies: coronagraph performance may require a specific telescope stability and the telescope stability that is actually feasible may levy requirements on the coronagraph technologies. Similarly, mirror coatings that enable FUV reflectivity may create polarization or amplitude aberrations that limit coronagraph performance, and cooling requirements for new detectors may have an impact on telescope stability.

\begin{figure}[htbp]
\centering
\includegraphics[width=0.9\linewidth]{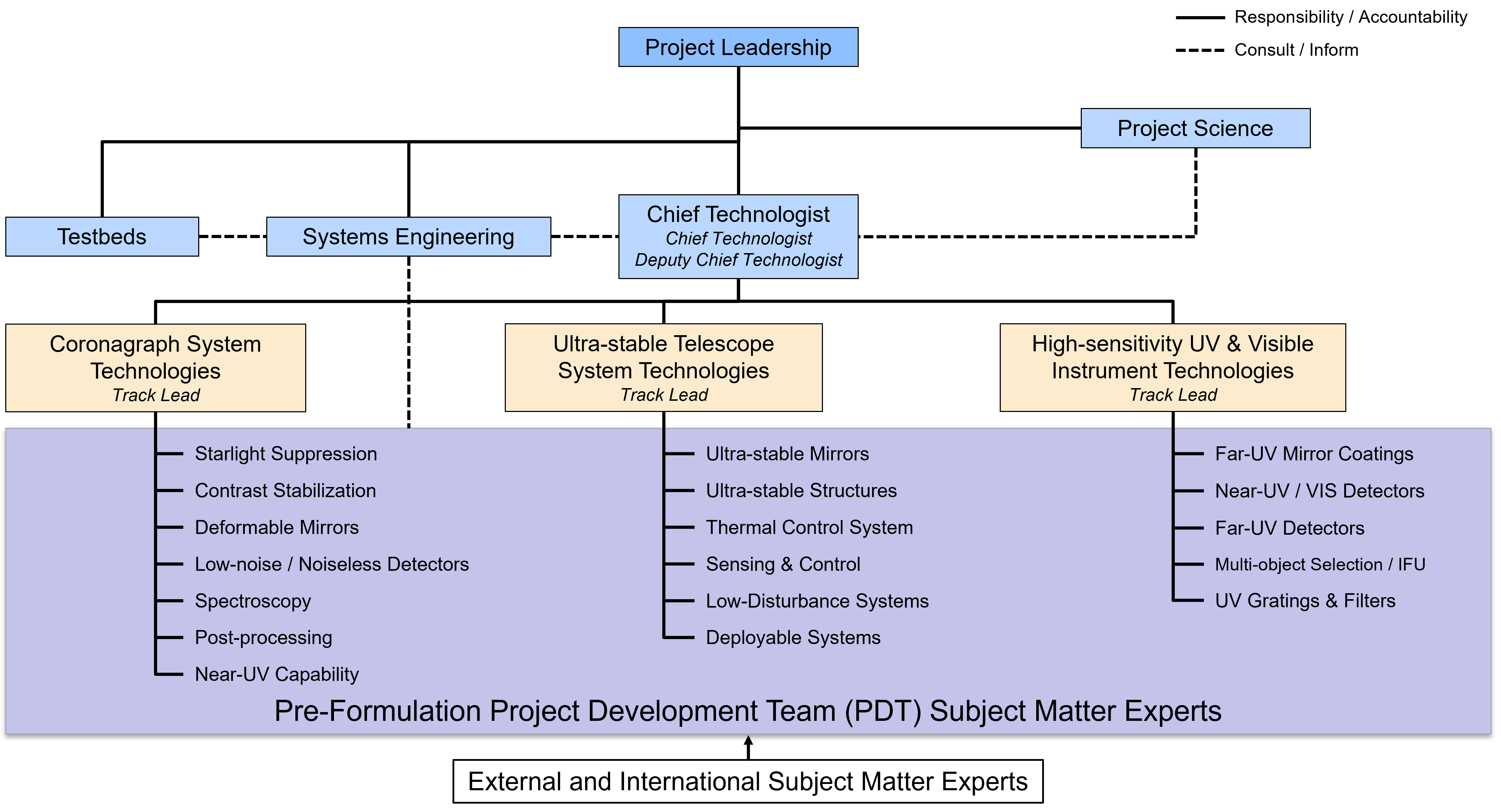}
\caption{HWO TMPO Organization}
\label{fig:2}
\end{figure}

Managing these interfaces is a key responsibility of the HWO TMPO technology development team, shown in Figure \ref{fig:2}.

\section{Technology Assessment Methodology}
The Habitable Worlds Observatory is very early in the project lifecycle: science objectives are still being defined, multiple architectures are under consideration, and the design trade space is still being explored. As such, stable well-defined functional and performance requirements do not yet exist. Yet HWO is the product of decades of concept studies that extend all the way back to initial Terrestrial Planet Finder (TPF\cite{TPF}) studies, the Advanced Technology Large Aperture Space Telescope (ATLAST\cite{ATLAST}) that was studied for the 2010 Decadal Survey and years following, and the Large Ultraviolet / Optical / Infrared Surveyor (LUVOIR\cite{LUVOIR}) and the Habitable Exoplanet Observatory (HabEx\cite{HABEX}) that were studied for the Astro2020 Decadal Survey. Following Astro2020, multiple technology roadmap efforts were commissioned by the Astrophysics Division’s program offices, including the Coronagraph Technology Roadmap (CTR) team\cite{CTR}, the Ultra-stable Optics Roadmap Team (USORT\cite{USORT}) and a comprehensive assessment of ultraviolet component technologies performed by the Cosmic Origins program UV science interest group (SIG\cite{UVSIG}).

These past efforts provide a foundation on which to build this technology development plan in parallel with ongoing science, architecture, and concept design studies. The process is necessarily iterative: as concepts mature and requirements become better defined, technical challenges also become clearer and new technologies get added to the technology development effort or existing development efforts are retired as no longer necessary. Conversely, as technologies mature and either succeed or fail at closing a technical gap, the science objectives and design concepts must be updated accordingly.

The primary objectives of this technology development plan are to:

\begin{enumerate}
\item Complete a technology assessment to identify technology gaps associated with the formulation and implementation of the Habitable Worlds Observatory, answering the following questions:
\begin{enumerate}
\item Of the identified gaps, which require ``new technology'' development and which are engineering?
\item Are there new modeling, facility, or testbed capabilities needed to complete the necessary development?
\item What is the priority of closing a gap, relative to the Mission Concept Review (MCR) and Preliminary Design Review (PDR) project milestones?
\item What candidate technologies fill a particular gap?
\item What is the current Technology Readiness Level (TRL) of each candidate technology?
\end{enumerate}
\item Develop roadmaps, milestones, and success criteria for the development of technologies, capabilities, and facilities to close the identified gaps.
\end{enumerate}
HWO TMPO will use these assessments, roadmaps, and milestones to determine the cost, schedule, and human resources needed to achieve TRL 5 by MCR and TRL 6 by PDR.

\subsection{Definitions}
HWO TMPO will use the TRL definitions from NASA’s Technology Readiness Assessment Best Practices Guide (SP-20205003605), as shown in Table 2.

{\footnotesize
\begin{xltabular}{\textwidth}{@{} >{\raggedright\arraybackslash}p{0.10\textwidth} >{\raggedright\arraybackslash}X >{\raggedright\arraybackslash}X >{\raggedright\arraybackslash}X >{\raggedright\arraybackslash}X >{\raggedright\arraybackslash}X >{\raggedright\arraybackslash}X >{\raggedright\arraybackslash}X @{}}
\caption{Technology Readiness Level (TRL) definition}\label{tab:2}\\
\hline
\textbf{TRL} & \textbf{Definition} & \textbf{Completion Criteria} & \textbf{Performance / Function} & \textbf{Fidelity of Analysis} & \textbf{Fidelity of Build} & \textbf{Level of Integration} & \textbf{Environ}\textbf{-}\textbf{ment} \\
\hline
\endfirsthead
\caption[]{(continued)}\\
\hline
\textbf{TRL} & \textbf{Definition} & \textbf{Completion Criteria} & \textbf{Performance / Function} & \textbf{Fidelity of Analysis} & \textbf{Fidelity of Build} & \textbf{Level of Integration} & \textbf{Environ}\textbf{-}\textbf{ment} \\
\hline
\endhead
\hline
\multicolumn{8}{r}{\textit{(continued on next page)}}\\
\hline
\endfoot
\hline
\endlastfoot
1 & Basic principles observed and reported & Peer-reviewed documented principles & Knowledge underpinning technology concepts / applications & Physics principles identified & N/A & N/A & N/A \\
2 & Technology concept and/or application formulated & Documented description addressing feasibility and benefit & Concept formulated & Feasibility presented & N/A & N/A & N/A \\
3 & Analytical proof-of-concept of critical function & Documented analytical / experimental results validating predictions & Proof-of-concept demonstrated analytically or experimentally & Low & N/A or low-fidelity breadboard & N/A & N/A \\
4 & Component / breadboard validated in a laboratory environment & Documented test performance agreeing with analysis; Definition of relevant environment & Basic functionality / performance demonstrated & Medium & Low-fidelity breadboard & Component / Assembly & Tested in lab; Relevant environment identified \\
5 & Component / brassboard validated in relevant environment & Documented test performance agreeing with analysis; Definition of scaling requirements & Basic functionality / performance maintained & Medium & Medium fidelity brassboard with realistic support elements & Component / Assembly & Tested in relevant environment; life-limiting mechanisms and failure modes characterized \\
6 & System / subsystem model or prototype demonstrated in a relevant environment & Documented test performance agreeing with analysis & Required functionality / performance demonstrated & Medium & High fidelity prototype that addresses critical scaling issues & Subsystem / System & Tested in relevant environment; Verify resiliency to life-limiting mechanisms \\
7 & System prototype demonstration in an operational environment & Documented test performance agreeing with analysis & Required functionality / performance demonstrated & High & High fidelity prototype or engineering unit that addresses all scaling issues & Subsystem / System & Tested in actual operational environment \\
8 & Actual system completed and ``Flight Qualified'' through test and demonstration & Documented test performance verifying requirements and analysis predicts & Required functionality / performance demonstrated & High & Final Flight product & System & Tested as part of project environmental test program; Completed life tests \\
9 & Actual system flight proven through successful operations & Documented mission results verifying requirements & Required functionality / performance demonstrated & High & Final Flight product & System & Operated in actual operational environment \\
\end{xltabular}
}
\subsubsection{Fidelity of Analysis}
Table 3 tailors the fidelity of analysis framework outlined in SP-20205003605 specifically to the HWO TMPO requirements and mission objectives. This tailoring establishes appropriate analysis standards for HWO's unique technical challenges and mission criticality.

{\small
\begin{xltabular}{\textwidth}{@{} >{\raggedright\arraybackslash}p{0.10\textwidth} >{\raggedright\arraybackslash}X >{\raggedright\arraybackslash}X >{\raggedright\arraybackslash}X @{}}
\caption{Fidelity of Analysis Definition}\label{tab:3}\\
\hline
\textbf{Fidelity Level} & \textbf{Model Content} & \textbf{Model Basis} & \textbf{Validation Approach} \\
\hline
\endfirsthead
\caption[]{(continued)}\\
\hline
\textbf{Fidelity Level} & \textbf{Model Content} & \textbf{Model Basis} & \textbf{Validation Approach} \\
\hline
\endhead
\hline
\multicolumn{4}{r}{\textit{(continued on next page)}}\\
\hline
\endfoot
\hline
\endlastfoot
Low & Key Performance Metrics (KPMs) that include critical parameters and physics contributing to analytical predictions. & Quantitative relationship between KPMs to predict values at one design point. Typically uses ``rules-of-thumb,'' ``empirical knowledge,'' or ``first order'' equations. & N/A \\
Medium & KPMs are traceable to error budgets in the test configuration. Incorporates critical parameters, physics, and life-limiting factors. Includes realistic components and interfaces. May represent multi-physics behavior and interactions. & Quantitative relationship between KPMs, critical parameters, and life-limiting factors based on established analytical principles, equations, and/or modeling tools. Capable of analyzing performance across various operational conditions and environmental factors. Includes understanding of parameter sensitivity to performance. Where applicable, defines deviations and variability from baseline properties. & Validate KPM predictions of the test configuration against test results in relevant environments. Demonstrate parameter sensitivity effects match analysis. Identify any relevant physics, parameters, and error contributors not present in the baseline model that affect the validation. Account for test measurement uncertainties. Quantify test-analysis deviations for error budget. Identify model applicability range and limitations. \\
High & Complete set of KPMs applicable to the test and allocated from the flight error budget. Incorporates critical parameters, physics, and life-limiting factors. Comprehensive system models of components and detailed interfaces based on flight designs. Represents integrated multi-physics behavior and interactions. & Quantitative relationship between all relevant KPMs, critical parameters, and life-limiting factors based on established analytical principles, equations, and/or modeling tools. Capable of analyzing performance across the full range of operational conditions and environmental factors. Provides statistical assessment of KPM and error contributor range of variability and performs uncertainty quantification (UQ). Incorporates model improvements, parameters, and interfaces measured at lower assembly levels along with observed variations. & Validate KPM predictions of the system against technology tests in relevant environments. Demonstrate parameter sensitivity effects match analysis. Confirm all relevant physics, parameters and error contributors are accounted for in the model. Validation accounts for contributions from all test measurement uncertainties including instrument calibrations, error sources, and environments. Ensure test-analysis correlation of technology model in flight configuration stays within flight error budget allocations derived from statistical variations and UQ. Document applicability range and limitations comprehensively. \\
\end{xltabular}
}
\subsubsection{Fidelity of Build}
Table 4 provides further detail on the ``Fidelity of Build'' column of Table 2.

{\footnotesize
\begin{xltabular}{\textwidth}{@{} >{\raggedright\arraybackslash}p{0.10\textwidth} >{\raggedright\arraybackslash}X >{\raggedright\arraybackslash}X >{\raggedright\arraybackslash}X >{\raggedright\arraybackslash}X >{\raggedright\arraybackslash}X @{}}
\caption{Fidelity of Build definitions}\label{tab:4}\\
\hline
\textbf{Unit} & \textbf{Unit} & \textbf{Purpose} & \textbf{Performance / Function} & \textbf{Form and Fit / Scaling} & \textbf{Environmental Requirements} \\
\hline
\endfirsthead
\caption[]{(continued)}\\
\hline
\textbf{Unit} & \textbf{Unit} & \textbf{Purpose} & \textbf{Performance / Function} & \textbf{Form and Fit / Scaling} & \textbf{Environmental Requirements} \\
\hline
\endhead
\hline
\multicolumn{6}{r}{\textit{(continued on next page)}}\\
\hline
\endfoot
\hline
\endlastfoot
Technology & Breadboard & Proof-of-concept for a potential design & Demonstrate performance / function & Not required & Tested in laboratory environment \\
Technology & Brassboard & Demonstrate feasibility of form and fit, environments & Demonstrate performance / function & Approximate, with scaling factors understood & Designed to meet relevant environment requirements \\
Technology & Prototype & Representative design, pathfinder, or demonstrator & Tested to meet performance / function requirements & Representative with scaling factors understood & Tested to meet relevant environment requirements \\
Engineering & Engineering Unit & Finalize detailed design & Tested to meet performance / function requirements & Exact as known at time of build & Tested to meet relevant environment requirements \\
Engineering & Qualification Unit & Qualify design & Tested to meet performance / function requirements & Exact as known at time of build & Tested to meet flight qualification environmental requirements \\
Engineering & Flight Unit & Final product & Tested to meet performance / function requirements & Exact & Tested to meet flight qualification environmental requirements \\
Engineering & Flight Spare & Final product & Tested to meet performance / function requirements & Exact & Tested to meet flight qualification environmental requirements \\
\end{xltabular}
}
\subsubsection{Gap Definition}
\begin{figure}[htbp]
\centering
% TODO: convert EMF to PDF/PNG for LaTeX: figures/fig030.emf
\includegraphics[width=0.9\linewidth]{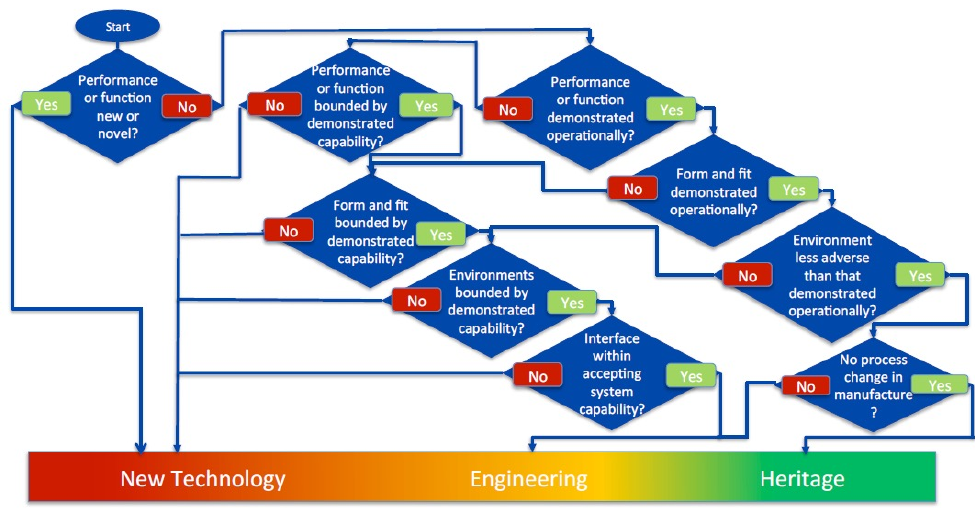}
\caption{Flow chart to determine the classification of a technology. Repeated here from SP-20205003605, Technology Readiness Assessment Best Practices Guide.}
\label{fig:3}
\end{figure}

The spectrum from new technology development to engineering to heritage is continuous, and there will always be a component of judgement associated with classifying a gap as one or another. However, clear definitions that can be applied consistently are critical to continually assessing progress towards achieving technical readiness. Figure \ref{fig:3} outlines the decision process for classifying a technology as new, engineering, or heritage.

\paragraph{New Technology Development}
A technology is defined as ``new'' if:

\begin{itemize}
\item Its application is new or novel, or
\item Its application exceeds its demonstrated performance or functional capability, or
\item Its application's fit or form exceeds previously demonstrated capability, or
\item Its application's integration needs exceed previously demonstrated capability.
\end{itemize}
\paragraph{Engineering Development}
A technology is defined as engineering if its development requires the use of existing, well understood components, techniques, and processes whose application is within design intention or demonstrated capability.

\paragraph{Heritage}
A technology is considered heritage if it has been used successfully in operation and:

\begin{itemize}
\item It is applied to its new use with no change to its fit, form, or function and:
\item The environments to which it will be exposed in its new application are no more adverse than those for which it was originally qualified, and
\item There have been no process changes in its manufacturing.
\end{itemize}
\subsubsection{Gap Prioritization}
\begin{figure}[htbp]
\centering
\includegraphics[width=0.7\linewidth]{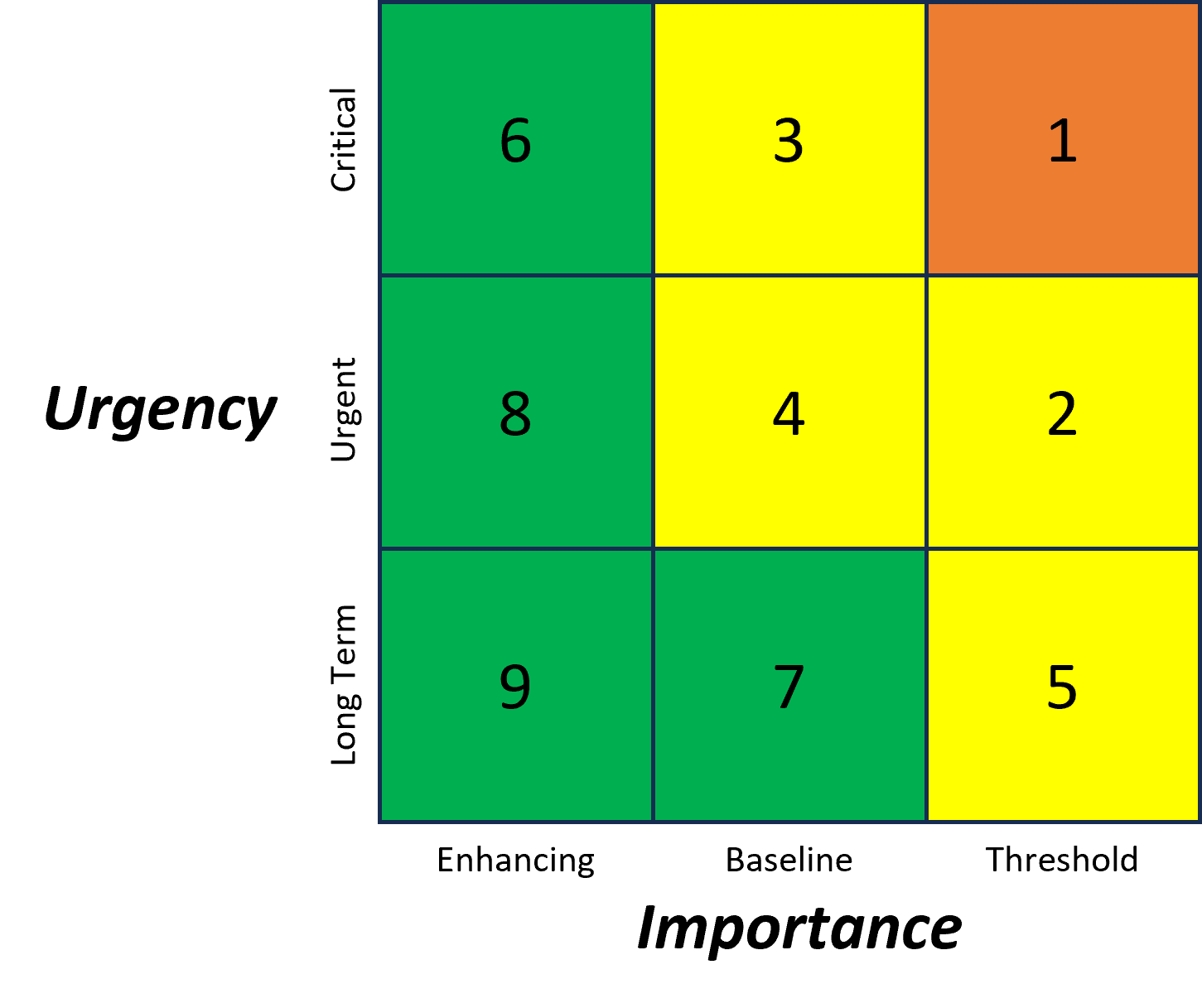}
\caption{Gap prioritization matrix. Each gap's overall priority is determined by the cell's numerical value. For example, a Threshold/Urgent gap is higher priority than a Baseline/Critical gap.}
\label{fig:4}
\end{figure}

In addition to classification as new technology, engineering, or heritage, gaps are also prioritized in terms of both importance and urgency. Importance gauges how enabling the technology is for HWO:

\begin{itemize}
\item \textbf{Threshold} technologies are those that are operationally critical to the HWO mission. If the threshold technologies are not matured, the HWO mission cannot proceed. Generally, these technologies do not have ``offramps,'' i.e., there are no suitably mature technologies with acceptably relaxed performance that could be adopted in their place and achieve HWO threshold science objectives.
\item \textbf{Baseline} technologies are those that enable the baseline HWO mission science objectives. Baseline technologies generally have offramps, such that if their development poses unacceptable cost or schedule risk, a state-of-the art backup could be used that would still enable HWO threshold science objectives. In such a scenario, the baseline technology would become a candidate for inclusion on a future servicing mission.
\item \textbf{Enhancing} technologies are those that are not fundamentally necessary to enable the HWO mission. Instead, enhancing technologies are those that would either substantially improve the science yield of the mission beyond the baseline objectives, or ease the development, implementation, test, and verification of the mission system.
\end{itemize}
Urgency gauges the impact of the technology's development to either the pre-formulation critical path (i.e., readiness of the technology by MCR), or to the mission design process:

\begin{itemize}
\item \textbf{Critical} technologies are those that require substantially complex development or consist of long duration activities that must begin early. Critical technologies are also those where the performance must be well understood early enough to inform the mission design activities being conducted during pre-formulation.
\item \textbf{Urgent} technologies are those that are at risk of not achieving TRL 5 by MCR or shortly thereafter. The specific performance capabilities of these technologies are not generally needed to inform pre-formulation mission design activities.
\item \textbf{Long-Term} technologies are those that can be matured later in the pre-formulation or formulation process as they pose low cost or schedule risk to the project.
\end{itemize}
Each technology gap's prioritization in each dimension is captured on a 3x3 matrix, shown in Figure \ref{fig:4}. The gap's overall prioritization is provided by the numerical value of its cell within the matrix. Technologies located in the Threshold/Critical cell (i.e., priority 1) are equivalent to Critical Technology Elements identified in the Technology Readiness Assessment Best Practices Guide.

\subsection{Technology Assessment Overview}
\begin{figure}[htbp]
\centering
\includegraphics[width=0.9\linewidth]{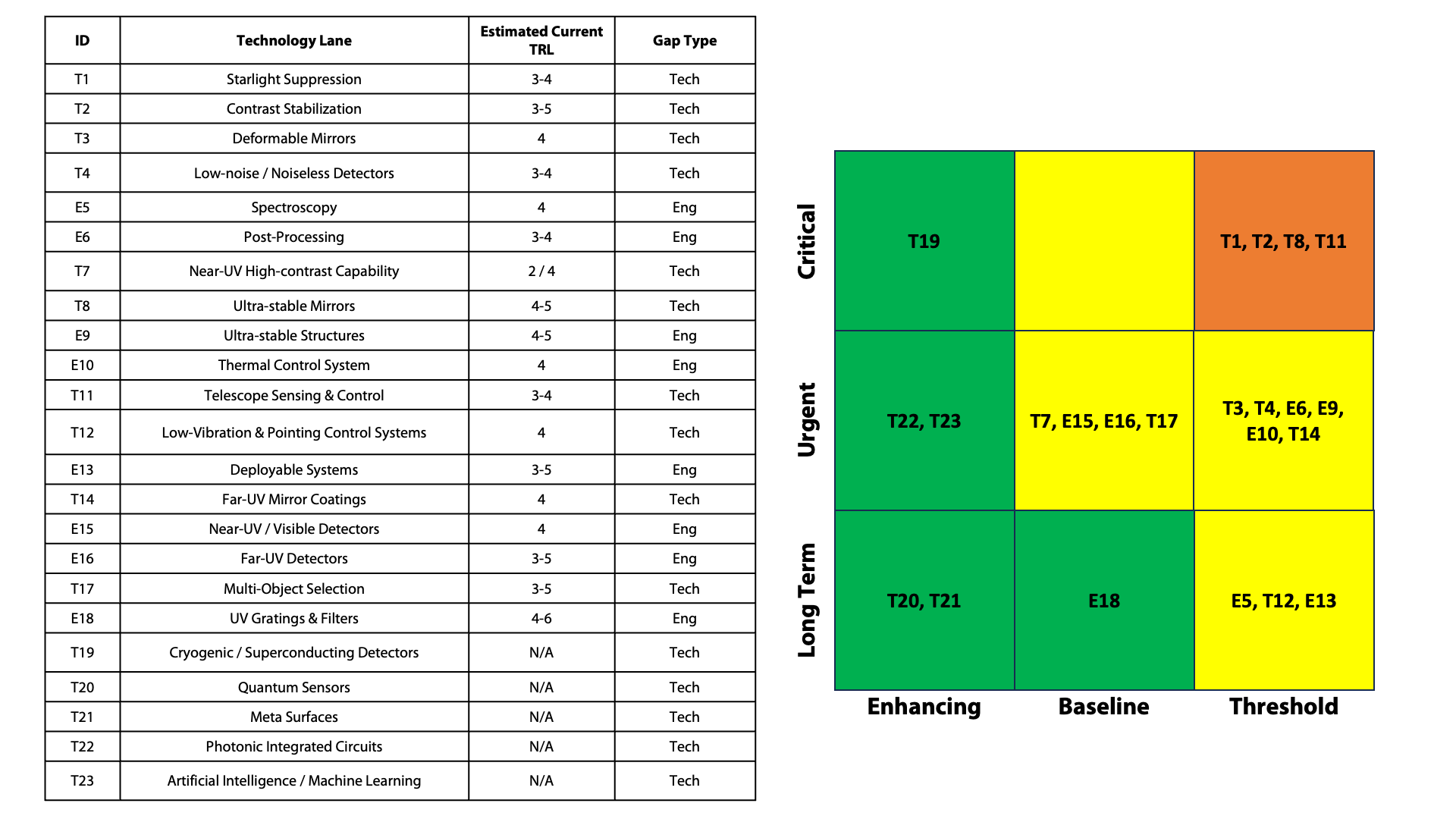}
\caption{Summary of all HWO Technologies and their prioritization.}
\label{fig:5}
\end{figure}

Figure \ref{fig:5} summarizes all of the technologies, including emerging enhancing technologies, discussed in this document along with their relative prioritization. A ``T'' prefix indicates a technology gap and an ``E'' prefix indicates an engineering gap.

%\section{<Insert Section on Roman \& CGI here>}
\section{Coronagraph System Technologies}
\subsection{Overview}
\begin{figure}[htbp]
\centering
\includegraphics[width=0.9\linewidth]{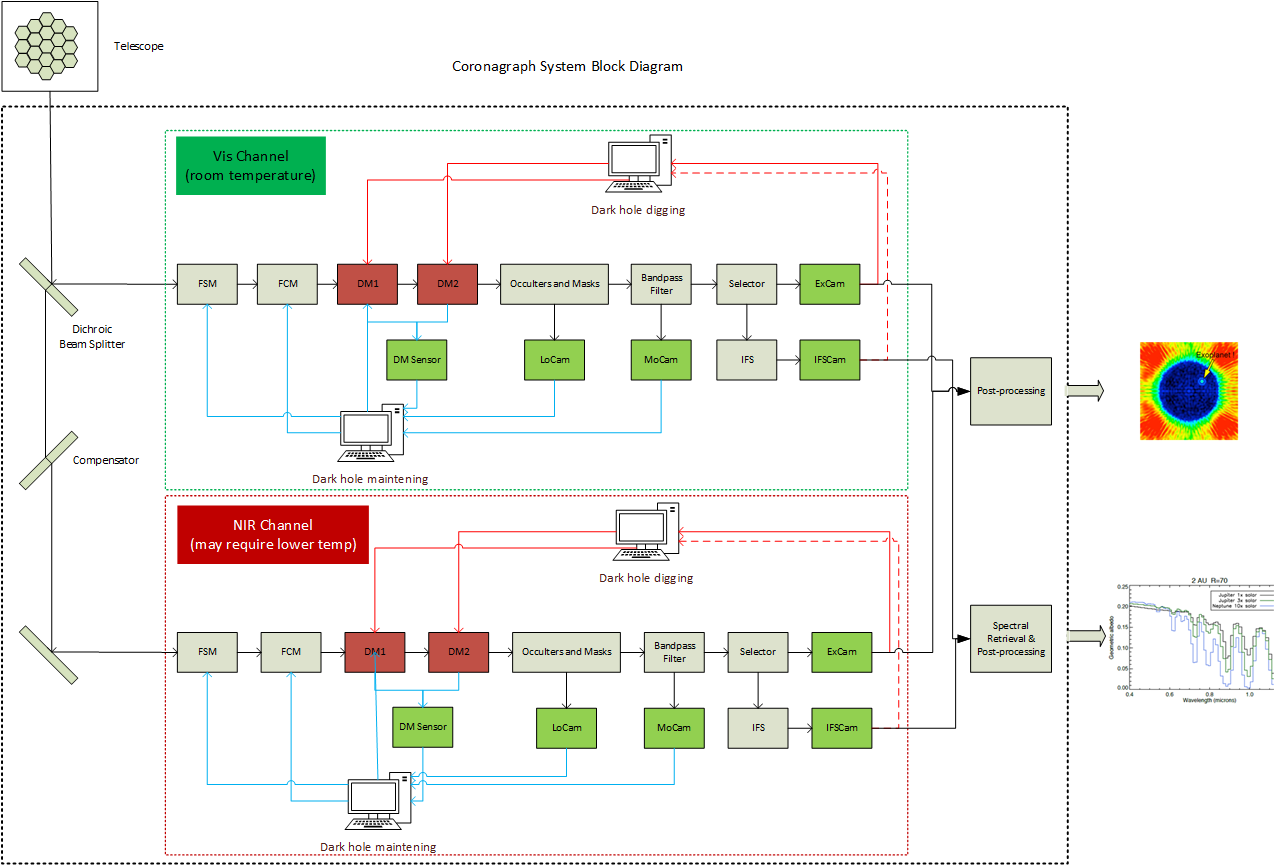}
\caption{Representative coronagraph system block diagram, showing (1) star-light suppression control loop (a.k.a., dark hole digging), and (2) dark hole maintenance control loop using wavefront sensing and control over low-to-mid spatial frequencies. Functionally, the VIS and NIR channels are identical.}
\label{fig:6}
\end{figure}

The Coronagraph System Track refers to the back-end coronagraph instrument (CI) suite that conducts exoplanet direct imaging and spectroscopic characterizations. A notional coronagraph system block diagram is shown in Figure \ref{fig:6}, in which light from the host star and exoplanets is collected by the HWO telescope and relayed to the CI. Current architectures for the HWO CI include at least two separate channels split by wavelength, often with one channel operating in the visible and one in the near infrared (NIR) to cover the desired wavelength range from 450 nm to 1.7 \textmu{}m. A set of dichroic beam splitters and compensators are used to split broadband light into visible and NIR channels. Each coronagraph channel has its own sets of coronagraph masks/occulters and associated deformable mirrors to perform the function of starlight suppression. This function is also known as high-order wavefront sensing and control (HOWFS), since it involves high actuator-count deformable mirror pairs to correct for high order wavefront errors of the transmitted starlight. In addition, each channel has its own low-to mid-order wavefront sensing and control (L/MOWFS) system to cancel time-varying wavefront error (disturbances and drifts) to maintain the dark hole. Because the exoplanets are extremely faint, we need high quantum efficiency, near-zero-noise photon-counting focal plane arrays to detect and measure the exoplanet spectrum.

\begin{figure}[htbp]
\centering
\includegraphics[width=0.9\linewidth]{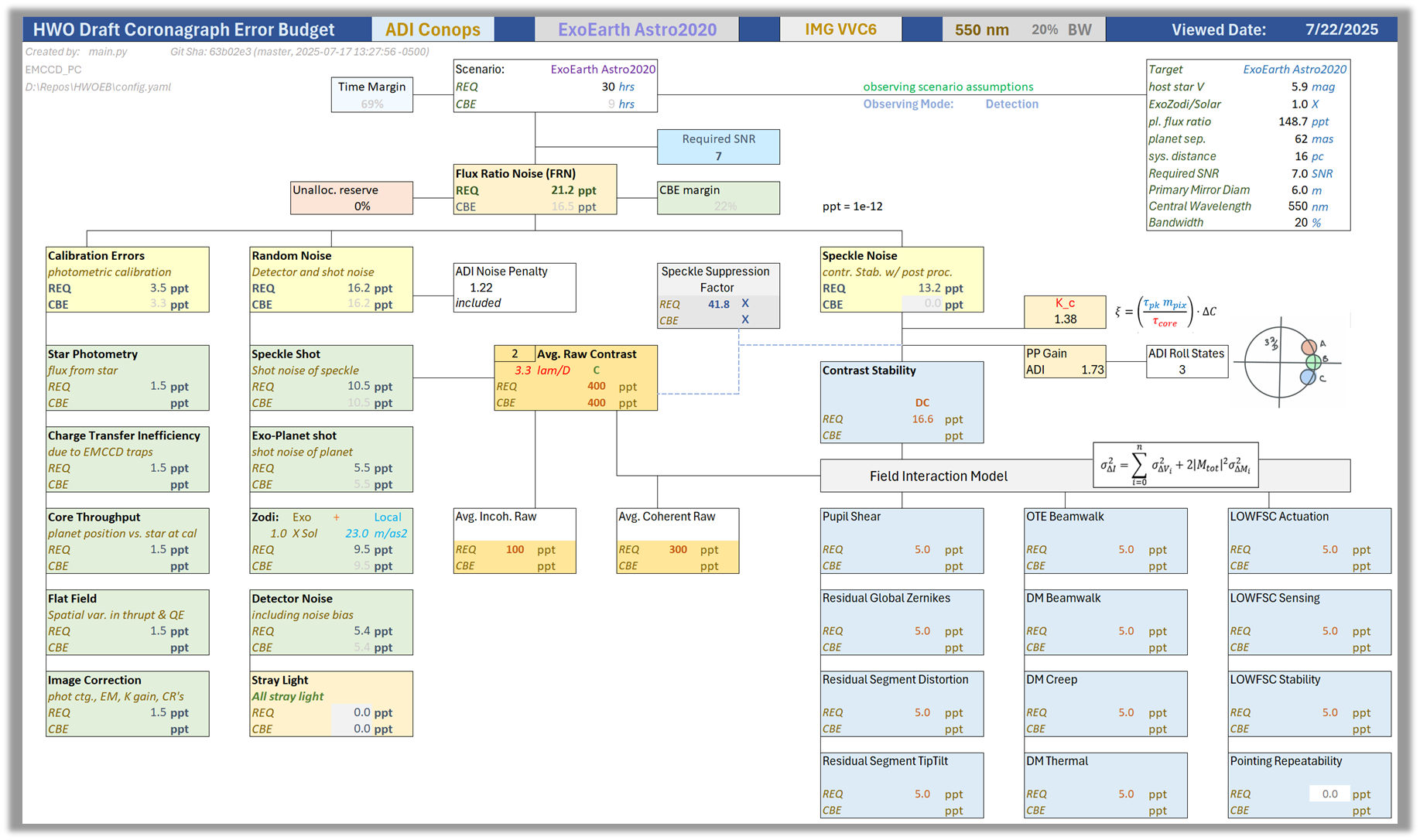}
\caption{Notional CI performance budget, sub-allocating contrast performance (in terms of flux ratio) to calibration, photometric, and contrast stability errors. Formal performance budgets are maintained by the HWO Systems Engineering Team.}
\label{fig:7}
\end{figure}

To help identify performance gaps and new technology items, we adopt a generic performance budget based on the Roman Space Telescope (``Roman'') coronagraph instrument (CGI) and large mission studies such as LUVOIR\cite{LUVOIR} and HabEx\cite{HABEX}. Figure \ref{fig:7} shows a notional HWO exoplanet imaging performance budget\cite{Nemati2024}. Note the top-level performance is broken down to branches that can be traced to lab demonstrations:

\begin{enumerate}
\item Static raw contrast. This is a measure of the residual starlight after suppression with the high-order wavefront sensing and control (HOWFS) loop using a pair of deformable mirrors and specially designed coronagraph masks and occulters. Typically, the raw contrast performance requirement for detecting an earth-like planet around a sun-like star is $\sim$10{\textsuperscript{-10}} at an angular separation $\sim$3 $\lambda$/D for visible wavelengths. This level of suppression must be kept stable for long periods to enable detection of faint planets.
\item Contrast stability. This is the stability of the residual starlight, enabling the isolation of much fainter exoplanets. The faintness of the exoplanet requires a long exposure or integration times to reach the desired exoplanet SNR. During an observation, other techniques (e.g., roll maneuvers or optical coherence measurements) are used to distinguish exoplanet light from the residual starlight. The contrast stability is directly traced to end-end optical beam train stability, from the telescope to the coronagraph instrument. The required stability in contrast is typically on the order of 10{\textsuperscript{-11}}, which translates to optical wavefront stability as small as a few picometers for the most critical spatial frequencies. The exposure/integration time for an exoplanet spectrum is multiple hours to days and perhaps even weeks. To achieve such tight stability over such long time, active wavefront sensing and control is used, in both the front-end telescope and the back-end coronagraph system. In the coronagraph, rejected starlight is used to sense the incoming wavefront pointing jitter and wavefront drifts over time. A separate low- and mid-order wavefront sensing and control (L/MOWFS) arm accomplishes this task.
\item Detector noise. Two science detectors are used in each coronagraph channel, one for direct imaging and one for measuring the exoplanet spectra using an integral field spectrograph (IFS) as shown in Figure \ref{fig:6}. Each coronagraph channel (visible, NIR) requires its own optimized photon-counting detectors with high quantum efficiency and near-zero noise performance. The IFS detector has a tighter SNR requirement because the exoplanet light is dispersed across several pixels. Based on current estimates of the spectral resolving power required, the minimum detector format is $\sim$4K $\times$ 4K. Engineering detectors used in the L/MOWFS loop may also require detectors beyond today's state-of-the art, resulting in new technology development needs. This technology development plan includes all detectors within the coronagraph instrument.
\item Post-processing gain. The post-processing gain is the improvement in planet sensitivity beyond the static raw contrast by data analysis. The achievable post-processing gain will ultimately be limited by the residual contrast stability after wavefront control, as well as the observing Concept of Operations (ConOps). Data processing approaches and algorithms have been developed to identify and subtract the residual starlight after wavefront control. Different ConOps and instrument hardware implementations provide data that can be used to estimate the residual starlight. Advanced ConOps and instrument calibrations are to be developed in parallel with observatory concept definition and science requirements. This post-processing development plan focuses on simulations with flight-like models and high-contrast laboratory demonstrations.
\item Spectroscopy. Many of the exoplanet science cases rely on reflected light spectroscopy at high contrast. The spectrograph needs to be capable of obtaining spectra at high contrast in the presence of the residual starlight and astrophysical backgrounds such as exozodiacal scattered light. Integral field spectroscopy, through traditional spectrograph designs or energy resolving detectors, may enable advanced wavefront sensing and control, post processing techniques, and simultaneous spectra of each spatial resolution element within the scene for scientific purposes. The post-processed spectrum must be calibrated to sufficient quality for spectral retrieval analysis.
\end{enumerate}
Initial Coronagraph System Track lanes are shown in Table 5. Similar new technology items have been identified and tracked by previous Exoplanet Program Office\footnote{https://science.nasa.gov/astrophysics/programs/exep/technology/gap-lists/} and LUVIOR/HabEx studies.

{\footnotesize
\begin{xltabular}{\textwidth}{@{} >{\raggedright\arraybackslash}p{0.10\textwidth} >{\raggedright\arraybackslash}X >{\raggedright\arraybackslash}X >{\raggedright\arraybackslash}X >{\raggedright\arraybackslash}X >{\raggedright\arraybackslash}X >{\raggedright\arraybackslash}X @{}}
\caption{Coronagraph System Technology Assessment. For importance and urgency ratings, refer to Section 3.1.4}\label{tab:5}\\
\hline
\textbf{ID} & \textbf{Lane} & \textbf{Driving Factor(s)} & \textbf{Est. Current TRL} & \textbf{Importance} & \textbf{Urgency} & \textbf{Gap Type} \\
\hline
\endfirsthead
\caption[]{(continued)}\\
\hline
\textbf{ID} & \textbf{Lane} & \textbf{Driving Factor(s)} & \textbf{Est. Current TRL} & \textbf{Importance} & \textbf{Urgency} & \textbf{Gap Type} \\
\hline
\endhead
\hline
\multicolumn{7}{r}{\textit{(continued on next page)}}\\
\hline
\endfoot
\hline
\endlastfoot
1 & Starlight Suppression & Component design and fabrication to achieve 10{\textsuperscript{-10}} raw contrast, bandwidth, inner working angle, etc. & 3-4 & Threshold & Critical & Tech \\
2 & Contrast Stabilization & Maintain 10{\textsuperscript{-11}} contrast stability during observations. & 3-5 & Threshold & Critical & Tech \\
3 & Deformable Mirrors & High actuator-count with stable, smooth surface. \newline Robust, precision electronics and interconnects. & 4 & Threshold & Urgent & Tech \\
4 & Low-noise / Noiseless Detectors & Photon-counting capability with low dark current and clock-induced charge. \newline Radiation hard. \newline High-QE at relevant biomarker wavelengths. & 3-4 & Threshold & Urgent & Tech \\
5 & Spectroscopy & Demonstrate wavefront control at high contrast through the spectrograph. \newline Preserve spectrophotometric precision in high dynamic range residual starlight field. \newline Maintain acceptable exposure times to achieve the desired SNR at required spectral resolution (R) value. & 4 & Threshold & Long Term & Eng \\
6 & Post-processing & Achieve desired planetary SNR in context of ConOps, sensing and control, and observatory stability. \newline Preserve spectrophotometric precision of post-processed planet image. & 3-4 & Threshold & Urgent & Eng \\
7 & Near-UV Capability & Achieve high-contrast photometry between 250-450 nm for critical ozone features. & 2 / 4 & Baseline & Urgent & Tech \\
\end{xltabular}
}
\subsubsection{Coronagraph Technology Development Philosophy}
Our general approach is to compress the technology development schedule by pursuing several development paths in parallel. The details of the work will be provided in subsequent sections that describe the scope, schedule, and key milestones, but the broad directions of our efforts are:

\begin{itemize}
\item Coronagraph instrument requirement definition, design, and modeling will evolve in response to the improved knowledge of the telescope and spacecraft environment, interfaces, and operational scenarios as the HWO Systems team continues to explore architectures and designs. Science output modeling will be continuously matured as well, along with our understanding of spectrum retrieval and post-processing algorithms and their capabilities. Science requirements and spacecraft constraints will determine coronagraph Level 1 requirements, which will flow down to the subsystem and component levels. The instrument design will evolve to meet these requirements, maximize the science output, and minimize cost and complexity.
\item Coronagraph testbed demonstrations and modeling will demonstrate the required performance in terms of raw contrast and throughput as a function of the inner working angle, bandwidth, aperture geometry, etc. They will also be used to validate the coronagraph instrument performance models. The coronagraph testbeds will be built and tested step by step to ensure traceability of demonstrated performance from testbeds to the flight instrument.
\item Key component development, maturation, and environmental testing. There are several coronagraph components that incorporate new technology and thus need to be matured. They include masks and apodizers with the associated mechanisms, the IFS detector, and  the deformable mirrors. Special optical elements such as dichroic or polarization beamsplitters, super-polished relay optics, polarization aberration compensators, high out-band rejection bandpass filters, low-scintillation glass, for example, need to be assessed and developed. Performance and environmental testing, along with reliability analysis, will be performed as needed to retire the greatest risks.
\item Coronagraph subsystems that will eventually become a part of the coronagraph instrument, such as the L/MOWFS and IFS will be built and have their key requirements validated in stand-alone testbeds, and later incorporated into the end-to-end coronagraph systems testbed.
\item Algorithm maturation is planned for both the DM wavefront control algorithms used to suppress the starlight and the post-processing algorithms that will allow contrast enhancement through means such as point-spread function (PSF) subtraction and data editing.
\end{itemize}
The following sections provide more detail on each coronagraph system technology lane.

\subsection{Starlight Suppression}
\textbf{Gap Type:} 	Technology

\textbf{Est. TRL:} 	3 -- 4

\textbf{Importance:}\textbf{	}  3 -- Threshold

\textbf{Urgency:}\textbf{	}  3 -- Critical

\subsubsection{Discussion}
\begin{figure}[htbp]
\centering
\includegraphics[width=0.9\linewidth]{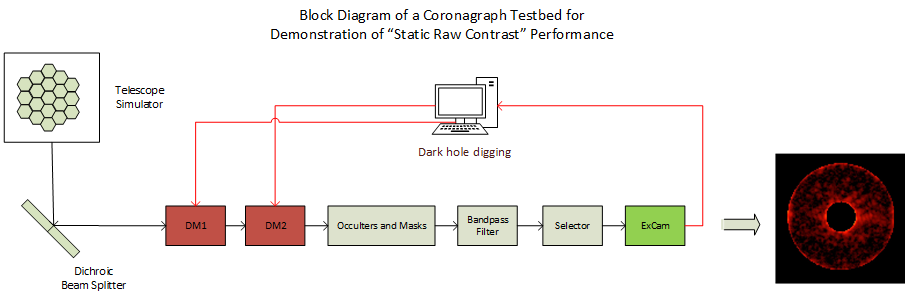}
\caption{A notional block diagram for the core starlight suppression system.}
\label{fig:8}
\end{figure}

Starlight suppression using HOWFS is a critical technology element for HWO. A notional block diagram representing starlight suppression is shown in Figure \ref{fig:8}. A set of coronagraph masks, occulters, and apodizers are inserted in the beam train using precision alignment mechanisms. Electric field estimation at the exoplanet science camera (ExCam) is used to estimate the high-spatial-order electric field errors (phase and amplitude). A pair of deformable mirrors (DMs) are used to control and compensate for the phase and amplitude errors of the incoming broadband star light from the telescope.

\subsubsection{TRL 5 Milestone Definition}
The TRL 5 milestone metric for the starlight suppression technology lane is to demonstrate the error budget allocation of static raw contrast is satisfied. Table 6 shows the expected key performance metrics for starlight suppression TRL 5 milestone:

{\small
\begin{xltabular}{\textwidth}{@{} >{\raggedright\arraybackslash}X >{\raggedright\arraybackslash}X >{\raggedright\arraybackslash}X >{\raggedright\arraybackslash}X @{}}
\caption{Key Performance Metrics for the Starlight Suppression TRL 5 Milestone}\label{tab:6}\\
\hline
\textbf{Performance metric} & \textbf{Value} & \textbf{Unit} & \textbf{Notes} \\
\hline
\endfirsthead
\caption[]{(continued)}\\
\hline
\textbf{Performance metric} & \textbf{Value} & \textbf{Unit} & \textbf{Notes} \\
\hline
\endhead
\hline
\multicolumn{4}{r}{\textit{(continued on next page)}}\\
\hline
\endfoot
\hline
\endlastfoot
Center Wavelength & 570 & nm &  \\
Spectral Bandwidth & $\geq 20\%$ at center wavelength &  &  \\
Inner Working Angle (IWA) & $\sim$3 & $\lambda$/D & Angular separation at which core throughput is 50\% of maximum \\
Outer Working Angle (OWA) & $\sim$9 & $\lambda$/D & Limited by existing 48$\times$48 actuator DMs in the testbed \\
Coherent Raw Contrast & $\leq$ 4$\times$10{\textsuperscript{-10}} average between 3-5 $\lambda$/D \newline $\leq$ 2$\times$10{\textsuperscript{-10}} average between 5-9 $\lambda$/D &  & Scaled with differences between testbed vs flight, for example 48$\times$48 actuator DMs in the testbed vs 96$\times$96 for flight \\
Max. Core Throughput & $>$10 & $\%$ & Amount of energy in planet PSF relative to energy incident on coronagraph aperture. \\
\end{xltabular}
}
\textbf{Significance:} This milestone demonstrates technical feasibility of the optical components and wavefront control system to achieve the required raw contrast. Raw contrast must be demonstrated at working angles consistent with coronagraph science requirements.

\textbf{Verification Method:} Starting with legacy testbeds from Roman CGI and the High-contrast Imaging Testbed (HCIT), re-use key hardware such as the 48$\times$48 actuator deformable mirrors made by AOA-Xinetics, use coronagraph models to map the performance predictions using flight-like 96$\times$96 actuator DMs.

We will use a step-by-step approach for this demonstration, starting with a monochromatic source and then expand the bandwidth step by step to $\geq$ 20\%. Each testbed dataset and its chromatic error will be delivered to the post-processing team who will use it to assess point-spread function (PSF) subtraction algorithms based on spectral diversity. The results achieved for this milestone will be used for optimizing wavefront control algorithms and for coronagraph model validation.

To demonstrate that reaching the required contrast is repeatable, this contrast demonstration will be performed at least 3 times, with DM voltages reset between each test.

A note on the planned testbed demonstration in the visible band: Past efforts on decadal survey testbeds, SATs, etc., have focused on visible wavelength performance, and there have been significant investments in visible wavelength testbeds. The choice of choosing a TRL 5 demonstration in visible is for both technical and cost reasons. If further HWO study shows that NIR channel performance becomes the driving performance, HWO TMPO will revisit the current plan and address the NIR channel requirements in a timely manner for TRL advancement.

\subsubsection{Development Strategy \& Roadmap}
\begin{figure}[htbp]
\centering
\includegraphics[width=0.9\linewidth]{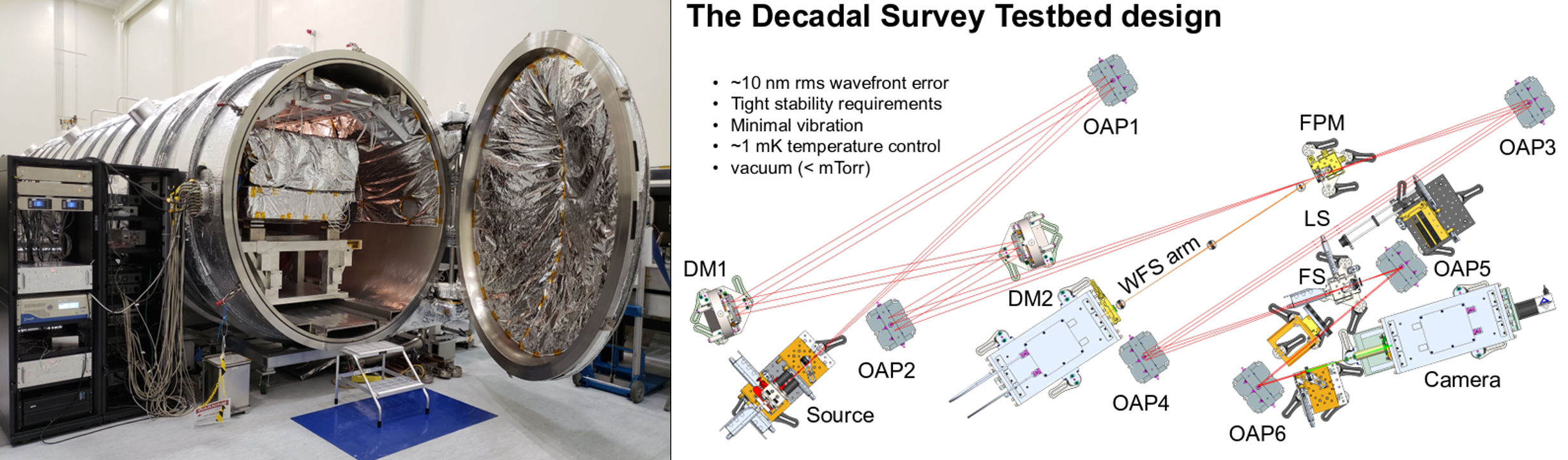}
\caption{The Decadal Survey Testbed (DST).}
\label{fig:9}
\end{figure}

HWO TMPO will use a dedicated testbed, the Exoplanet Imaging Coronagraph for TRL 5 (EPIC-5), for this demonstration. EPIC-5 will be based on earlier investment in decadal testbeds such as HCIT and the Decadal Survey Testbed (DST), shown in Figure \ref{fig:9}.

The plan is based on a step-by-step strategy to achieve the TRL 5, with key steps listed below in a sequential order::

\begin{enumerate}
\item Unobscured pupil testbed noise floor improvement -- demonstrate required testbed noise floor for later demonstration of an HWO-traceable coronagraph design.
\item HWO coronagraph mask design, fabrication and characterization -- design and produce testbed-worthy masks and occulters for testbed demonstration with a segmented aperture.
\item Narrowband (1\%) testbed demonstration -- demonstrate narrowband sensing capability required for later broadband wavefront control.
\item Broadband (20\%) testbed demonstration -- demonstration of wavefront control to overcome chromatic issues due to segmented telescope pupil.
\end{enumerate}
\subsection{Contrast Stabilization}
\textbf{Gap Type:} 	Technology

\textbf{Est. TRL:}\textbf{	}  3 -- 5

\textbf{Importance:}\textbf{	}  3 -- Threshold

\textbf{Urgency:}\textbf{	}  3 -- Critical

\subsubsection{Discussion}
\begin{figure}[htbp]
\centering
\includegraphics[width=0.9\linewidth]{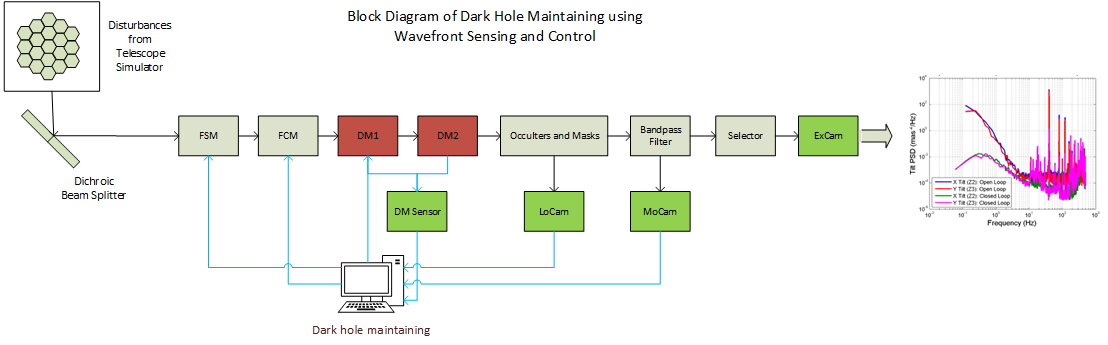}
\caption{A notional block diagram for the coronagraph contrast stabilization sub-system.}
\label{fig:10}
\end{figure}
The contrast stabilization subsystems sense and correct wavefront error disturbances and drifts due to spacecraft vibrations and temperature changes over time. A notional block diagram for the subsystem is shown in Figure \ref{fig:10}. At higher temporal frequencies (above $\sim$50 Hz), the line-of-sight jitter will likely be addressed mechanically with isolation, damping, and low-disturbance attitude control systems. Residual wavefront jitter and drifts from a controlled ultra-stable telescope will be addressed inside the coronagraph instrument with its own wavefront sensing and control strategy. For example, rejected starlight can be directed to a Zernike wavefront sensor for low-order wavefront sensing (LOWFS)\cite{Shi2016}. Residual line-of-sight pointing jitter can be controlled with a Fast-Steering Mirror (FSM). And low- to mid-order wavefront drift terms at lower temporal frequencies can be corrected using DMs and actuated focus control optics. All of these techniques are currently used on the Roman CGI.

For slower thermal wavefront drift, the concept of Dark Zone Maintenance (DZM) can be adopted, using the science camera as a null sensor with feedback to the DMs to maintain the contrast \cite{Rigg2016, Redmond2022}. The in-band signal for DZM at the science camera is expected to be very faint, therefore a wavefront sensor using rejected starlight needs to be able to maintain the dark hole to within the DZM temporal control bandwidth. The TRL 5 milestone aims to demonstrate contrast stability consistent with a performance budget defined by the Systems Engineering Team. Since the instrument detection limit is driven by both the residual coherent raw contrast and contrast stability (multiplier to each other), the final performance requirement for both will be evaluated together. For now, we use a contrast stability goal of 1$\times$10$^{-11}$ between 3-5 $\lambda$/D to guide contrast stability technology planning.

\subsubsection{TRL 5 Milestone Definition}
The TRL 5 milestone metric for the contrast stabilization technology lane is to demonstrate the error budget allocation of contrast stability is satisfied. Table 7 shows the expected key performance metrics for starlight suppression TRL 5 milestone:

{\small
\begin{xltabular}{\textwidth}{@{} >{\raggedright\arraybackslash}X >{\raggedright\arraybackslash}X >{\raggedright\arraybackslash}X >{\raggedright\arraybackslash}X @{}}
\caption{Key Performance Metrics for the Contrast Stability TRL 5 Milestone}\label{tab:7}\\
\hline
\textbf{Performance metric} & \textbf{Value} & \textbf{Unit} & \textbf{Notes} \\
\hline
\endfirsthead
\caption[]{(continued)}\\
\hline
\textbf{Performance metric} & \textbf{Value} & \textbf{Unit} & \textbf{Notes} \\
\hline
\endhead
\hline
\multicolumn{4}{r}{\textit{(continued on next page)}}\\
\hline
\endfoot
\hline
\endlastfoot
Line-of-sight pointing jitter (Z2/Z3) loop & $\leq$ 0.5 & Milli-arcsec over 20 Hz & V=5 star. The final pointing jitter requirement is coronagraph dependent. \\
Contrast stability & $\leq$ 1$\times10^{-11}$ over 3-5 $\lambda$/D &  & Both tests and analysis will be used to roll up to the overall contrast stability \\
\end{xltabular}
}
\textbf{Significance:} This milestone demonstrates the technical feasibility of maintaining the wavefront stability (thus the dark hole stability) by using the wavefront sensing and control subsystem. This demonstration is in parallel to the ultra-stable telescope technology demonstration, where the ``front-end'' wavefront stability is demonstrated. The contrast stability testbed will use a telescope simulator to create wavefront disturbances (residual from the ``front end'') and demonstrates that the ``back end'' inside the coronagraph is capable of sensing and correcting residual wavefront errors from the front end. Demonstrating both the front-end and back-end wavefront stability performance constitutes the achievement of the ``contrast stability'' technology milestone.

\textbf{Verification Method:} This milestone will be validated in a dedicated contrast stability testbed. The validated system will include an optical telescope assembly (OTA) simulator that injects realistic pointing jitter and low- to high-order wavefront errors in the optical path. Coronagraph WFS\&C performs sensing and closed-loop suppression of pointing jitter and low order wavefront drift to levels that enable the coronagraph to meet its science requirements. DZM will use the in-band light on the science camera \cite{Redmond2022}. For the TRL 5 demonstration, we will focus on the technical feasibility demonstration using a bright star and available photon counting detectors in the testbed. For flight observations, we expect the dark zone maintenance bandwidth to be limited by the science star brightness, which will establish the stability duration upper bond for the rest of wavefront sensing and control loops. This dark zone maintenance demonstration may be carried out at the EPIC-5 testbed.

Sensing information is needed to enable a variety of data post-processing (PSF subtraction, data editing) algorithms. This task will work very closely with the post-processing team on required telemetry and verification tests.

\subsubsection{Development Strategy \& Roadmap}
The development strategy for the coronagraph contrast stabilization subsystem is  The process begins with coronagraph design and mask fabrications. Note the mask design includes necessary features for LOWFS and MOWFS. The Dark Hole Maintenance Testbed activities involve implementing the LOWFS and MOWFS through $\sim$year \#2. This leads to a closed-loop demonstration with an OTA simulator, aiming to achieve TRL-5 by end of this decade. Ultimately, these systems are integrated into the EPIC-6 testbed for system-level dynamics and stability demonstrations. Throughout the entire period, a Performance Model Validation track runs continuously, with major validations milestones following each testbed demonstrations.

\subsection{Deformable Mirrors}
\textbf{Gap Type:} 	Technology

\textbf{Est. TRL:}\textbf{	}  4

\textbf{Importance:}\textbf{ }	3 -- Threshold

\textbf{Urgency:} 	2 -- Urgent

\subsubsection{Discussion}
\begin{figure}[htbp]
\centering
\includegraphics[width=0.9\linewidth]{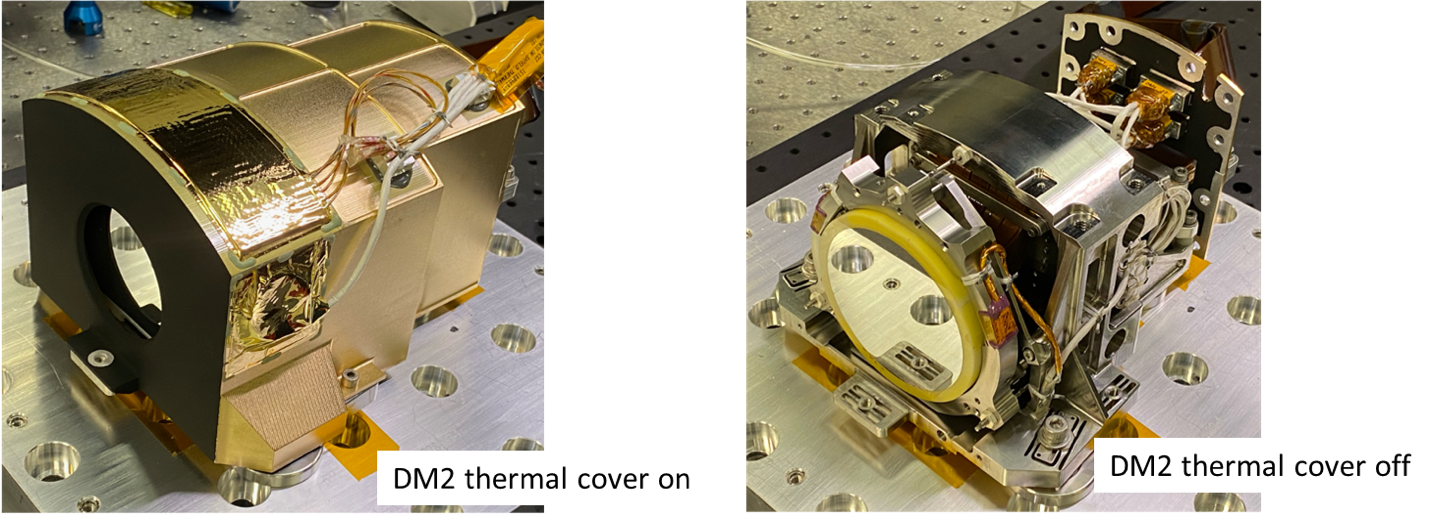}
\caption{Roman CGI AOA-Xinetics DM, with and without thermal cover.}
\label{fig:11}
\end{figure}

\begin{figure}[htbp]
\centering
\includegraphics[width=0.9\linewidth]{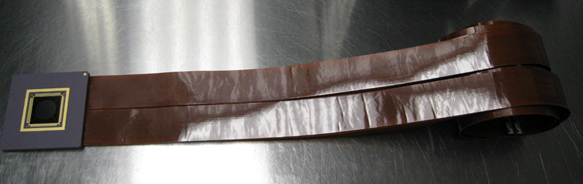}
\caption{BMC DM with flight-like interconnect used in environmental testing.}
\label{fig:12}
\end{figure}

High actuator count (up to 96 $\times$ 96) deformable mirrors are a key critical technology that enables high contrast coronagraphs for exo-Earth imaging. The Roman CGI technology program carried development for both AOA-Xinetics (AOX) DMs with 48 $\times$ 48 actuators and Boston Micromachines (BMC) DMs with 50 $\times$ 50 actuators. Figure \ref{fig:11} shows a flight Roman CGI DM with and without thermal cover, Figure \ref{fig:12} shows a BMC DM with flight-like interconnect that passed all the environmental tests. Larger-format DMs such as 64 $\times$ 64 have been used successfully in ground-based observatories. HWO likely requires even higher actuator count in order to correct for higher spatial wavefront error for larger dark holes and improved HOWFS performance.

\begin{figure}[htbp]
\centering
\includegraphics[width=0.9\linewidth]{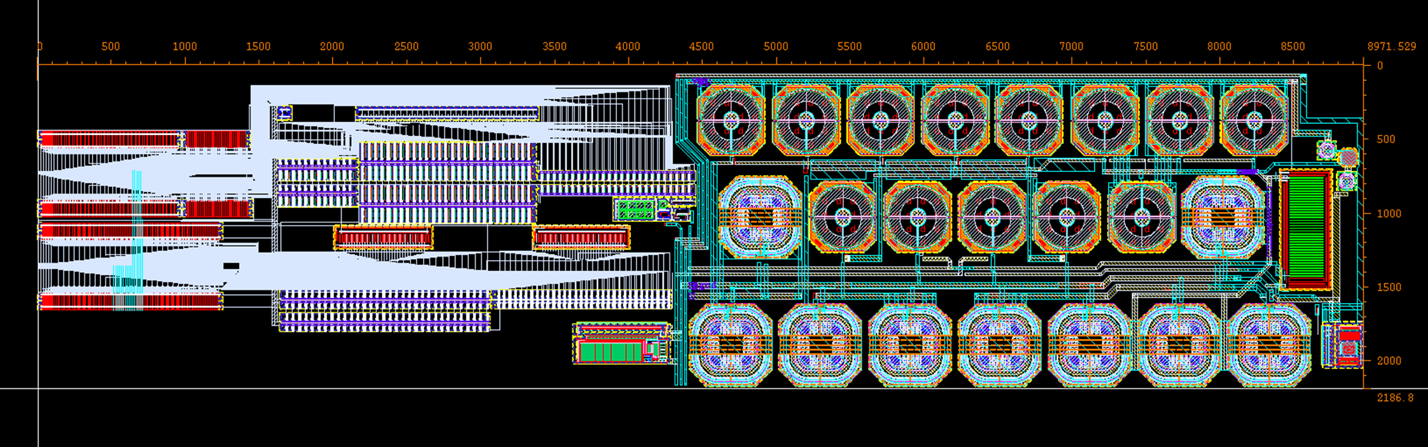}
\caption{Example DM ASIC driver designed to work with a Roman CGI DM.}
\label{fig:13}
\end{figure}

In addition to large format, HWO also requires precision control to be better than 1 picometer. To achieve this level of resolution, 20-bit digital-to-analog converters (DACs) are required, with a high-voltage amplifier with max voltage of $\sim$100 V \cite{Groff2024}. The DM driver electronics for Roman CGI used discrete components, an implementation that is large in volume and mass. Low mass, small volume, and highly reliable DM driver electronics based on application specific integrated circuits (ASICs) are required for HWO. Figure \ref{fig:13} shows 1-channel 16-bit DAC and 120V high voltage amplifier preliminary design for Roman CGI ASIC DM driver (2019). The ASIC is ``rad-hard by design'' to meet stringent radiation and reliability requirements. The 1-channel layout is about 2 mm $\times$ 9 mm, which allows packaging 16-channels into a standard 64-pin package. For HWO applications, the DAC would be modified from 16-bit to 20-bit to enable finer DM actuator control for deeper contrast.

Low-order and parabolic deformable mirrors also show promise in improving or alleviating observatory requirements. Closed-loop control concepts that stabilize the telescope wavefront using the high order deformable mirrors risk destabilizing their starlight suppression performance due to mid-spatial frequency residuals, inducing creep instability, and/or inducing gain variability in the actuators. A low-order flat or parabolic DM would stabilize the telescope wavefront in closed loop using separate sensing loops (such as Zernike or LOWFS) without imposing these challenges on the high-order DM. Parabolic DMs with sufficiently high actuator count (but $<$ 96 $\times$ 96) could additionally eliminate the need for an out-of-pupil 96 $\times$ 96 DM, or augment performance of HOWFS and LOWFS loops with the baseline 2-DMs in series \cite{Groff2016, Subedi2019, Subedi2021}.

\subsubsection{TRL 5 Milestone Definition}
The TRL 5 milestone definition for HWO high-actuator-count deformable mirror is summarized in Table \ref{tab:8}.

{\small
\begin{xltabular}{\textwidth}{@{} >{\raggedright\arraybackslash}X >{\raggedright\arraybackslash}X >{\raggedright\arraybackslash}X >{\raggedright\arraybackslash}X @{}}
\caption{Key Performance Metrics for the Deformable Mirror TRL 5 Milestone}\label{tab:8}\\
\hline
Performance Metric & Value & Unit & Notes \\
\hline
\endfirsthead
\caption[]{(continued)}\\
\hline
Performance Metric & Value & Unit & Notes \\
\hline
\endhead
\hline
\multicolumn{4}{r}{\textit{(continued on next page)}}\\
\hline
\endfoot
\hline
\endlastfoot
Actuator count & $\geq$ 64 $\times$ 64 \newline 96 $\times$ 96 (goal) &  & May be reduced based on parametric studies \\
Actuator max stroke & 500 & nm & Includes DM electronics contribution \\
Actuator pitch & $\leq$ 1 & mm &  \\
Actuator yield & 100\% within the inscribed circular clear aperture as defined by the telescope. &  & Includes DM electronics contribution over lifetime. \\
Actuator resolution & 2 & pm & Includes DM electronics contribution \\
Actuator stability \& drift & 5 & pm RMS per control cycle (target 1 hr) &  \\
Wavefront Error (WFE) & TBD & nm & At nominal bias voltage \\
Residual WFE & $\leq$1 & nm & WFE such as quilting that cannot be corrected by the DM itself \\
Mirror reflectivity & $>$98\% & Over each channel's bandwidth & Prefer protected Ag coatings for the visible channel; protected Au coatings for NIR channel. \\
\end{xltabular}
}
Low-order and parabolic DMs would carry many of the same requirements for a TRL 5 milestone with some changes in performance targets. Stability and residual WFE would have to hold the same target, but potentially only in closed-loop. Other requirements would have relaxed targets, specifically lower actuator count and larger actuator pitch (by design). With the low-order/parabolic DM operating directly in closed loop with a LOWFS and/or other observatory metrology inputs there is a credible path for stability (by far one of the most challenging requirements) to only be required in closed-loop operation.

\subsubsection{Development Strategy \& Roadmap}
The Exoplanet Program Office (ExEP) completed a deformable mirror industry survey in 2023 \cite{Groff2024} that surveyed the top DM suppliers based on experience and current capability. Under the HWO TMPO, we adopt the recommended path from ExEP and have issued study contracts to the top three vendors for their plan to develop DMs that meet HWO requirements. All three DM vendors delivered their final 96 $\times$ 96 DM Development and Manufacturing Plans. A kick-off meeting was held on February 18, 2025, with a multi-center DM Assessment Team to review and discuss the assessment process. The next step is to issue a request for proposals and have at least two vendors selected for further prototype development. From development risk perspective, it is too early to commit to only one DM vendor.

NASA will partner with industry to bring expertise that DM vendors don't have, such as flight interconnect development, optical metrology for DM performance testing To mitigate DM actuator drift and creep, a DM drift sensor is also being developed to close-loop control the DM actuators.

\subsection{Low-noise / Noiseless Detectors}
\textbf{Gap Type:} 	Technology

\textbf{Est. TRL:} 	3 -- 4

\textbf{Importance:}\textbf{	}  3 -- Threshold

\textbf{Urgency:}\textbf{	}  2 -- Urgent

\subsubsection{Discussion}
The detection of faint exoEarths within coronagraph dark zones also requires detectors with exquisite sensitivity and noise properties. Expected count rates from these objects imply the need for photon-counting sensitivity, and the specific biomarker signatures of interest set requirements on where in the spectral band quantum efficiency (QE) must be maximized. Fundamental to these detector performance characteristics is the detector's concept of operations (ConOps), or specifically how the detector is read out. Therefore, throughout this document the word ``detector'' is used generically to include the entire system chain, from the sensor head to the readout electronics that produce the digital signal image.

Science yield modeling is ongoing and will ultimately inform the threshold performance requirements for coronagraph imaging and spectroscopy. However, Table 9 summarizes the expected key performance metrics. Generally, these requirements hold regardless of whether the coronagraph visible or NIR channel is being considered.

\begin{xltabular}{\textwidth}{@{} >{\raggedright\arraybackslash}X >{\raggedright\arraybackslash}X @{}}
\caption{Notional HWO Coronagraph Detector System Requirements}\label{tab:9}\\
\hline
\textbf{Parameter} & \textbf{Performance Needed} \\
\hline
\endfirsthead
\caption[]{(continued)}\\
\hline
\textbf{Parameter} & \textbf{Performance Needed} \\
\hline
\endhead
\hline
\multicolumn{2}{r}{\textit{(continued on next page)}}\\
\hline
\endfoot
\hline
\endlastfoot
Format (pixels) & $\geq$1K $\times$ 1K (Imaging) \newline $\geq$4K $\times$ 4K (Spectroscopy) \\ \hline
Read noise & $<$ 0.1 e- (i.e., photon counting) \\ \hline 
Total Noise Rate (includes dark current, clock-induced charge, etc.) & $<$ 1$\times$10{\textsuperscript{3}} e-/p/s (detection) \newline $<$ 1$\times$10{\textsuperscript{5}} e-/p/s (spectroscopy) \\ \hline
QE & $\geq$ 90\% over detection band \\ \hline
Lifetime & $\geq$ 5 years at L2 environment \\ 
\end{xltabular}
A key discriminator in detector technologies under consideration is their operating temperature: those that can be operated with passive or thermal electric cooling (TEC), and those that require cryogenic, $<$1 K temperatures, i.e., super-conducting quantum sensors. The former category trades easier implementation on a near-room-temperature picometer-stable observatory for higher noise performance, potentially making the exoEarth detection and characterization problem more challenging. Conversely superconducting detectors are truly zero-noise devices with several options adding the benefit of energy resolution, negating the need for separate spectrograph channels within the coronagraph. But these detectors are generally less mature and come with significant implementation challenges including thermal isolation and multi-stage cooling that not-only drives size, weight, and power (SWaP) considerations, but also provides a source of significant dynamic instability in a system that requires picometer-level stability.

\paragraph{Current State-of-the-Art}
Table 10 summarizes several detector technologies under consideration and their current state-of-the-art performance. The first three columns (TES, MKID, SNSPD) are three superconducting technologies. MAS-CCD, SiSeRo CCD, EMCCD, and scientific CMOS are evolutions of silicon-based detectors that use special readout channels to perform photon counting. Any of these technologies can be ``UV Enhanced'' via processes such as delta-doping to push their operational bandwidth into the near-UV, however all generally cut-off around 1 micron at the red end. The last two columns, H4RG and LM-APD, represent NIR detector technologies, with only the LM-APD devices providing photon-counting performance.

\begin{landscape}
{\footnotesize
\begin{xltabular}{\linewidth}{@{} >{\raggedright\arraybackslash}p{0.12\linewidth} >{\raggedright\arraybackslash}X >{\raggedright\arraybackslash}X >{\raggedright\arraybackslash}X >{\raggedright\arraybackslash}X >{\raggedright\arraybackslash}X >{\raggedright\arraybackslash}X >{\raggedright\arraybackslash}X >{\raggedright\arraybackslash}X >{\raggedright\arraybackslash}X @{}}
\caption{State-of-the-Art Performance of Potential Detector Technologies}\label{tab:10}\\
\hline
 & \textbf{Transition Edge Sensor (TES)} & \textbf{Microwave Kinetic Inductance Detectors (MKIDs)} & \textbf{Super-conducting Nanowire Single-Photon Detector (SNSPD)} & \textbf{P-channel Multi-Amplifier Sensing (MAS) ``Skipper'' CCD} & \textbf{P-channel Single-Electron Sensitive Readout SiSeRo ``Skipper'' CCD} & \textbf{N-channel EMCCD} & \textbf{Single-photon Scientific CMOS} & \textbf{H4RG} & \textbf{Linear Mode Avalanche Photodiode (LM-APD)} \\
\hline
\endfirsthead
\caption[]{(continued)}\\
\hline
 & \textbf{Transition Edge Sensor (TES)} & \textbf{Microwave Kinetic Inductance Detectors (MKIDs)} & \textbf{Super-conducting Nanowire Single-Photon Detector (SNSPD)} & \textbf{P-channel Multi-Amplifier Sensing (MAS) ``Skipper'' CCD} & \textbf{P-channel Single-Electron Sensitive Readout SiSeRo ``Skipper'' CCD} & \textbf{N-channel EMCCD} & \textbf{Single-photon Scientific CMOS} & \textbf{H4RG} & \textbf{Linear Mode Avalanche Photodiode (LM-APD)} \\
\hline
\endhead
\hline
\multicolumn{10}{r}{\textit{(continued on next page)}}\\
\hline
\endfoot
\hline
\endlastfoot
\textbf{Form Factor (pixels)} & 2K $\times$ 2K & 256 $\times$ 256 & 500 $\times$ 800 & 1K $\times$ 1K & $<$ 1K $\times$ 1K & 4K $\times$ 4K & 2K $\times$ 2K -- 8K $\times$ 8K & 4K $\times$ 4K & 4K $\times$ 4K \\\hline
\textbf{Pixel Pitch (\textmu{}m)} & 6--20 & 150 & 10 & 9--15 & 9--15 & 4 & 7 & 10 & 18 \\
\textbf{OpTemp (K)} & 70--100 $\times$ $10^{-3}$ & 30 $\times$ $10^{-3}$ & 0.8 & $>$ 135 & $>$ 135 & 163--188 & 163--188 & 89 & $<$ 60 \\\hline
\textbf{Waveband (nm)} & 200--2500 & 400--1000 (200--1800 possible) & 250 nm -- 29~\textmu m & 450--980; down to 200~nm possible & 450--980; down to 200~nm possible & 120--1100 & 200--1100 & 480--2500 & 800--2500 \\\hline
\textbf{Spectral Res.} & UV: $R{=}180$, VIS: $R{=}90$, NIR: $R{=}30$ & $R{\sim}30$ (visible), $R{>}50$ (IR) & X & X & X & X & X & X & X \\\hline
\textbf{QE} & UV: 80--89\%, VIS: 99\%, NIR: 99\% & $>$70\% & 98\% (1550~nm); 80\% (370~nm) & $>$ 80\% & Unclear & 40\% (350~nm); 30\% (200~nm) & $>$ 60\% (200--400~nm) & $>$ 60\% (480--2500), $>$ 80\% (600--2400) & 85--90\% \\\hline
\textbf{Read Noise ($e^-$)} & 0 & 0 & 0 & 0.08 & $<$ 0.15 & $<$ 0.02 & $<$ 1 & $<$ 6.5~$e^-$/180~s & $<$ 10 \\\hline
\textbf{Dark Current ($e^-$/pix/s)} & 0 & $1.8 \times 10^{-3}$ (visible / NIR) & $6 \times 10^{-6}$ & $< 6.82 \times 10^{-9}$ & $< 6.82 \times 10^{-9}$ & $< 1.5 \times 10^{-5}$ & $< 1 \times 10^{-3}$ & $< 1 \times 10^{-3}$ & $\sim$ 0.01 \\\hline
\textbf{CIC ($e^-$/pix/frame)} & 0 & N/A & N/A & $1.52 \times 10^{-4}$ & $1.52 \times 10^{-4}$ & $6.9 \times 10^{-4}$ & Unknown & (included in read noise) & Unknown \\\hline
\textbf{Photon Counting} & $\checkmark$ & $\checkmark$ & $\checkmark$ & $\checkmark$ & $\checkmark$ & $\checkmark$ & $\checkmark$ & X & X \\\hline
\end{xltabular}
}
\end{landscape}

%\begin{figure}[htbp]
%\centering
%\includegraphics[width=0.9\linewidth]{figures/fig034.png}
%\caption{State-of-the-Art Performance of Potential %Detector Technologies}
%\label{fig:14}
%\end{figure}
The Systems Team is currently performing systems trade studies and evaluations to determine the viability of implementing a superconducting detector (with the associated cooling requirements) on HWO, either as part of the baseline mission or a future servicing mission. Until those evaluations are complete, this Technology Plan treats those options as potential enhancing technologies (see Section 7) and focuses the near-term development strategy on the technologies in columns 4 through 9.

\begin{figure}[htbp]
\centering
\includegraphics[width=0.9\linewidth]{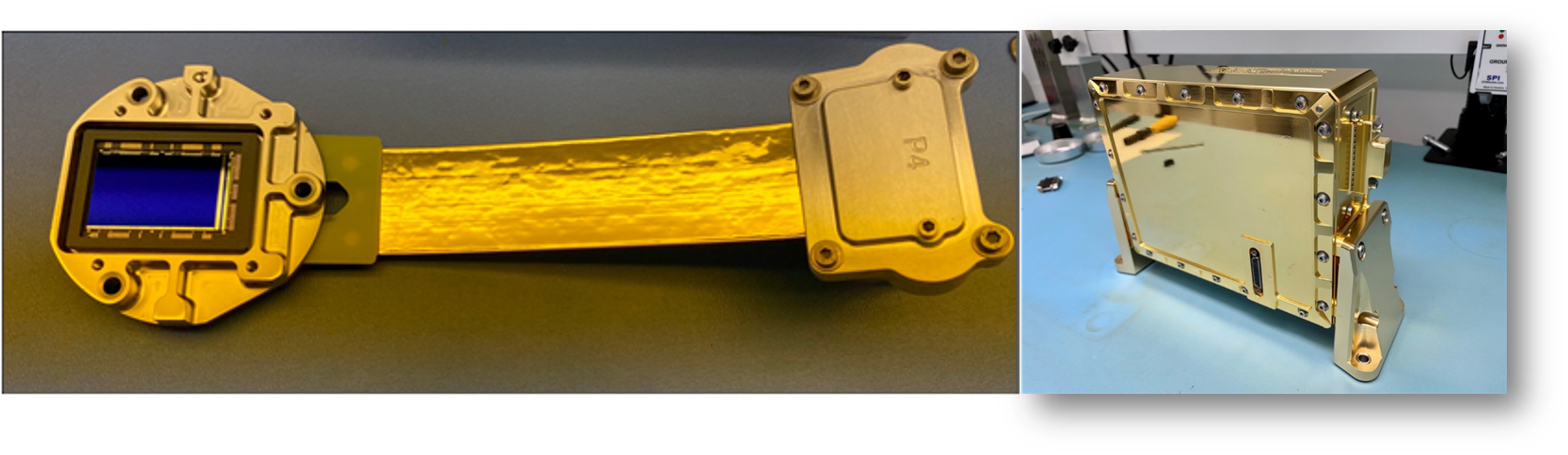}
\caption{CGI EMCCD (left) and proximity readout electronics (right).}
\label{fig:14}
\end{figure}

The Electron-Multiplying CCD (EMCCD) camera on the Roman CGI is currently the highest TRL photon-counting device in the visible band. Figure \ref{fig:14} shows the 1K $\times$ 1K flight device along with the proximity readout electronics. The same device is used for both the Exoplanet science Camera (ExCam) and the low-order wavefront sensor (LoCam) within CGI. Table 11 summarizes the expected CGI camera performance\cite{Morrissey2023, Bush2025}.

{\footnotesize
\begin{xltabular}{\textwidth}{@{} >{\raggedright\arraybackslash}p{0.34\textwidth} >{\centering\arraybackslash}X >{\centering\arraybackslash}X >{\centering\arraybackslash}X @{}}
\caption{CGI EMCCD Beginning-of-Life (BOL) performance specifications}\label{tab:cgi_emccd_bol}\\
\hline
\textbf{Measurement} & \textbf{Requirement} & \textbf{LoCam} & \textbf{ExCam} \\
\hline
\endfirsthead
\caption[]{(continued)}\\
\hline
\textbf{Measurement} & \textbf{Requirement} & \textbf{LoCam} & \textbf{ExCam} \\
\hline
\endhead
\hline
\multicolumn{4}{r}{\textit{(continued on next page)}}\\
\hline
\endfoot
\hline
\endlastfoot
Read Noise ($e^-$, with EM Gain) & $<$ 0.2 & $<$ 0.2 & $<$ 0.2 \\ \hline
Image Area Full Well Capacity (ke$^-$) & 50 & 90 & 88 \\\hline
Serial Register Full Well Capacity (ke$^-$) & 90 & 100 & 100 \\\hline
Parallel Charge Transfer Inefficiency (CTI) & $< 5 \times 10^{-5}$ & $1.04 \times 10^{-6}$ & $1.29 \times 10^{-6}$ \\\hline
Serial CTI & $< 5 \times 10^{-5}$ & $1.40 \times 10^{-5}$ & $1.45 \times 10^{-5}$ \\\hline
Mean Thermal Dark Current ($e^-$/pix/s) \newline (Req.\ applies to 95\% of pixels at BOL) & --- (req.\ applies to distribution) & $2.8 \times 10^{-3}$ & $1.0 \times 10^{-3}$ \\\hline
Unity Gain Non-linearity (\%) & $<$ 4\% & $<$ 4 & $<$ 4 \\\hline
Clock Induced Charge ($e^-$/pix/frame) & 0.01 & Less Applicable -- 0.020 & 0.008 \\
\end{xltabular}
}

For NIR detectors, the state-of-the-art devices are the Roman Wide-field Instrument (WFI) Teledyne H4RG detectors with custom ASIC for Control and Digitization of Imagers for Astronomy (ACADIA) readout electronics. While these devices achieve exquisite performance, they do not possess photon counting abilities, with read noise $>$ 6 e- per 180 second integration. Analysis is underway to determine the viability of these detectors for the Coronagraph Instrument NIR channel, however it is expected that a photon-counting solution such as the LM-APD will be required. Therefore, this Technology Plan focuses early development efforts on addressing key the key challenges of that technology, including reducing persistence, reducing readout integrated circuit (ROIC) glow, and increasing operational temperatures.

%\begin{figure}[htbp]
%\centering
%\includegraphics[width=0.9\linewidth]{figures/fig027.png}
%\caption{CGI EMCCD Beginning-of-Life (BOL) performance specifications\cite{Morrissey2023}}
%\label{fig:16}
%\end{figure}
\subsubsection{TRL 5 Milestone Definition}
The TRL 5 milestone demonstration of a detector system (sensor and readout electronics) consists of a device meeting the notional specifications in Table 9 with a readout ConOps consistent with an expected coronagraph observing scenario. Separate demonstrations are expected for visible and NIR devices.

\subsubsection{Development Strategy \& Roadmap}
The development strategy for the coronagraph detectors will follow a similar approach as we as done for the deformable mirrors, described in Section 4.4. Based on a survey of potential devices (including but not necessarily limited to those identified in Table 10), HWO TMPO will partner with the appropriate industry, academic, or international partners for a vendor study. Each vendor will deliver a report on the device's performance specifications, ConOps, and development plan (to include budget) to achieve the TRL 5 milestones prior to MCR. From this study, HWO TMPO will down-select to the most promising visible and NIR devices to execute the provided development plans, contingent upon available funding. Depending on the provided plans, additional milestone or down-selects may also be performed.

\subsection{Spectroscopy}
\textbf{Gap Type:}\textbf{	}  Engineering

\textbf{Est. TRL:} 	4

\textbf{Importance:}\textbf{	}  3 -- Threshold

\textbf{Urgency:}\textbf{	}  1 -- Long Term

\subsubsection{Discussion}
A variety of optical and near-infrared spectrograph architectures have been developed for observations of directly imaged exoplanets \cite{Claudi2008, Delorme2021, Zhang2024}. The preliminary design of the Nancy Grace Roman Space Telescope Coronagraph Instrument included a lenslet integral field spectrograph (IFS) channel \cite{Groff2019}. Both the HabEx and LUVOIR mission concept studies baselined visible and near-infrared IFS channels \cite{HABEX, LUVOIR}. Each of these IFS concepts use a lenslet array to spatially sample the image search region around the occulted host star \cite{Groff2015}. A prism disperses the light from the lenslets into an array of low-resolution spectra on a detector (R $\sim$ 40–140) which are then remapped by a data processing pipeline into a three-dimensional data cube (axes x, y, and $\lambda$) \cite{Zimmerman2011}.

Although the IFS channel of the Roman CGI was eventually descoped, the project’s technology maturation efforts established the laboratory performance baseline of a combined coronagraph and IFS system using the Prototype Imaging Spectrograph for Coronagraphic Exoplanet Studies (PISCES) \cite{McElwain2016}. Electric field conjugation experiments with PISCES under vacuum achieved a mean contrast of 1$\times 10^{-8}$ inside a 3 – 9 $\lambda$/D bowtie-shaped control region over an 18\% fractional ($\Delta\lambda / \lambda$c) bandpass \cite{Groff2017}. This milestone corresponded to TRL 4 in the context of the Roman CGI system requirements. The next iteration in maturity would have corresponded to a test configured with flight-like pointing errors and wavefront disturbances \cite{Cady2017}.

\subsubsection{TRL 5 Milestone Definition}
The TRL 5 milestone metric for the high-contrast spectroscopy technology lane is to demonstrate spectrograph wavefront control below 10{\textsuperscript{-}}{\textsuperscript{9}} raw contrast over a 20\% bandpass with a systematic error less than 5\% of the continuum per spectral bin. Additional desired performance metrics such as end-to-end throughput and sensitivity are currently being studied.

\textbf{Significance:} Spectroscopic observations of directly imaged exoplanets at visible and near-infrared wavelengths will enable the measurement of atmospheric absorption features in their disk-integrated albedos. Among the gases with remotely observable abundances on Earth and other Solar System planets, H$_{2}$O, CO, CO$_{2}$, CH$_{4}$, NH$_{3}$, and O$_{2}$ all present prominent absorption features in the 0.5–2.0 \textmu{}m wavelength range. Among these, molecular oxygen, most prominently seen in the O2 A-band absorption at 760 nm, is challenging due to its relatively narrow width, approximately 10 nanometers full width at half-maximum depth\cite{Marais2002}. Since molecular oxygen is of particular significance as a potential biomarker when observed in combination with other gases like CH4, its narrow width is likely to drive the choice of spectral resolution of a visible-wavelength IFS camera \cite{Meadows2018}. Previously, several authors have numerically modeled the retrieval of atmosphere parameters from an Earth-analog exoplanet spectrum \cite{Brandt2014, Feng2018, Damiano2022}. Their results suggested that a spectrograph would need a resolving power (R $= \lambda/\Delta\lambda$) in the range 140–150 at 760 nm to detect the O$_{2}$ feature and potentially constrain the atmosphere’s oxygen mixing ratio. Other prominent absorption features in the visible require lower spectral resolution, R $\sim$ 70 or lower, such that the visible IFS need not maintain R=140 over the entire visible bandpass.

\textbf{Verification Method:} End-to-end starlight suppression demonstration combining DM wavefront control, coronagraphy, and spectroscopy in a vacuum chamber. Contrast metrics will be evaluated as a function of wavelength in the extracted spectroscopic images or spatial samples.

\subsubsection{Development Strategy \& Roadmap}
Leveraging the Roman PISCES IFS instrumentation \cite{Groff2017}, HWO TMPO will develop an updated IFS design and instrument breadboard to be paired with one or several starlight suppression demonstrations discussed in Section 4.2. This effort will include the development of specific components (including lenslet arrays and high-throughput, low-scintillation optics). As the spectral resolution and sensitivity requirements are expected to be more stressing in the visible band, the IFS demonstration will be targeted to operate with a visible coronagraph instrument and impacts due to scaling the performance to the NIR documented. Alternate technology demonstrations for NIR spectroscopy, including photonic integrated circuits (PICs) will also be considered.

\subsection{Post-Processing}
\textbf{Gap Type:} 	Engineering

\textbf{Est. TRL:}\textbf{	}  3 -- 4

\textbf{Importance:} 	3 -- Threshold

\textbf{Urgency:}\textbf{	}  2 -- Urgent

\subsubsection{Discussion}
The direct detection and characterization of exo-Earths requires spectroscopy in the 10{\textsuperscript{-10}} regime at the requisite SNR. The starlight suppression system will provide a raw contrast image using a combination of coronagraphy and wavefront sensing and control. A post-processing gain factor beyond the raw contrast can be achieved via calibration and subtraction of residual starlight within the dark hole, the intent of which is to minimize the variance of the residual starlight. The achievable gain factor will ultimately be limited by contrast stability and measurement noise.

Post-processing gains have been demonstrated using a variety of different approaches on ground and space telescopes, as well as on contemporary high-contrast testbed facilities. The Roman CGI currently estimates its coherent on-orbit raw contrast to be 3.3 $\times10^{-8}$, with a factor of $\sim8.5\times$ attributed to post-processing beyond the gain factors from angular differential imaging (ADI) and reference star differential imaging (RDI) \cite{Ygouf2016}.

Reference images can be obtained simultaneously (e.g., polarimetric, spectral processing, coherence, telemetry-aided model PSFs) or with a delayed measurement (e.g., RDI, ADI). Simultaneous measurements are preferable in principle to minimize wavefront changes that affect contrast stability. Effectively, the differential images must be well-correlated in the dimension of the introduced diversity (e.g., polarization, spectral content, or wavefront error)\cite{McElwain2025}. Several different forms of post-processing have been proposed for HWO, including coherent differential imaging (CDI)\cite{Mann2024}, ADI, RDI, etc. Wavefront drift can impact all of these methodologies\cite{Sanchez2026}.

In the case of the Roman CGI, a reference estimate of the residual starlight speckle image is obtained with a delayed measurement through a combination of reference star observations and different angular positions. No post-processing modes based on simultaneous measurements are currently baselined by Roman CGI.

\subsubsection{TRL5 Milestone Definition}
The post-processing algorithm shall achieve a post-processing factor, defined as the ratio of the raw contrast to the standard deviation of the post-processed exoplanet contrast $\geq$ 5 starting from a raw contrast $<$ 10{\textsuperscript{-9}} over a 20\% bandpass in a dynamic environment using observing strategy traceable to the on-orbit ConOps. The post-processing algorithm shall also achieve systematic photometric error in the extracted planet flux $<$5\% relative to ground truth in each spectral bin for a planet with contrast equal to that of the raw contrast.

\textbf{Significance:} To date, post-processing techniques have never been demonstrated at or near the 10{\textsuperscript{-10}} contrast regime. Performing post-processing experiments at the contrast levels achievable by current high-contrast testbed facilities is crucial to exploring the trade space between raw contrast, contrast stability, and post-processing techniques. Furthermore, it is paramount to demonstrate not only improvement in contrast, but also the ability of post-processing techniques to maintain accurate photometry in the resultant planet spectrum.

\textbf{Verification Method:} End-to-end starlight suppression demonstration combining DM wavefront control, coronagraphy, and spectroscopy in a vacuum chamber. A planet and/or exozodi with known brightness and spectrum should be injected into the dark hole region. If the planet and exozodi are injected into the data synthetically, they must be representative of the flight condition and well documented. Raw and post-processed contrast metrics will be evaluated as a function of wavelength in the extracted spectroscopic images or spatial samples. For the TRL 5 milestone, dynamic telescope wavefront disturbances will be introduced and compensated.

\subsubsection{Development Strategy \& Roadmap}
Post-processing strategies and algorithms will be developed as part of the Systems Team's architecture and systems trade studies conducted as part of each Exploratory Analytic Case (EAC). This will ensure that the post-processing approach is consistent with the expected ConOps of each EAC, as well as assumptions about the active wavefront sensing and control system and residual contrast stability.

Initial post-processing demonstrations will be performed on simulated data while early testbeds are developed. Once data from the TRL 5 testbeds for starlight suppression, contrast stabilization, and ultra-stable sensing and control are available, they will be incorporated in the post-processing demonstrations consistent with EAC assumptions.

\subsection{Near-UV Capability}
\textbf{Gap Type:}\textbf{	}  Technology

\textbf{Est. TRL: }\textbf{	} 2 (coronagraph), 4 (starshade)

\textbf{Importance:} 	2 -- Baseline

\textbf{Urgency:}\textbf{	}  2 -- Urgent

\subsubsection{Discussion}
\begin{figure}[htbp]
\centering
\includegraphics[width=0.9\linewidth]{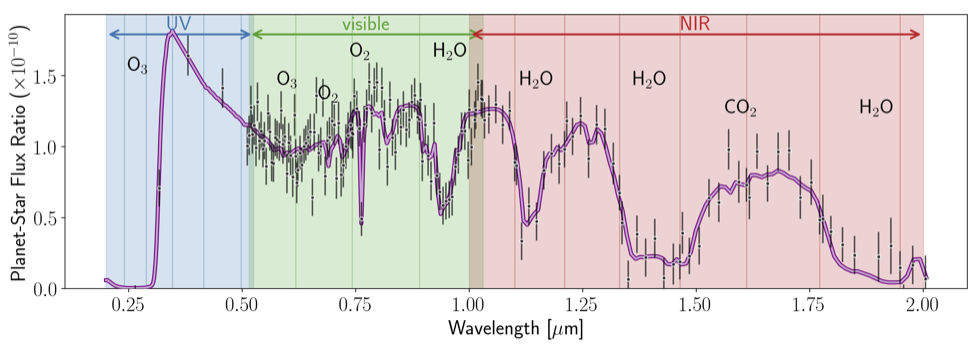}
\caption{Spectrum of an Earth-like exoplanet from the LUVOIR Final Report. Note the deep ozone absorption line in the NUV. Image credit: J. Lustig-Yaeger.}
\label{fig:15}
\end{figure}

One of the most prominent biosignatures, ozone, is present in the Near-UV (NUV, 250 – 400 nm). Identifying and measuring the abundance of O3 requires low spectral resolution (R $\sim$ 5 – 10) and requires sensitivity down to 250 nm, as shown in Figure \ref{fig:15}. Other biosignatures, such as oxygen, are crucial atmospheric constituents for potentially habitable planets in the NUV. UV spectroscopy can be used to detect and measure the mixing ratios of ozone and improve oxygen gas measurements independently of near-infrared data \cite{Damiano2022}. Detection of the full ozone line requires covering the 250 – 350 nm bandpass (33\% bandwidth). Sequential imaging through smaller bandwidth filters ($\sim$10 – 20\%) enables a deep contrast characterization of the ozone line. 

The NUV coronagraph should be a stand-alone instrument (not integrated into the rest of the coronagraph channels) to maximize throughput; silver coatings that may be used for the visible and near-infrared channels have negligible reflectance at wavelengths shorter than 350 nm. The NUV channel will probably be picked off separately from the visible/NIR channel(s) after the telescope secondary mirror, so operating in parallel with the visible and NIR channels is expected to be impossible. The NUV channel should also only be used to characterize exoplanets previously detected at visible wavelengths. This effectively relaxes requirements on the inner working angle (IWA{\textsubscript{UV}} $\sim$ 2 $\times$ IWA{\textsubscript{VIS}}).

Some progress has been made on the conceptual design of a near UV coronagraph, including creating candidate optical layouts, identifying coronagraph mask designs capable of achieving $<$ 10$^{-10}$ contrast in simulations, and performing aberration sensitivities with possible wavefront sensing and control algorithms \cite{JuanolaParramon2026, VanGorkom2025}. Several key items remain unclear in the development of a NUV coronagraph, including adequate materials (for dichroics and deformable mirrors, for example), coatings (surface roughness, uniformity, and polarization-induced aberrations), and the impact of scattered light (due to cleanliness level on all optics, including deformable mirrors, etc.). Additional study of the NUV coronagraph will be necessary to ensure requirements can be met.

An alternative approach to a NUV coronagraph is to use a starshade or external occulter. To achieve the desired contrast over a NUV bandwidth between 250-500 nm will require a starshade $\sim$ 35 m in diameter, operating at a separation distance from a 6-m diameter HWO of 55.5 Mm. Such a starshade is consistent with an architecture developed under the ExEP’s “S5” or Starshade TRL 5 program\footnote{See: \url{https://exoplanets.nasa.gov/exep/technology/starshade/}}. While the S5 program achieved all the milestones originally defined for a TRL 5 demonstration of a visible-band starshade compatible with a notional Roman rendezvous mission, additional work would be necessary to achieve TRL 5 for a UV-band starshade for HWO.

\subsubsection{TRL 5 Milestone Definition}
Demonstration of near-UV coronagraph high-contrast (raw contrasts $\leq$10{\textsuperscript{-10}}) photometry at R$\sim$5 - 7 between 250 and 450 nm in a relevant dynamic environment.

\textbf{Significance:} Ozone absorption line detection and characterization requires low (R$\sim$5--10) spectral resolution between 250 -- 350 nm. This can be accomplished via photometry on high-contrast (10{\textsuperscript{-10}} raw contrast) images with $\sim$3 broadband ($\sim$10 -- 20\% bandwidth) filters. Near-UV coronagraphy to date has not been demonstrated in any environment, including in air. For the NUV coronagraph to reach TRL 5, high-contrast photometry will need to be demonstrated at adequate contrasts levels in a relevant environment. Fabrication techniques, adequate coatings to obtain the throughput levels necessary to successfully obtain high-contrast photometry in a reasonable amount of time, and contamination need to be studied and have a clear path to flight.

\textbf{Verification Method:} End-to-end starlight suppression demonstration combining DM wavefront control and coronagraphy in a vacuum chamber. For the TRL 5 milestone, pointing jitter and telescope wavefront disturbances will be introduced and compensated.

Testbeds like those at University of Arizona (PI: Van Gorkom) are critical for advancing this technology to TRL 5.

\subsubsection{Development Strategy \& Roadmap}
HWO TMPO will explore NUV high-contrast instruments that meet the necessary capabilities. Close interaction with the groups studying NUV coronagraphy, including the University of Arizona, GSFC, and JPL should develop requirements on the testbeds to ensure successful demonstrations.

\section{Ultra-Stable Telescope System Technologies}
\subsection{Overview}
{\footnotesize
\begin{xltabular}{\textwidth}{@{} >{\raggedright\arraybackslash}p{0.10\textwidth} >{\raggedright\arraybackslash}X >{\raggedright\arraybackslash}X >{\raggedright\arraybackslash}X >{\raggedright\arraybackslash}X >{\raggedright\arraybackslash}X >{\raggedright\arraybackslash}X @{}}
\caption{Ultra-stable telescope system technologies}\label{tab:12}\\
\hline
\textbf{ID} & \textbf{Lane} & \textbf{Driving Factor(s)} & \textbf{Estimated Current TRL} & \textbf{Impor}\textbf{-t}\textbf{ance} & \textbf{Urgency} & \textbf{Gap Type} \\
\hline
\endfirsthead
\caption[]{(continued)}\\
\hline
\textbf{ID} & \textbf{Lane} & \textbf{Driving Factor(s)} & \textbf{Estimated Current TRL} & \textbf{Impor}\textbf{-t}\textbf{ance} & \textbf{Urgency} & \textbf{Gap Type} \\
\hline
\endhead
\hline
\multicolumn{7}{r}{\textit{(continued on next page)}}\\
\hline
\endfoot
\hline
\endlastfoot
1 & Ultra-stable Mirrors & Integrated mirror assembly that meets required stability and optical performance. & 4-5 & Threshold & Critical & Tech \\
2 & Ultra-stable Structures & Composite beams, sheets, and joints with low creep and high stiffness. & 4-5 & Threshold & Urgent & Eng \\
3 & Thermal Control System & Milli-Kelvin control with compact flight electronics. \newline Low-vibration thermal control or conduction systems. & 4 & Threshold & Urgent & Eng \\
4 & Telescope Sensing \& Control & Sense and control segment-level and global telescope alignment at picometer level over relevant time scales. & 3-4 & Threshold & Critical & Tech \\
5 & Low-Vibration \& Pointing Control Systems & Active and passive isolation systems. \newline Low-disturbance attitude control (e.g., microthrusters). & 4 & Threshold & Long Term & Tech \\
6 & Deployable Systems & Large deployable barrel assembly that is robust to micrometeoroids. & TBD & Threshold & Long Term & Eng \\
\end{xltabular}
}
\subsection{Ultra-stable Mirrors}
\textbf{Gap Type:}\textbf{	}  Technology

\textbf{Est. TRL:}\textbf{	}  4 -- 5

\textbf{Importance:}\textbf{	}  3 -- Threshold

\textbf{Urgency:} 	3 -- Critical

\subsubsection{Discussion}
A critical and enabling part of achieving wavefront stability of a few picometers over durations of minutes or hours over a range of temporal frequencies based on the optical error budget, is the ultra-stable mirror technology. Early system-level engineering design and modeling studies consistently concluded that wavefront error budgets for telescope alignment and figure stability will have allocations at the single-digit picometer level. While the HWO telescope optical error budgets are under development, earlier relevant industry-led studies \cite{Coyle2019} have shown that the optical budgets will be categorized into spatial and temporal frequencies. Although some of the rigid body motions of mirror assemblies may be controlled or managed through optical wavefront sensing and control technology as described in Section 5.5, the mirror optical surface figure error must be maintained to a handful of picometers in absence of any active optical surface figure control.

The HWO optical telescope element  will be operating nominally at a room temperature ($\sim$293 K) and builds off of heritage from past concepts \cite{Eisenhower2019}. Thermal control systems will be required to achieve those mirror temperatures while maintaining the stability of the thermal control surfaces to a milli-K scale over a range of temporal frequencies as described in Section 5.4. Furthermore, there are only a few mirror substrate materials (either Corning ULE\textsuperscript{\textregistered} or Schott Zerodur\textsuperscript{\textregistered}) with extremely low coefficient of thermal expansion (CTE) as possible candidates for an ultra-stable mirror. The CTE of those candidate materials are at near-zero at room temperature with a variation of roughly $\pm$10-30 ppb\footnote{\url{https://tinyurl.com/wb26pb7c}}\cite{Jedamzik2020}.

A second consideration for the choice of material is the mirror substrate areal density (expressed in kg/m{\textsuperscript{2}}). The processes used to manufacture each raw material and then subsequently form the raw material into a mirror substrate led to different options in the mirror substrate design, such as open vs. closed back, core rib width and depth, and mass vs. stiffness \cite{FeinbergSCDA}. Given overall mass constraints imposed by available launch vehicles, areal density will be a significant driver of the mirror substrate design and material choice.

A design balance must be achieved between a fast temperature response with a lower thermal mass mirror design. The former requires a higher frequency thermal sensing and control loop but lighter overall mirror, while the latter relaxes the thermal sensing and control requirements but requires a heavier mirror that is inherently more stable. Again, the allowable mass allocation for the mirror assembly may drive the type of mirror segment architecture.

It is highly desirable to achieve a `stiff' mirror segment design to minimize optical surfaces distortion from induced forces from actuators and observatory dynamics and thermal disturbances. A `stiff' mirror substrate with a first free-free mode $>$ 200 Hz will also reduce uncertainties associated with the gravity sag induced during ground testing.

While the mirror substrate is a critical component of ultra-stable optical performance, picometer stability at the mirror segment assembly level also must be thoroughly considered as part of a TRL 5 demonstration. A mirror segment assembly must ultimately include bonds, mounts, flexures, actuators, and the interface structure(s) between the mirror substrate and the backplane that supports the mirror segment assembly.

\begin{figure}[htbp]
\centering
\includegraphics[width=0.9\linewidth]{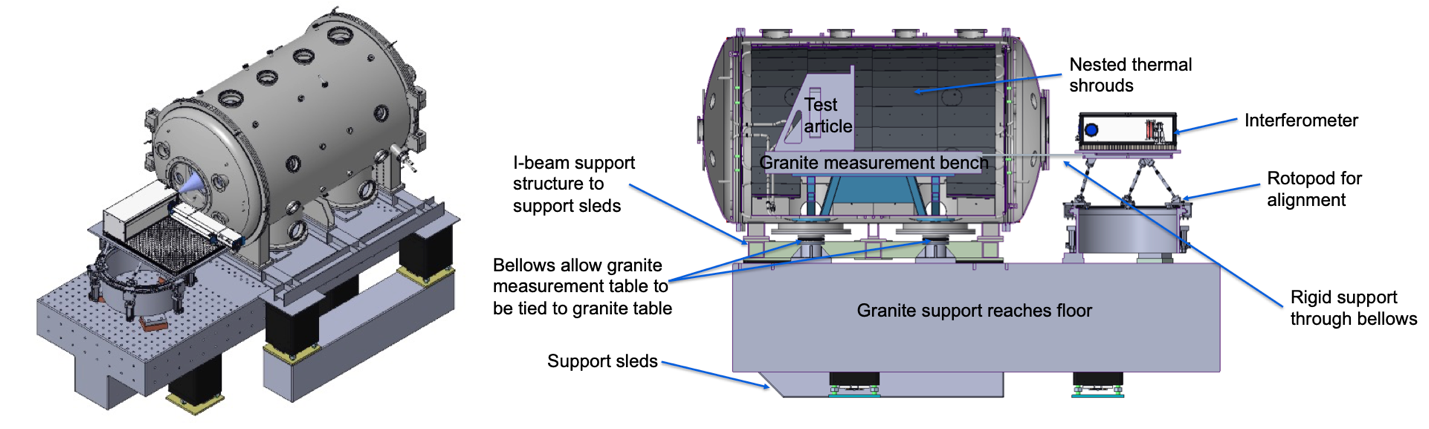}
\caption{Two views of the Mini-MUST system.}
\label{fig:16}
\end{figure}

\paragraph{Ultra-Stable Structures Laboratory (USSL) and the Mini-Metrology and Ultra-Stable Testbed (Mini-MUST)}
The Goddard Space Flight Center's Ultra-Stable Structures Lab (USSL) has a history of making picometer accuracy dynamic tests over the spatial extent of test articles \cite{Saif2017, Saif2019}, and has been actively working towards achieving stability measurements at the required levels over the past few years. The laboratory uses high-speed, phase-shifting optical interferometers originally created for testing critical components of the James Webb Space Telescope \cite{Saif2018, Saif2021}, like the primary mirror segments and the Backplane Stability Test Article (BSTA) \cite{Saif2008}. These interferometers optically measure specular and diffuse targets in the Mini-Metrology and Ultra-Stable Testbed (Mini-MUST; PIs: Sitarski and Hadjimichael), a 1.75 m diameter $\times$ 2.5 m long thermal vacuum chamber designed to host $\leq$ 1 m-class optics. Mini-MUST features enhanced $\pm$1 mK/hour stable thermal control enabled by two nested thermal shrouds and operates at $\sim$10$^{-6}$  Torr levels. It contains a stable granite optical bench embedded with optical inserts and rails for easy access to test articles, and the granite bench will be supported by a series of isolators. The vacuum chamber rests on a large, granite, T-shaped bench that measures 14' long $\times$ 6' wide, and is $\sim$ 40" thick. The vacuum chamber sits on a bridge that is suspended over the large granite bench and floats around the test article bench. The whole setup sits on its own concrete slab inside an acoustic-isolated and thermally controlled ($\pm$0.5 K/24 hours) laboratory space with crane access. Figure \ref{fig:16} shows the Mini-MUST setup.

Several critical components, including mirror and composite materials, laser metrology trusses, edge sensors, actuators, and combinations of those systems, will be supplied by industry partners that have been funded by ROSES D.19 proposals. Other institutions are welcome to use the facility to test their HWO-related work for TRL 5 advancement.

\subsubsection{TRL 5 Milestone Definition}
A TRL 5 demonstration of an ultra-stable mirror system consists of:

\begin{itemize}
\item A fully integrated meter-class mirror cell, including substrate, support structure, actuation, thermal control hardware, and mounted metrology elements;
\item In a dynamically isolated TVAC environment representative of the expected operational environment;
\item Demonstrating active thermal control and wavefront stability performance;
\item Consistent with assumed sensing and control and ConOps scenarios;
\item Achieving ``medium fidelity of analysis'' definition.
\end{itemize}
Impacts of scaling from meter-class to actual segment dimensions and geometry shall be understood via analysis. Flight qualification of the fully assembled mirror cell is not required, though individual component qualification may be completed as appropriate.

\subsubsection{Development Strategy \& Roadmap}
\begin{figure}[htbp]
\centering
\includegraphics[width=0.9\linewidth]{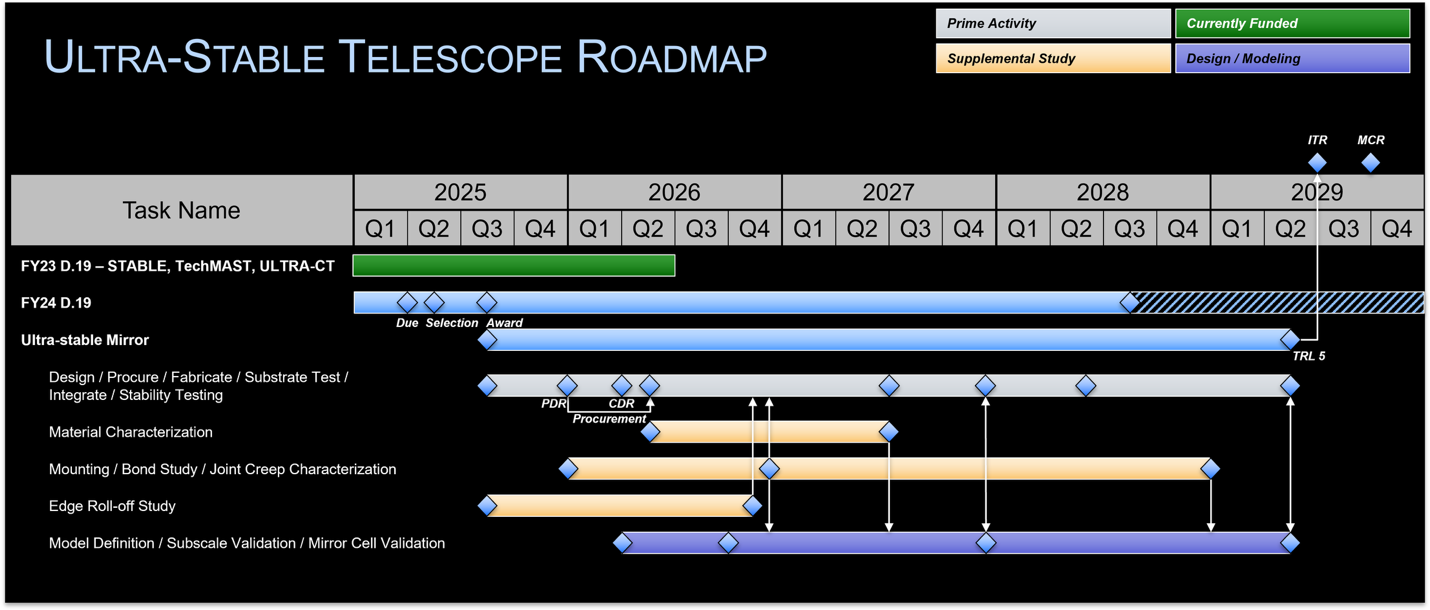}
\caption{Roadmap for Development of the Ultra-stable Mirror technology. Contingent upon available funding.}
\label{fig:17}
\end{figure}

The general development strategy for the ultra-stable mirror technology is shown in Figure \ref{fig:17}, and will leverage investments made through the ROSES Appendix D.19 programs. The FY23 D.19 efforts are currently underway and address many of the technologies in the Ultra-stable Telescope Systems Track.

\paragraph{Primary Development Path}
The primary development path will commence with a study to design the TRL 5 test article. Once a preliminary design review is completed, procurement of materials and hardware can proceed. Following a critical design review, fabrication and integration of the mirror cell will begin.

Component-level testing, including the mirror substrate, actuators, and metrology elements\cite{Coyle2024}, should be completed to verify TRL 4 performance metrics. Upon completion of integration, ambient testing can be completed to verify basic functionality is consistent with expectations. A final TVAC stability test meeting the TRL 5 milestone criteria completes the demonstration.

\paragraph{Supplemental Studies}
Several supplemental studies will be completed to answer critical scaling or performance limiting questions.

For HWO, it is expected that the mirror surface figure quality will need to meet specifications to within $\sim$1 mm of the physical mirror edge to limit diffraction effects within the coronagraph. Prior to completing polishing of the mirror cell substrate, an edge roll-off study should be completed to inform both the practical limitations associated with achieving such small edge roll-off, as well as best practices for implementation on the full mirror substrate. Bevels as low as 0.5 mm have been shown on segmented mirror segments for ground-based observatories \footnote{\url{https://www.glyndwrinnovations.co.uk/wp-content/uploads/2016/10/Overview-of-TMT-Project-101116b.pdf}}.

Scaling a meter-class segment to full-size segment of appropriate geometry will largely be limited by our understanding of how the material characteristics of the parent boule, in particular the distribution of the CTE, are applied to larger mirror segments, whether they be cut from larger boules, flowed from smaller boule sections, or fused from multiple individual pieces. A detailed material characterization study to understand these relationships should be completed to inform the model validation activities.

After the mirror substrate, mirror cell stability will be driven by the stability of its mount and the mounting interface. Prior to mirror cell integration, a mounting and bond study should be completed on subscale test articles to inform optimal bonding geometries and best practices. Once completed, these bonded joints should be regularly monitored and tested for long term creep or sudden stress release (e.g., lurch).

Finally, the mirror surface roughness is a driving consideration for scattered light in both the coronagraph and ultra-violet instruments. A study on subscale test articles should be completed to characterize limitations on achievable surface roughness and quantify surface power spectral densities (PSDs).

\paragraph{Model Validation Activities}
Upon completion of the design study, a model definition and validation plan should be completed to define what models are necessary and at what level of fidelity to achieve the defined TRL 5 milestone. Subscale validation efforts should be completed in parallel with component-level development and testing. A fully integrated structural-thermal-optical model of the TRL 5 test article should also be completed and validated by the final stability test.

\subsection{Ultra-stable Structures}
\textbf{Gap Type:} 	Engineering

\textbf{Est. TRL:} 	4 -- 5

\textbf{Importance:}\textbf{	}  3 -- Threshold

\textbf{Urgency:} 	2 -- Urgent

\subsubsection{Discussion}
The performance of the ultra-stable structures comprising the optical telescope and instrument metering path is key to the eventual performance of HWO.

First, the quasi-static rigid body motion of the optical elements (SM to PM spacing, PM segment rigid body control, etc.) from room-temperature, 1g-aligned conditions to on-orbit conditions must be predictable, with any residual motion correctable within the capture range of alignment actuators. Segment-level actuators in particular are a concern due to the large disparity in significant range required for initial alignment and the fine resolution needed for wavefront control during operations.

Second, the quasi-static figure distortion that is induced by any structural change between ambient and operational conditions must be accounted for, including uncertainty, within the system wavefront error budget. The system-level allocation of quasi-static wavefront error as seen by the instruments are challenging, so any increased suballocation to room-to-operational distortion comes at a cost to other areas of the budget (optic-level ambient surface figure, alignment success criteria, etc.) and can drive cost/complexity in these areas. These first two concerns have largely been tackled by several recent space telescope missions (e.g., JWST and Roman) and are treated here as challenging, but within the state of the art.

Finally, and most pertinent to HWO technology development, the final aligned optical system must be sufficiently stable (i.e. within the capture range of active wavefront sensing) to not only support the contrast stability required over long integration times during exoplanet observations, but to also deliver this level of stability during the reference observations needed to support the post-processing requirements in Section 4.7. This stability can be suballocated to changes in rigid-body/metering (within the capture range of active wavefront sensing) and to changes in surface figure of each optical element (within the capture range of instrument compensators, i.e. DMs). Given the current understanding of these allocations, the ultra-stable structures needed for HWO are outside of the state-of-the-art and require technology development.

NASA's Roman Space Telescope Optical Telescope Assembly (OTA) and Instrument Carrier (IC) represent a recent design reference point for ultra-stable structures. These structures were designed for near zero CTE at pre-determined temperature ranges (room temperature to $\sim$220K) \cite{Smith2018, Miller2022, Whitman2022}. As a design example, the secondary mirror support tubes for Roman were designed for near-zero CTE at 270 K. The as-manufactured tubes, complete with harnessing and multi-layer insulation (MLI) blanketing, demonstrated a zero CTE point of $\sim$268 K (or an offset of $\sim$14ppb/K from prediction). In other areas of the assembly, materials were selected and arranged to compensate for the higher than desired CTE for interface and high strength materials. While this was highly effective, reasonably small uncertainties in the assumed material CTEs and the detailed assumptions made within the distortion model caused a higher-than-expected sensitivity to both quasi-static and transient temperature changes; unaccounted for changes results in an additional 275 pm/K primary mirror piston motion due to the design of the PM struts. While Roman is predicted to easily meet the stability requirements of the mission, including those for the demonstration of the CGI, the stability needs for HWO dictate technology investment in ultra-stable structures as well as in the associated thermal control architecture as outlined in Section 5.4.

\subsubsection{TRL 5 Milestone Definition}
A TRL 5 demonstration of the ultra-stable structure technology consists of:

\begin{itemize}
\item A set of at-scale structural components and joints;
\item Qualified for strength;
\item Demonstrate the necessary thermal stability, dynamic stability, and damping;
\item In an actively control thermal-vacuum environment.
\end{itemize}
The intent is to understand the stability of structural elements such as beams, composite-composite joints, and composite-metal joints of various geometries to validate structural models at the picometer level. Detailed material characterization (coefficients of thermal and moisture expansion, damping coefficients, etc.) and coupon testing should be completed to facilitate model validation.

\subsubsection{Development Strategy \& Roadmap}
The general development strategy for the ultra-stable structure technology will leverage investments made through the ROSES Appendix D.19 programs. The FY23 D.19 efforts are currently underway and address many of the technologies in the Ultra-stable Telescope Systems Track.

\paragraph{Primary Development Path}
Ongoing D19 activities are already studying aspects of the ultra-stable structures technology, including improved material testing and screening capabilities, joint and bondline geometry studies, damping characterization, and long-term dynamic effects such as creep or lurching. It is expected that these studies will evolve into the development of new metrology facilities to perform the necessary material characterization (CTE, moisture desorption) before fabrication of the structural demonstration articles begins.

As structural element fabrication completes, thorough strength, stability, damping, and dynamic testing will be completed using the Ultra-stable Structures Laboratory (USSL) or similar capabilities.

\paragraph{Supplemental Studies}
In addition to developing structural test articles, comprehensive coupon testing of various material layups, bonds, etc. should be completed to help facilitate the model validation activities.

\paragraph{Model Validation Activities}
The large-scale structures comprising HWO will require analysis to verify picometer-level stability in the Flight environment. Model validation will require a crawl-walk-run strategy where first the most basic elements of a structure can be modeled and validated through test: coupons of material layups or single-interface bonds. These models can then progress to more complicated geometries the represent typical interfaces in much larger structures: joints where two or more composite beams may interface to one another, or a metal node. Each layer of complexity builds on the model validation framework, eventually leading to a TRL 6 demonstration that includes a full-scale representative structure, similar to the JWST BSTA\cite{Saif2008} that is small enough to still perform picometer-level model validation tests on, yet large enough to capture critical scaling considerations.

\subsection{Thermal Control System}
\textbf{Gap Type:} 	Engineering

\textbf{Est. TRL:}\textbf{ }	4

\textbf{Importance:}\textbf{ }	3 -- Threshold

\textbf{Urgency:}\textbf{	}  2 -- Urgent

\subsubsection{Discussion}
Thermal control of optical elements and metering structures is crucial to the optical stability of the HWO optical telescope and instruments. In combination with the ultra-stable structures outlined in Section 5.3, this system serves four main functions.

First, the thermal control system must target room temperature for the telescope optics. This is important from both a programmatic and technical point of view. The difference between operational and ambient (manufacturing) temperature, both in bulk quasi-static temperature as well as in static spatial temperature gradients, effect the program in important ways. When this difference is large, the figuring of the PM segments must either use at-temperature testing between figuring runs to develop the next figuring hit map (sometimes referred to as ``cold figuring'' or ``at-temperature figuring'') or use analytical predictions for the difference between ambient and operational environments to adjust the ambient hit map. The former would be highly costly for the large number of segments and optics envisioned for HWO. The latter may be unavoidable at the surface precision required for HWO, so the magnitude of the analytical adjustment must be minimized as the uncertainty on the adjustment is typically a scalar multiple and is booked against the wavefront error budget until late in the manufacturing flow when it can be validated by test. Regardless, a room-temperature active thermal control system implies the need for many control channels; early estimates are in the many hundreds. Minimizing critical static spatial gradients implies minimizing heat flux applied by the heaters by reducing the total amount of power needed as well as distributing that power in a uniform way on the structure.

Second, the thermal control system must be designed such that critical optics and structures may operate at temperatures that minimize their as-built CTE. This will allow the reduction of the range of CTE uncertainty carried in the stability budget. Typically, the shape of the ``CTE vs. Temperature curve'' is well-known from previous testing prior to flight unit manufacturing. While designers will attempt to target room temperature as discussed previously, the as-built CTE curve of the flight hardware, and therefore the desired operating point, can often shift by several degrees. Optimizing the operational temperature ($\pm\sim$2 K) can have a 10$\times$ improvement in dimensional stability. This implies an active heater system with adjustable control temperatures and a targeted heater scheme capable of isolated control of components.

Third, thermal control stability while in a stable environment must be optimized to the greatest extent possible; controlling self-induced temperature instability due to heater actuation is critical. This implies several design requirements for the active heater system, notably the control algorithms, frequency, the output resolution/accuracy, sensor noise/accuracy, drift, etc.

Finally, the thermal control system must work (through a combination of passive and active means) to control temperature changes of critical optics and structures in opposition to disturbances from attitude changes, instrument operational modes, telescope PM phasing, etc. to better than 10s of mK. This implies a focus on passively minimizing environmental changes and rigorous heater controller tuning and sensitivity. Feed-forward algorithms could also be explored to further enhance temperature control.

NASA's Roman OTA Telescope Control Electronics (TCE) offers a recent design reference to gauge the state of the art. Early subassembly testing of the TCE demonstrated sub-mK stability as measured by a control sensor in a stable environment\cite{Miller2022}. The completed TCE is equipped with 94 heater 1 Hz control channels, each with programmable setpoints, proportional-integral-derivative (PID) control parameters, variable sensor mapping, and high output resolution. Temperature feedback is provided by sensor channels sensitive to $\sim$1 mK. The combined system was tuned in the test environment with the resulting sensors showing stability to $\sim$10-50 mK during a simulated worst case cold to hot slew (on-orbit tuning is expected to show even better performance).

\subsubsection{TRL 5 Milestone Definition}
TRL 5 will be demonstrated in concert with the ultra-stable mirror and ultra stable structure demonstrations described in Sections 5.2 and 5.3. It is anticipated such a demonstration will include the following in a relevant environment:

\begin{itemize}
\item Demonstrate ultra-stable thermal control ($\sim$1-100 mK stability) of the hardware under test, i.e. optical elements and structures, in the test environment
\item Use brassboard electronic board(s) with flight-equivalent parts and architecture including:
\begin{itemize}
\item SWaP estimates for a flight system
\item SWaP scalability to a flight system serving hundreds of heaters
\item Demonstrate the following functionality:
\begin{itemize}
\item Commandable PID control parameters and control sensor mapping
\item 0.1-1 Hz update frequency
\item High output resolution ($\sim$10 bit)
\item $<$ 2 mK sensor noise/accuracy
\item $<$ 1 mk/6 hours drift
\end{itemize}
\end{itemize}
\end{itemize}
The brassboard electronics must show a path to flight with no known susceptibility to the launch vibration or on-orbit radiation environments.

\subsubsection{Development Strategy \& Roadmap}
The general development strategy for the thermal control technology will leverage investments made through the ROSES Appendix D.19 programs. The FY25 D.19 proposals are currently under evaluation.  When those selections are made, this section will be updated to show more details of the chosen path forward.

\subsection{Telescope Sensing \& Control}
\textbf{Gap Type:} 	Technology

\textbf{Est. TRL:} 	3 -- 4

\textbf{Importance:} 	3 -- Threshold

\textbf{Urgency:}\textbf{	}  3 -- Critical

\subsubsection{Discussion}
Realizing a high contrast system capable of achieving 10$^{10}$ contrast imposes unprecedented requirements on stability, precision metrology, and wavefront sensing and control, and it is essential to develop advanced physical optics models of high-contrast systems, assess the impacts of important phenomena not captured by existing models, and validate these predictions against a testbed traceable to HWO\cite{Redding2024}. Several testbeds and demonstrations are planned to facilitate these activities.

\begin{figure}[htbp]
\centering
\includegraphics[width=0.9\linewidth]{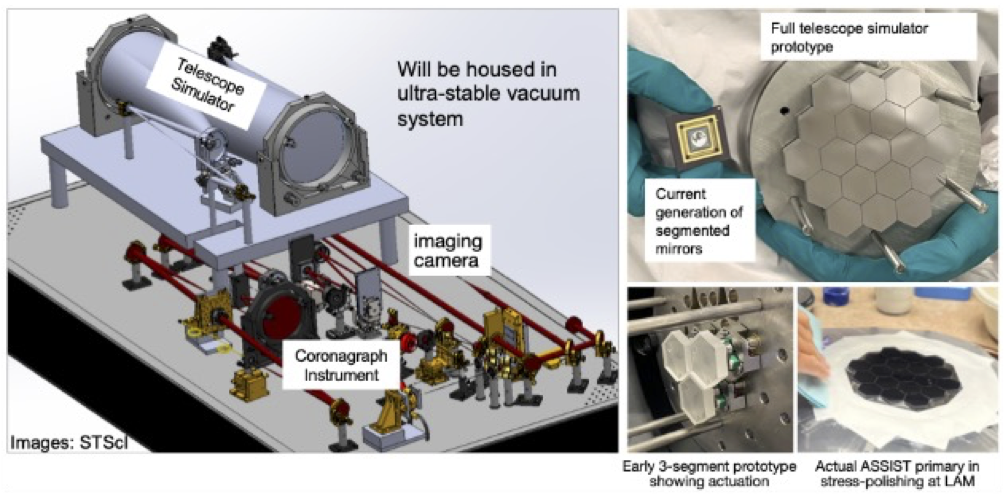}
\caption{Concept of the Habitable Worlds Observatory Systems Testbed (HOST). A segmented telescope simulator will feed a high-contrast coronagraph instrument with wavefront sensing and control algorithms traceable to HWO.}
\label{fig:18}
\end{figure}

\paragraph{Habitable Worlds Observatory Systems Testbed (HOST)}
HOST (PI: Soummer, Space Telescope Science Institute) is a dedicated testbed that will demonstrate system-level performance and functionality through a segmented telescope simulator representative of HWO, coronagraph instrument, and integral field spectrograph. It will have wavefront sensing and control aspects, including continuous dark zone maintenance, and will be used for testing post-processing algorithms. HOST is a collaboration between NASA GSFC, the Space Telescope Science Institute (STScI), and JPL and builds on the success of the High-contrast imager for Complex Aperture Telescopes (HiCAT) testbed\cite{Soummer2024}, the Active Segmented Surrogate for Integrated System Tests (ASSIST) telescope prototype, and the PISCES and DST2 integral field spectrographs \cite{McElwain2016, Zimmerman2026}. The testbed is designed to be ultra-stable on relevant timescales and will have a matched integrated model that evolves with the testbed. The primary objective of HOST is to achieve system-level performance and validate end-to-end system models at a few $\times$10{\textsuperscript{-10}} level. A conceptual view of the testbeds is shown in Figure \ref{fig:18}.

\begin{figure}[htbp]
\centering
\includegraphics[width=0.9\linewidth]{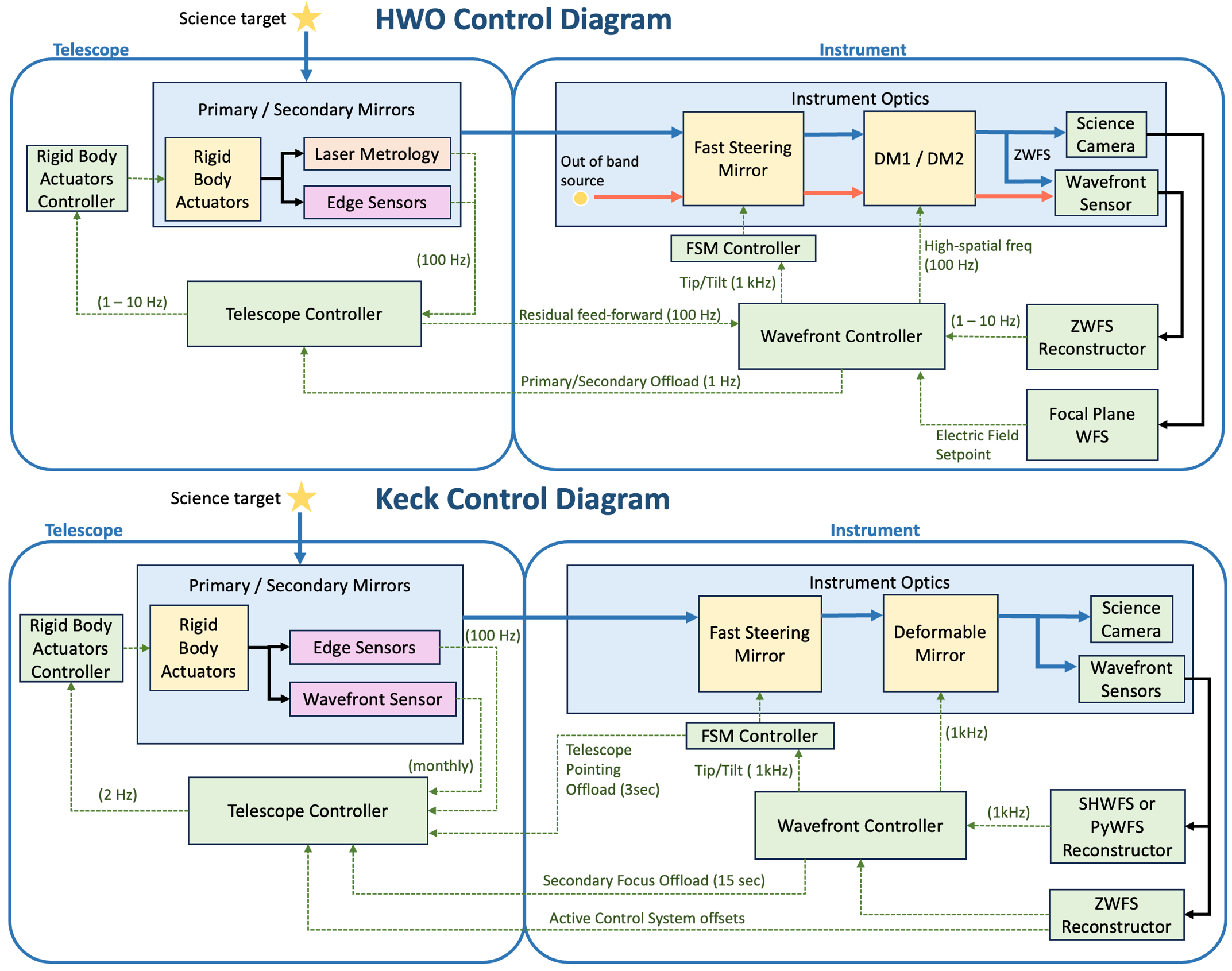}
\caption{The HWO wavefront control architecture (top) is extremely similar to the current Keck wavefront control architecture (bottom). Image: M. Salama.}
\label{fig:19}
\end{figure}

\paragraph{Keck as an HWO Testbed}
W. M. Keck Observatory has two premier, 10-meter class segmented ground-based telescopes that have state-of-the-art adaptive optics systems that were used to test JWST algorithms\cite{Albanese2006}. It is the only facility in the world with all hardware components necessary for validating HWO segment phasing and system-level architectures and capture full effects of realistic primary mirrors. It has a segmented primary mirror, capacitive edge sensors, a high-order deformable mirror, a Zernike wavefront sensor, and high-contrast science instruments. The HWO control diagram and Keck control diagrams are very similar, albeit they run at different update rates. The idea is to not use Keck to get to the picometer-level, but to test and validate models and wavefront control algorithms with readily available hardware on a realistic system that has a similar control architecture to HWO (see Figure \ref{fig:19}). A demonstration is underway to explore wavefront control interactions and strategies for segment mirror phasing (PI: M. Salama, UCSC)\cite{Salama2025}. Keck can also be used for testing post-processing algorithms, exploring polarization aberrations, etc.

\begin{figure}[htbp]
\centering
\includegraphics[width=0.9\linewidth]{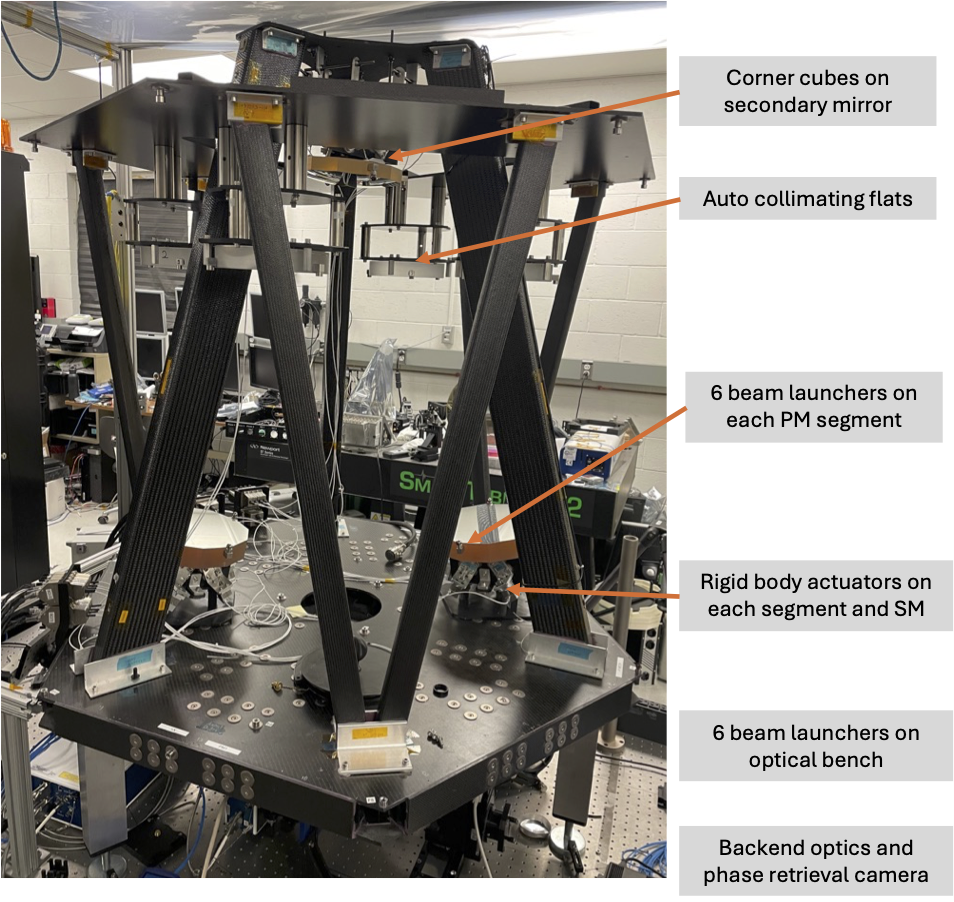}
\caption{IRIS testbed in the laboratory at JPL. The major subsystems are labeled.}
\label{fig:20}
\end{figure}

\paragraph{Image Retrieval in Segments (IRIS) Testbed}
The IRIS testbed (PI: Jonathan Tesch, JPL)\footnote{\url{https://assets.science.nasa.gov/content/dam/science/astro/programs/exep/technology/tedm-awards/files/ExoTAC-Tesch-2022SAT-whitepaper-v4.pdf}} provides telescope disturbance rejection using laser metrology, rigid body actuators, and a deformable mirror. It has adopted fault-tolerant telescope wavefront sensing and control to demonstrate ``on-board'' phase retrieval and to maintain absolute alignment. The testbed will be used to mature active telescope rigid body control to TRL 4 before future testbeds become available. Currently, IRIS has two primary mirror segments, a secondary mirror, and 18 functional laser metrology and rigid body actuator channels. A phase retrieval camera is used in the backend to provide additional wavefront sensing, and control is provided by the rigid body actuators. A heritage Xinetics DM will be integrated into the system to provide a disturbance feed-forward loop. An image of the IRIS testbed is shown in Figure \ref{fig:20}.

\subsubsection{TRL 5 Milestone Definition}
A TRL 5 demonstration of the telescope sensing and control technology consists of:

\begin{itemize}
\item A minimum of three subscale mirror segments with traceable degrees of freedom;
\item Actively maintaining required alignment tolerances over required control frequencies with out-of-control-band drifts characterized;
\item Incorporating all metrology systems (laser metrology, edge sensors, etc.) to be used in the baseline control approach;
\item Actuators that have a clear path to Flight qualification and operation.
\end{itemize}
Scale factors associated with controlling a larger number of segments and/or additional optics (e.g., active secondary mirror control) shall also be understood. Alignment of the segments to the required tolerances will be verified by an independent metrology system (e.g., high-speed interferometry or similar).

\subsubsection{Development Strategy \& Roadmap}
The general development strategy for the telescope sensing \& control technology will leverage investments made through the ROSES Appendix D.19 programs. The FY23 D.19 efforts are currently underway and address many of the technologies in the Ultra-stable Telescope Systems Track.

\paragraph{Primary Development Path}
The primary development path consists of designing, procuring, fabricating, and integrating the components necessary to implement the HOST testbed telescope simulator, or similar. The effort will commence with a study to design the TRL 5 testbed. Once a preliminary design review is completed, procurement of materials and hardware can proceed. Following a critical design review, fabrication and integration of the testbed elements.

Component-level testing, including the subscale segments, metrology components, and rigid body actuators, should be completed to verify TRL 4 performance metrics. Upon completion of integration, ambient testing can be completed to verify basic functionality is consistent with expectations. A final TVAC stability test meeting the TRL 5 milestone criteria completes the demonstration.

\paragraph{Supplemental Studies}
To ensure rigid body actuators have a path to Flight qualification, a supplemental study to investigate integrated electro-mechanical actuators that are capable of achieving the necessary stroke (a few millimeters) with required resolution (picometers). While the actual actuators used in the testbed do not need to be Flight qualified, it is critical to understand mounting geometry, load paths, and life-limiting factors of the proposed actuator components.

Multiple metrology systems are being explored to sense mirror alignment errors as part of the control system, and several of these technologies have made substantial advancements in their sensing accuracy, bandwidth, and SWaP considerations. Laser metrology systems\cite{Guzman2026, Wirth2022} incorporating PICs, and capacitive and optical edge sensors\cite{Cromey2022} should all continue to be matured until a final control system architecture is selected. Image-based techniques such as phase retrieval\footnote{\url{https://ntrs.nasa.gov/citations/20250007345}} or Zernike sensors\footnote{\url{https://assets.science.nasa.gov/content/dam/science/astro/programs/exep/technology/tedm-awards/files/SAT22-Trauger-Whitepaper-20240603.pdf}} may also be used to maintain global alignment of the system or sense slow drifts that are outside the control bandwidth of the metrology systems. These components should be tested in a relevant environment to achieve TRL 5.

A critical aspect of metrology systems is understanding the error introduced between what the reference plane the metrology system measures against, and the mirror surface position that is actually relevant for high-contrast imaging. Understanding how metrology elements are mounted to the mirror segments they measure is a critical component of the sensing and control error budget.

\paragraph{Model Validation Activities}
Finally, sensing and control algorithms must be developed that can adequately ``nest'' control loops of differing bandwidth and sensitivity to ensure uniform control over all spatial and temporal frequencies of concern\cite{Redding2024}. The Keck and IRIS testbeds will be used to help validate such models early in the development process, before transitioning activities to the TRL 5 demonstration testbeds such as HOST.

\subsection{Low-Vibration and Pointing Control Systems}
\textbf{Gap Type:} 	Technology

\textbf{Est. TRL:} 	4

\textbf{Importance:}\textbf{	}  3 -- Threshold

\textbf{Urgency:}\textbf{	}  1 -- Long Term

\subsubsection{Discussion}
Precise attitude control and mitigation of high frequency optical error (vibration) are critical to meeting coronagraph contrast stability requirements. During observations, line-of-sight (LOS) error must be maintained below $\sim$0.2 milli-arcsecond (mas) RMS, while WFE change must remain $\sim$1 picometer RMS. These stringent stability requirements necessitate a combination of active pointing control loops and vibration isolation and damping systems.

\begin{figure}[htbp]
\centering
\includegraphics[width=0.9\linewidth]{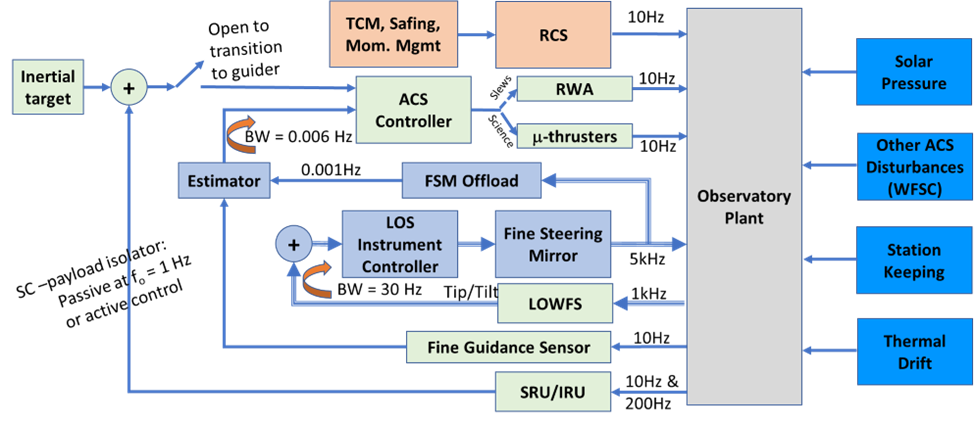}
\caption{ACS Block Diagram}
\label{fig:21}
\end{figure}

The typical nested pointing control system used to achieve several milli-arcsecond closed-loop stability includes the Attitude Control System (ACS), which maintains or changes the observatory orientation, and the fine guidance loop, which rejects LOS disturbances at higher temporal frequencies. Figure \ref{fig:21} shows one possible configuration where reaction wheel assemblies (RWAs) are used for slewing, then hand off to micro-thrusters during observations, controlling LOS with a 0.006 Hz bandwidth. Residual pointing error is corrected using the FSM, with a bandwidth of 30 Hz. Several alternative architectures are being considered to demonstrate viable options and to help derive technology requirements.

\begin{figure}[htbp]
\centering
\includegraphics[width=0.9\linewidth]{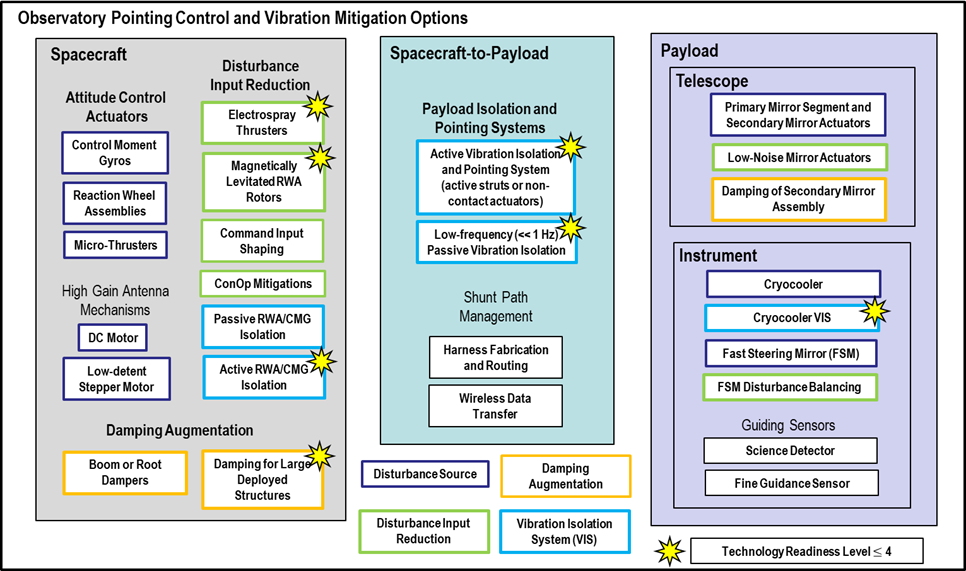}
\caption{LOS Stability Architecture Options}
\label{fig:22}
\end{figure}

Active pointing control loops alone are often insufficient to reduce LOS error to the required level due to their limited bandwidth. Additional mitigation approaches such as active/passive vibration isolation and damping systems are necessary to achieve acceptable LOS and WFE jitter at frequencies above the control system bandwidths. Figure \ref{fig:22} provides a `menu' of potential technologies that could be used to build the pointing and vibration mitigation architectures for HWO. The disturbance inputs identified in this diagram are associated with sources that operate continuously (such as ACS actuators and cryocoolers) or require significant settling time (such as the high gain antenna (HGA) gimbal system).

Many of the potential jitter mitigation options shown in Figure \ref{fig:22} are already at high TRL with flight-proven experience. However, some options will need to be appropriately scaled to meet HWO requirements and lifetime needs, which may require further engineering development and/or technology maturation efforts. Specifically, the approaches that require technology maturation are:

\begin{itemize}
\item Disturbance Input Reduction: Technology that enables extremely low induced vibration and/or novel disturbance cancellation techniques applicable to various control actuators and cryocoolers for HWO.
\item Vibration Isolation System (VIS): Either passive systems with a very low corner frequency ($<<$1 Hz) or active systems that incorporate technologies such as non-contact actuators or active strut systems.
\item Damping Augmentation: Technologies that significantly increase structural/material damping for large structures and/or low-frequency deployed assemblies.
\end{itemize}
The performance requirements for these technologies are being evaluated through system-level architecture and trade analyses. The final system architecture will also consider cost, schedule, feasibility, and accommodation complexities to meet the stringent HWO jitter requirements. For items that are not currently TRL 5 (marked with yellow stars in Figure \ref{fig:22}), subsequent sections provide a plan for maturing these technologies to this level.

\subsubsection{TRL 5 Milestone Definition}
A TRL 5 demonstration of the low-vibration and pointing control system technology requires:

\begin{itemize}
\item Ground testbed demonstrations that validate the proposed technology performance;
\item Test measurements with accuracy appropriate for evaluating induced vibration, isolation system performance across the required frequency range, and/or achieved damping in flight-like environments;
\item Measured performance that agrees with predicted performance within expected model uncertainty and test measurement noise limits;
\item For active systems, validation of electrical components/brassboards with flight-like computers running control algorithms that drive relevant hardware in a closed-loop system under appropriate environmental conditions;
\item Incorporation of realistic shunt paths to evaluate performance degradation due to mechanical shorts in the VIS and/or damping systems;
\item Achievement of "medium fidelity of analysis" standards.
\end{itemize}
The primary objective is to test the hardware assembly, along with electronics and software components for active systems, in a relevant environment to demonstrate system performance (e.g., transmissibility, damping, and/or induced vibration). The test model should have sufficient fidelity to accurately predict test measurement results. Any deviations between test and model should be evaluated and incorporated into the error budget. Beyond the test demonstration, additional analysis should identify and characterize potential failure modes and any life-limiting factors.

\subsubsection{Development Strategy \& Roadmap}
The general development strategy for the low-vibration and pointing control systems technology will leverage investments made through the ROSES Appendix D.19 programs, including the Disturbance-Free Payload (DFP) concept\cite{Zeledon2026}. Currently, FY23 D.19 efforts are underway addressing many technologies in the Ultra-stable Telescope Systems Track. Simultaneously, the EACs design and analysis efforts are progressing to demonstrate the feasibility of meeting pointing control and jitter requirements across multiple architecture configurations. This work will inform and support the proposed FY25 D.19 proposals. Once selections are made from the FY25 proposal call, this section will be updated with more detailed information about the chosen development path.

\subsection{Deployable Systems}
\textbf{Gap Type:} 	Engineering (TBD)

\textbf{Est. TRL:} 	TBD

\textbf{Importance:} 	3 -- Threshold

\textbf{Urgency:}\textbf{	}  1 -- Long Term

\subsubsection{Discussion}
The Systems Team is currently working through several architecture and systems trade studies to define the EACs. These studies are aimed at developing an overall HWO concept that achieves the science objectives defined in the 2020 Decadal Survey in Astronomy and Astrophysics, subject to mission constraints, including available launch vehicle capabilities, the need for micrometeoroid protection, operating temperature, etc. Deployable systems will play a critical role in the HWO system, including but not limited to a deployable barrel assembly, solar arrays, sunshields, structural hinges and latches, etc.

At the current point in the systems trade studies, it is not clear what specific components and/or sub-systems are beyond the state-of-the-art. For now, this technology lane is a placeholder to capture any future technology or engineering development activities that may be identified by the Systems Team. The most likely area of focus will be on deployed systems' stability, particularly associated with latching mechanism stability, lurch, and creep. While JWST's deployment systems represent the state-of-the-art, dynamic stability requirements at the picometer level may drive these assemblies beyond the state-of-the-art.

\section{High-Sensitivity Ultraviolet and Visible Instrumentation Techniques}
\subsection{Overview}
{\footnotesize
\begin{xltabular}{\textwidth}{@{} >{\raggedright\arraybackslash}p{0.10\textwidth} >{\raggedright\arraybackslash}X >{\raggedright\arraybackslash}X >{\raggedright\arraybackslash}X >{\raggedright\arraybackslash}X >{\raggedright\arraybackslash}X >{\raggedright\arraybackslash}X @{}}
\caption{High-sensitivity UV and Visible Instrument Technologies}\label{tab:13}\\
\hline
\textbf{ID} & \textbf{Lane} & \textbf{Driving Factor(s)} & \textbf{Est. Current TRL} & \textbf{Importance} & \textbf{Urgency} & \textbf{Gap Type} \\
\hline
\endfirsthead
\caption[]{(continued)}\\
\hline
\textbf{ID} & \textbf{Lane} & \textbf{Driving Factor(s)} & \textbf{Est. Current TRL} & \textbf{Importance} & \textbf{Urgency} & \textbf{Gap Type} \\
\hline
\endhead
\hline
\multicolumn{7}{r}{\textit{(continued on next page)}}\\
\hline
\endfoot
\hline
\endlastfoot
1 & Far-UV Mirror Coatings & Broadband coverage with high-reflectivity down to 100 nm. \newline High uniformity and low near-angle scattering. \newline Polarization properties characterized. & 4 & Threshold & Urgent & Tech \\
2 & Near UV/VIS Detectors & Large-format, low noise, high QE. & 4 & Baseline & Urgent & Eng \\
3 & Far-UV Detectors & Large-format and high-QE. \newline Solar-blind. & 3-5 & Baseline & Urgent & Eng \\
4 & Multi-object Selection & Micro-shutters, micro-mirrors, or slicers for multi-object and/or integral field spectroscopy. & 3-5 & Baseline & Urgent & Tech \\
5 & UV Gratings \& Filters & High out-of-band rejection. \newline Curved substrates for aberration control. & 4-6 & Baseline & Long Term & Eng \\
\end{xltabular}
}
\subsection{Far-UV Mirror Coatings}
\textbf{Gap Type:} 	Technology

\textbf{Est. TRL:} 	4

\textbf{Importance:} 	3 -- Threshold

\textbf{Urgency:}\textbf{ }	2 -- Urgent

\subsubsection{Discussion}
The mirror coatings necessary for HWO to be sensitive down to 100 nm involve the use of materials and deposition techniques that can deliver the appropriate reflectivity at the appropriate level of uniformity with knowledge about both polarization aberration terms and the resilience and stability for a long mission lifetime at the conditions typical of L2.

Based on analysis with the community, the GOMAP Technical Assessment Group (TAG) committee identified the requirements in Table 14 for mirror coatings for the necessary performance metrics for the baseline mission. Note that the performance of the mirror coatings is regarded as a threshold capability since it enables the wavefront control necessary for coronagraphy while still providing the throughput down to 100 nm needed for transformational astrophysics prioritized by the Decadal Survey 2020.

\begin{xltabular}{\textwidth}{@{} >{\raggedright\arraybackslash}X >{\raggedright\arraybackslash}X @{}}
\caption{Notional HWO Telescope Mirror Coating Requirements}\label{tab:12}\\
\hline
\textbf{Parameter} & \textbf{Performance Needed} \\
\hline
\endfirsthead
\caption[]{(continued)}\\
\hline
\textbf{Parameter} & \textbf{Performance Needed} \\
\hline
\endhead
\hline
\multicolumn{2}{r}{\textit{(continued on next page)}}\\
\hline
\endfoot
\hline
\endlastfoot
Single mirror coating reflectance & Reflectance at 102 nm: $>$ 60\% \newline Reflectance at 103 nm: $>$ 75\% \newline Average reflectance for 105 nm - 250 nm: $>$ 80\% \newline Average reflectance for 250 nm - 2.5 microns: $>$ 90\% \\
Resilience (Ground) & $<$ 1\% degradation of reflectance at 103 nm with 1 month exposure to relative humidity (RH) 50\%. \\
Resilience (Space) & $<$ 1\% degradation of reflectance at 103 nm per year under L2 environment conditions (radiation, etc.) \\
Scalability to $>$ 1 m class optics & Successful demonstration on a mirror larger than 50 cm, and up to 2 m. \newline Develop a facility capable of measuring the reflectance of optics of this size from 100 nm to 2.5 \textmu{}m. \\
Care and Handling & Procedure for optic cleaning in the event of contamination during integration \& test (I\&T) / handling \\
Reflectance Uniformity & $<$ 1\% reflectance variation across the surface of one mirror blank ($\sim$1.8m) for wavelengths $>$ 200 nm (this parameter only applies to wavelengths used for high-contrast imaging). \newline $\sim$3\% reflectance uniformity between successive mirror blanks coated. \\
Polarization Aberrations & Measure the complex index of refraction and polarization aberrations (including cross-terms) for various angles of incidence and optic curvature. This data must be made publicly available to evaluate any on coating for HWO. \\
Facilities & Metrology facilities for these measurements do not currently exist. Significant TRL advancement is also possible through direct demonstration of UV-reflective coatings in a HWO high-contrast imaging system \\
\end{xltabular}
\paragraph{Current State-of-the-Art}
The current state-of-the-art can be found in Scowen et al. (2026)\cite{Scowen2026}.
\subsubsection{TRL 5 Milestone Definition}
A TRL 5 demonstration of a Far-UV Mirror Coating consists of:

\begin{itemize}
\item Coating deposition on a meter-class substrate with material relevant to HWO (e.g., polished ULE or Zerodur)
\item Complete performance characterization of coating reflectivity, surface roughness, and polarization properties demonstrating uniformity over the full optic area
\item Run-to-run repeatability of coating deposition demonstrated via smaller substrates distributed over meter-class area a minimum of three times
\end{itemize}
If the meter-class substrate is less than the full size expected for HWO (i.e., $<$ 1.5 m), then process scalability from the demonstrated size to full size must be thoroughly documented and justified to be facility-limited only. That is, the process must be limited by chamber size only, with no other significant changes to the deposition process required.

Additional lifetime testing of the coating, subject to expected environments during ground integration and test and on-orbit operation, should also be performed in support of eventual TRL 6 validation.

\subsubsection{Development Strategy \& Roadmap}
The primary effort involves upgrading existing chamber facilities at NASA or industry partners to accommodate larger optics (up to at least 2-meters in diameter). Once appropriate facilities are available, trial runs using small scale coupons will be performed to validate the new facilities and provide preliminary demonstration of the desired coating properties. Full-scale coating runs and repeatability tests will then follow.

Two additional objectives will be pursued in parallel to determine the best approach for achieving coating uniformity over the larger area:

\begin{itemize}
\item Perform a comprehensive study to develop new correction masks\cite{Villa2000} for thickness uniformity by using a single deposition source (of LiF or MgF$_{2}$) in large-area thin films that may achieve the required uniformity in reflectance (< 1\%). The use of a mask will be combined with a rotation scheme to deliver reflectance uniformity over a mirror substrate area around 1.3-meter.
\item Add a second precursor source of trimethylaluminum (TMA) to enable a full Atomic Layer Deposition (ALD) process of AlF {\textsubscript{3}} to protect the Al to form Al+AlF$_{3}$ in the existing 2-meter chamber.
\end{itemize}
The goal in this effort is to demonstrate whether the source mask and rotation or ALD schemes will be required to achieve uniformity of protected Al-mirror substrates that are comparable to the size propose for HWO segments.

\subsection{Near UV/VIS Detectors}
\textbf{Gap Type:} 	Engineering

\textbf{Est. TRL:}\textbf{	}  4

\textbf{Importance:}\textbf{	}  2 -- Baseline

\textbf{Urgency:}\textbf{	}  2 -- Urgent

\subsubsection{Discussion}
For the baseline astrophysics instruments (UVI and HRI), there are applications that include both ``traditional'' integrating devices, where sensitivity is prioritized over background noise, and photon-counting devices, where low-noise is prioritized over sensitivity. For this reason, many of the devices outlined in Section 4.5 for low-noise coronagraph detectors are applicable to this section as well, where the operational mode can be switched based on the readout circuit that is used. For example, EMCCDs can operate in either traditional CCD mode by simply bypassing the electron-multiplying readout register.

The discussion in this section is therefore focused on the two key discriminating factors between the coronagraph detector needs and those of the NUV/visible instruments: NUV sensitivity, and form factor.

Both the UVI and HRI instrument are expected to require sensitivity at wavelengths as low as 200 nm over fields-of-view (FOV) greater than 2' $\times$ 2'. The former requirement implies thick, fully-depleted sensors leveraging delta-doping or custom anti-reflection (AR) coating processes to push the average QE above 60\% over the 200 -- 400 nm band. The latter requirement implies large format arrays (i.e. 4K $\times$ 4K or 8K $\times$ 8K pixels) that can be tiled into larger mosaics with minimal gaps between sensors (similar to the Roman WFI focal plane array).

\paragraph{Current State-of-the-Art}
The current state-of-the-art can be found in Scowen et al. (2026)\cite{Scowen2026}.
\subsubsection{TRL 5 Milestone Definition}
%\begin{figure}[htbp]
%\centering
%\includegraphics[width=0.9\linewidth]{figures/fig013.png}
%\caption{28}
%\label{fig:25}
%\end{figure}
For traditional integrating devices, mature devices already exist and demonstrations of enhanced UV sensitivity via delta-doping and AR coating have been performed. For the devices discussed in Section 4.5, the TRL 5 milestone consists of demonstrating UV-enhanced sensitivity via thickening, delta-doping, AR coating or a combination thereof, including both integrating and photon-counting readout modes of the device. The TRL 5 milestone should include a description of the path to either larger format devices or tiling capabilities.

\subsubsection{Development Strategy \& Roadmap}
The development strategy will leverage the same process as described in Section 4.5.3 whereby the vendors will also be solicited for plans to include ``NUV enhanced'' devices as part of their reports. Selected proposals will include options for delivering devices for both coronagraph high-contrast imaging and large-form NUV enhanced devices for the UVI and/or HRI, contingent upon available funding.

\subsection{Far-UV Detectors}
\textbf{Gap Type:}\textbf{	}  Engineering

\textbf{Est TRL:}\textbf{	}  3 -- 5

\textbf{Importance:}\textbf{	}  2 -- Baseline

\textbf{Urgency:}\textbf{	}  2 -- Urgent

\subsubsection{Discussion}
FUV detectors are fundamentally different technologies than the more conventional solid state detectors used in the previous section.  FUV detectors are a long-standing technology that usually involves micro-channel plate (MCP) detectors, which is a 40+ year old mission proven technology. Modern variants using ALD deposited materials and borosilicate glass enable large customizable, and stable, form factors.  Photocathode choices are used to tailor the spectral response. These devices are typically operated at room temperatures. Open face FUV devices that operate down to below 120nm are more sensitive to contamination than sealed window NUV detectors, but have proven stable on long term planetary missions.  MCP detectors are ``photon counting'' asynchronous devices allowing images to be built using the photon X and Y positions and arrival time with intrinsic time resolution of less than a microsecond, they do not have any frame read time. We discuss in this section a couple of potential approaches that can be used to provide sensitivity down to 100nm that the HWO science drivers are asking for in a low-risk but high performance way.

Again, based on analysis with the community, the GOMAP TAG committee identified the notional requirements in Table 16 for FUV detectors for the baseline astrophysics mission.

\begin{xltabular}{\textwidth}{@{} >{\raggedright\arraybackslash}X >{\raggedright\arraybackslash}X @{}}
\caption{Notional HWO Far-UV Detector Requirements}\label{tab:13}\\
\hline
\textbf{Parameter} & \textbf{Performance Needed} \\
\hline
\endfirsthead
\caption[]{(continued)}\\
\hline
\textbf{Parameter} & \textbf{Performance Needed} \\
\hline
\endhead
\hline
\multicolumn{2}{r}{\textit{(continued on next page)}}\\
\hline
\endfoot
\hline
\endlastfoot
Form Factor (Number of Elements) & A 2' FOV, assuming 50 mas resolution requires $>$2400 resolution elements in the cross-dispersion axis. \newline R = 30,000 at 120 nm and 3$\times$ grating modes to cover the 100 - 200 nm bandpass requires $\sim$8,750 resolution elements (resels) in the dispersion direction. \\
Form Factor \newline (Physical Size) & Assuming 25 micron resolution elements requires focal plane arrays of at least 60 mm (cross-dispersion axis) x 220 mm (dispersion axis). \newline Longer (dispersion axis) arrays are desirable to fully sample the multi-object spectrograph FOV in the spectral bandpass. \\
Pixel Size/Resolution & $\sim$ 15-25 \textmu{}m resolution elements (5-8 \textmu{}m pixels at 3$\times$ pixels / resel) \\
Quantum Efficiency (QE) & HWO science concepts baseline photon-counting capability for the FUV to maximize detection potential for low surface-brightness sources. \newline QE (103 nm): $>$30\% \newline QE ($>$110 nm): $>$40\% \\
Background: & Total of all intrinsic sources (read, dark, CIC, etc.): $<$ 5$\times$10{\textsuperscript{6}} e-/pixel/s \\
Photon Counting Device Rate Limit (Global): & $>$ 1 M count/s per device with $<$ 5\% Dead Time \\
Photon Counting Device Rate Limit (Local): & $>$ 100 counts/s/resel \\
Radiation Hardness: & Tolerant to an L2 radiation environment or a has a viable shielding solution \\
Contamination Mitigation: & FUV-sensitive detectors that operate at cold temperatures require strategies for hydrocarbon contamination mitigation (see e.g. HST-WFPC2; Koekemoer et al., 2002) \\
\end{xltabular}
\paragraph{Current State-of-the-Art}
The current state-of-the-art can be found in Scowen et al. (2026)\cite{Scowen2026}.
\subsubsection{TRL 5 Milestone Definition}
\paragraph{Large format ALD MCPs}
MCPs are electron multiplying devices that have been used in suborbital and satellite imaging particle and photon detectors for decades, including aboard the Hubble Space Telescope. Recent improvements, including high temperature glass substrates and applying functional coatings via ALD, have demonstrated increased detector QE, reduced background rates, improved gain stability. Borosilicate substrate ALD MCPs have been successfully used in sub-orbital flights for open face formats of 20$\times$20 cm{\textsuperscript{2}} and 10$\times$10 cm{\textsuperscript{2}} in pore sizes down to 10 \textmu{}m. ALD MCPs have also been used in sealed tube detectors with 5$\times$5 cm{\textsuperscript{2}} format with 100 events resel-1 sec-1 which have been qualified to TRL 5/6. Implementation in detectors for HWO qualification requires a supply of 10 $\times$ 10 cm{\textsuperscript{2}} ALD MCPs with 10 \textmu{}m capillaries and state of the art ALD coatings. These are needed both for photocathode work and demonstration of sealed tube devices in 10 $\times$ 10 cm$^{2}$ format. High performance borosilicate substrates have been made/flown but manufacturers need to work on consistency and yields through improvements in the process. Alternative substrates show promise for improved performance (e.g., larger open area ratio), but need continued development to evaluate if they are a viable option. Continued development of the resistive and emissive ALD coatings will also realize performance improvements. Our expectations for HWO TRL 5 demonstrations include 10$\times$10 cm{\textsuperscript{2}} format sealed tube compatible MCPs supporting high QE photocathodes, intrinsic noise levels of $<$ 4$\times$10{\textsuperscript{-6}} events resel{\textsuperscript{-1}} sec{\textsuperscript{-1}}, 100 events resel{\textsuperscript{-1}} sec{\textsuperscript{-1}} capability, and high gain stability with lifetime of $>$ 1014 events cm{\textsuperscript{-2}}.

\paragraph{Photocathodes}
Photocathodes are employed for MCP based detectors to detect and convert incoming photons to photoelectrons. Photocathodes can be opaque (these are deposited directly onto the surface of the MCP to intercept incoming radiation) or semitransparent and deposited on the underside of the optical entrance window of a sealed device. Alkali halide, Alkali Telluride and multialkali FUV/NUV photocathodes have been used for sub-orbital, and satellite payloads numerous times with some formats up to 100$\times$100 mm{\textsuperscript{2}}. For HWO at TRL 5 we wish to target the QE and format size appropriate for HWO implemented with the MCPs and appropriate device configuration. Open face devices with CsI photocathodes up to 100 mm format with high QE ($>$ 40\% at 103 nm, out of band rejection 1$\times$10{\textsuperscript{7}}) are proven in flight. Additionally sealed tube Bialkali photocathodes (QE of $\sim$25\% at $>$ 120 nm) have been made and flown in 50 mm MCP detector formats (many have been made in 200 mm format but not flown). For TRL 5 demonstration for HWO we are envisaging $>$ 40\% QE 110 nm - 140 nm with optimizations of alkali halide photocathodes (CsI, KI, CsBr) deposited on state-of-the-art ALD MCP 100 mm open face formats suitable for HWO. Above 120 nm sealed tubes with bialkali photocathodes have demonstrated $>$ 30\% QE and in some cases $>$ 40\%. We propose to employ these optimized deposition techniques to demonstrate 35\% QE at 140 nm - 200 nm, while preserving good out-of-band rejection, on 100 mm sealed tube demonstration devices. Additionally, both photocathode types can be used simultaneously in the same sealed tube device(s) to further enhance QE.

\paragraph{Readout Systems}
Cross-Strip (XS) readouts are charge collection and position encoding systems for MCP based detectors. XS anodes with SMT type readout systems have flown successfully on sub-orbital payloads in formats up to 100 $\times$ 100 mm{\textsuperscript{2}}. These have performance characteristics ($<$ 20 \textmu{}m resolution, 1 MHz with $\sim$10\% deadtime) close to HWO goals. One of our pathways to TRL 5 HWO compatibility is updating and qualification of the high TRL SMT electronics. Preamplifiers, analog-to-digital converters (ADCs), and field-programmable gate arrays (FPGAs) with adequate radiation tolerance exist to accomplish this allowing preservation of the resolution and imaging performance whilst improving deadtime to 5\% at 1 MHz. A second initiative is to bootstrap on an existing effort to produce ASIC devices to perform the same processing of XS signals. The GRAPH ASIC has shown promise in early testing under an APRA but needs further development/iteration/qualification to meet HWO goals. We anticipate that this would result in power/mass/volume savings while achieving rad hardness, and resolution goals for HWO FUV/NUV even exceeding the global rate/deadtime goal.

Timepix ASICs are programmable general-purpose readouts developed at CERN primarily for high-resolution particle tracking and imaging applications. The latest version, Timepix4 (TPX4), has a 28 $\times$ 25 mm$^{2}$ active area consisting of 448$\times$512 pixels with a 55 \textmu{}m pixel pitch and features 4-side abutment for scaling to larger formats. An MCP detector with a single TPX4 chip and 2 $\times$ 2 arrays of earlier versions have been demonstrated in sealed tubes. Timepix-based sensors have demonstrated the potential to extend the capabilities of photon counting detector technology for much higher counting rates (exceeding 107 events cm{\textsuperscript{-2}} sec{\textsuperscript{-1}}), with high spatial resolution (15 \textmu{}m centroiding), high dynamic range (thousands of simultaneous events), long lifetime (because of very low gain operation of $<$ 105). For HWO, further development is proposed to achieve larger arrays of TPX4 and advance the technology to higher TRLs in both sealed tube and open face detectors.

\paragraph{Large format sealed tube and open face devices}
We anticipate incorporation of the readout, ALD MCP, and photocathode work into representative HWO 100 mm format detectors raising the TRL to 5 / 6, while establishing a sealed tube processing facility to enable HWO-type devices to be built thereafter. Initial efforts for 100 mm sealed XS tubes is currently funded by APRA will conclude in $\sim$1 year. A follow on for development of fully functional detectors with HWO compatible configuration would be needed, including performance evaluation and environmental tests.

\subsection{Multi-Object Selection}
\textbf{Gap Type:}	 Technology

\textbf{Est. TRL:} 	3 -- 5

\textbf{Importance:}\textbf{	}  2 -- Baseline

\textbf{Urgency:}\textbf{	}  2 -- Urgent

\subsubsection{Discussion}
Multi-object spectroscopy (MOS) is a technology development priority of both the Decadal Survey: Pathways to Discovery in Astronomy and Astrophysics for the 2020s (PDAA) and the current NASA Cosmic Origins Program. Aperture control methods that are popular in ground-based MOS applications (e.g., robotically configured fibers and punch plates) are not practical options for spaceflight.

The main target for the MOS capability is the HWO UVI which is considered a primary instrument for the mission. The science drivers for the mission are being defined at this time by HWO TMPO and call for the ability to take multiple spectra in sparsely populated fields and across both extended objects and point like targets, ranging in context from cosmological targets to regions of star formation across galaxies to individual resolved extended objects to solar system targets. In all cases a wide field of view (several arcminutes) is required with individual apertures subtending a few hundred milli-arcseconds, while delivering contrast performance between adjacent spectral channels of $\sim$1 part in 10{\textsuperscript{5}} for stellar sources. The MOS assembly would need be located at an intermediate point in the optical path of a telescope system to enable this high contrast.

Based on analysis with the community, the GOMAP TAG committee identified the following requirements in Table 18 for multi-object selection for the baseline astrophysics mission.  Note that the requirements below make reference to a parallel IFU spectroscopy mode -- since this remains an open discussion and had no precursor in either the LUVOIR or HabEx studies, the requirements are still being determined. What is presented here are the technologies for provision of the MOS mode itself.

\begin{xltabular}{\textwidth}{@{} >{\raggedright\arraybackslash}X >{\raggedright\arraybackslash}X @{}}
\caption{Notional HWO MOS Requirements}\label{tab:15}\\
\hline
\textbf{Parameter} & \textbf{Performance Needed} \\
\hline
\endfirsthead
\caption[]{(continued)}\\
\hline
\textbf{Parameter} & \textbf{Performance Needed} \\
\hline
\endhead
\hline
\multicolumn{2}{r}{\textit{(continued on next page)}}\\
\hline
\endfoot
\hline
\endlastfoot
Form Factor (Number of Elements) & Anticipated HWO UV MOS designs cover 2' $\times$ 2' with a resolution of 50 mas, with any given spectral element no more than 5$\times$ the size of the telescope PSF ($\sim$ 250 mas) to maintain spectral resolution for extended sources. \newline An IFS would be a smaller FOV (a few arcseconds) and unlikely to meet both the field-of-view and resolution goals -- more likely no greater than R $\sim$ 5000 and could be a strong science enhancement as an additional channel. \\
Form Factor: \newline (Physical Size) & Minimum 80 mm length/width for a MOS (Assumes element size of 0.1 mm to capture > 98\% of the light from a 20 micron PSF at 50 mas resolution, 2' FOV) \\
Fill Factor (Open area ratio within FOV): & $>$ 50\% \\
Yield: & $>$ 90\% functionality, two-side buttable if form factor cannot be met by a single device \\
Targeting: & Re-programmable \newline Actuation timescales < 3 seconds for tracking moving objects. \newline Does not apply to an IFS. \\
Contrast: & Sufficient to reduce light leak to at least a factor of 2$\times$ less than typical detector backgrounds. \newline For interplanetary Lyman-a at R$\sim$1500, an element FOV of 150 mas\textasciicircum{}2, and an effective area of 10,000 cm\textasciicircum{}2, the required contrast to reduce stay light to half an MCP background level at L2 of 4$\times$10{\textsuperscript{8}} /micron is $>$ 20,000 for extended emission. \newline Contrast $>$ 100,000 for stellar sources. \newline Does not apply to an IFS. \\
Source Multiplexing: & Spectroscopic measurement of > 200 objects simultaneously in any grating mode/resolution. \\
Lifetime: & Demonstrated to meet HWO lifetime goals $\sim$ 10\textasciicircum{}6 actuations \\
Bandpass: & Source multiplexing over the entire UV instrument bandpass, including to 100 nm. \\
\end{xltabular}
\paragraph{Current State-of-the-Art}
The current state-of-the-art can be found in Scowen et al. (2026)\cite{Scowen2026}.

\subsubsection{TRL 5 Milestone Definition}
\paragraph{Microshutter Arrays (MSAs)}
Laying out the plans to achieve TRL 5, the following tasks shown in Table 19. Note that some of the tasks listed are already supported by existing SAT funds.

\begin{figure}[htbp]
\centering
\includegraphics[width=0.9\linewidth]{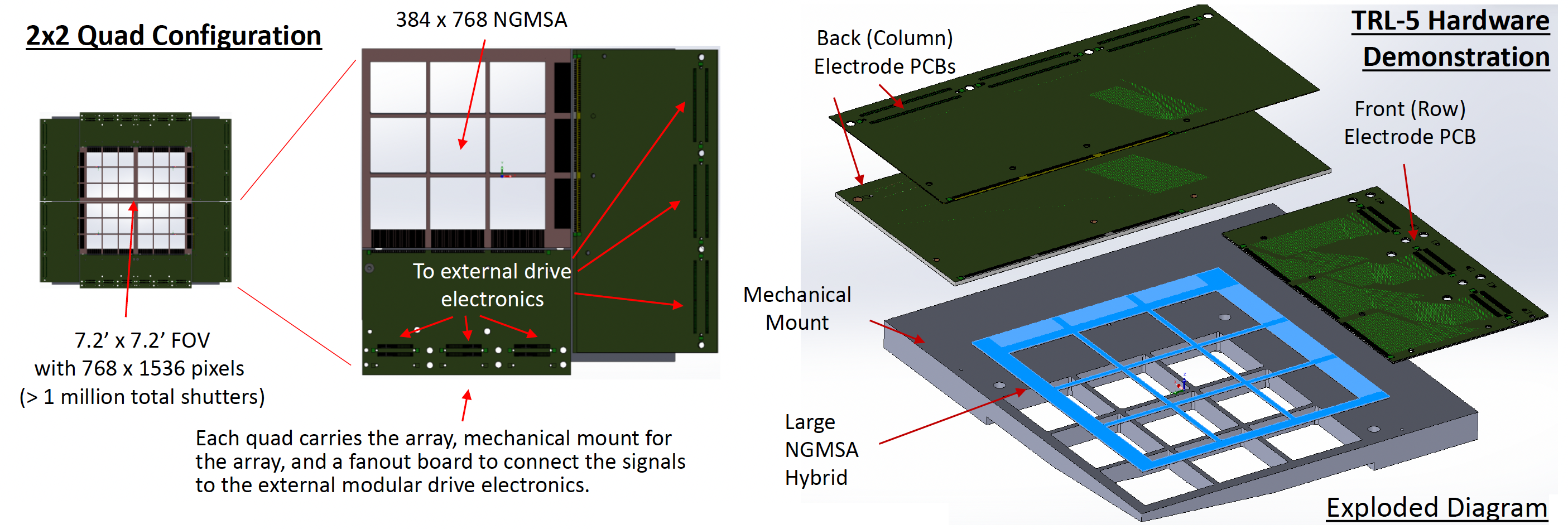}
\caption{NGMSA TRL 5 demonstration unit.}
\label{fig:29}
\end{figure}

{\footnotesize
\begin{xltabular}{\textwidth}{@{} >{\raggedright\arraybackslash}p{0.08\textwidth} >{\raggedright\arraybackslash}X @{}}
\caption{NGMSA TRL 5 Milestones. Tasks 5.1 - 5.7 is supported by NNH22ZDA001N-SAT (FY24-26);
Task 5.8 needs support from a new SAT (FY26-28) and HWO PO (FY26-FY29).}
\label{tab:fig014_ng_msa}\\
\hline
\textbf{5.1} & The final blade shape shall be down selected from (8) 128 X 64 designs, a new large format photomask set generated and arrays fabricated with light shields. \\
\hline
\textbf{5.2} & A new silicon fanout board and array mechanical mount enabling 2-side buttability shall be designed, analytically modeled and fabricated. \\
\hline
\textbf{5.3} & Arrays shall be mounted to the fanout boards, installed and operated in the 2D vacuum actuation system. \\
\hline
\textbf{5.4} & Life testing shall occur in the 2d vacuum actuation system, consisting of 100,000 cycles on all of the shutters \\
\hline
\textbf{5.5} & Radiation testing on the NGMSA module shall be performed to the TID level expected at L2 with 2.5mm Al shielding. \\
\hline
\textbf{5.6} & An array shall be mounted to a substrate and vibrated to GEVS levels. Shock and acoustics testing shall be performed. \\
\hline
\textbf{5.7} & The open/closed contrast ratio shall be measured. \\
\hline
\textbf{5.8} & Performance and yield shall be improved to meet the HWO requirement. \\
\hline
\end{xltabular}
}
To demonstrate TRL 5, a two-side buttable entirely electrically actuated NGMSA quad, shown in Figure \ref{fig:29} with specifications in Table 20, will be demonstrated as a multi object selector with $>$ 105 UV contrast through a series of tests demonstrating HWO requirements are met in a relevant environment. This TRL 5 demonstration hardware will include an NGMSA assembly (an array hybridized with a substrate) and printed circuit board (PCB) fanout boards, all supported by an array mechanical mount. An auxiliary set of drive electronics, not part of the TRL 5 hardware, will be used for array actuation.

{\footnotesize
\begin{xltabular}{\textwidth}{@{} >{\raggedright\arraybackslash}p{0.22\textwidth} >{\raggedright\arraybackslash}X @{}}
\caption{NGMSA TRL 5 demonstration specifications}\label{tab:fig033_ngmsa}\\
\hline
\textbf{Attribute} & \textbf{Spec/Attribute Value} \\
\hline
\endfirsthead
\caption[]{(continued)}\\
\hline
\textbf{Attribute} & \textbf{Spec/Attribute Value} \\
\hline
\endhead
\hline
\multicolumn{2}{r}{\textit{(continued on next page)}}\\
\hline
\endfoot
\hline
\endlastfoot
FOV & $>$ 7.2$' \times$ 7.2$'$ for 2 $\times$ 2 butted arrays \\
Form Factor & Two side buttable with 76.8~mm length/width for each quad \\
Pixel Numbers & 384 $\times$ 768 (294,912 shutters) for each quad; $>$ 1 million total shutters for 2 $\times$ 2 butted arrays \\
Fill Factor & 50\% -- 60\% \\
Yield & $>$ 90\% functionality \\
Contrast & $>$ 20,000 for emission sources; $>$100,000 for stellar sources \\
Source Multiplexing & $>$ 256 objects simultaneously \\
Lifetime & $>$ $10^6$ actuations \\
Bandpass & Wavelength independent within UV/Optical range (100~nm -- 1000~nm) \\
Actuation Method & Entirely electrical without macro mechanical components \\
Operation Voltage & $-40$ to 80 DC Volts \\
Size & $<$ 180~mm $\times$ 180~mm for each quad \\
Power & $\sim$ 40 Watt (peak) for each quad (driver dependent) \\
Mass & $<$ 1 Kg for each quad \\
\end{xltabular}
}

\paragraph{Digital Micromirror Devices}
DMDs involve the use of a MEMs device that needs to be maintained in a clean environment. TI provides such containment by fusing a borosilicate glass to a Kovar frame. The window is hermetically sealed to the Kovar lid via a glass-to-metal seal in a controlled, clean environment, using specialized equipment. The lid assembly is then seamwelded to the Kovar platform surrounding the micromirror array.  Borosilicate glass is transparent down to 300 nm but opaque to UV radiation. Replacing the standard borosilicate window with a UV-transmissive material (e.g., MgF{\textsubscript{2}}, Sapphire, fused silica) extends the operational range of the DMD to approximately 200 nm. To reach wavelengths as short as 120 nm one has to look at LiF, MgF{\textsubscript{2}}, CaF{\textsubscript{2}} or, ideally, a solution where the window is displaced when the DMD is operated in a controlled environment.

Replacing the DMD window with a custom one is not a trivial task. With TI uninterested in customizing DMD windows, window replacement is a procedure that has attracted the attention of a few US companies. A commercial firm, L1 Standards and Technologies Inc., has done so, supplying a handful of such devices. These windows have been replaced by simply machining away the seam-welded joint and epoxying an oversized window to the Kovar platform; an approach followed by other researchers interested in rewindowing DMDs. However, functional testing of these devices showed that the reliability of these re-windowed DMDs was adversely affected. Of the six window-replaced DLP7000 DMDs processed by L1 Standards and Technology, only 50\% operated successfully through three cryogenic cycles. This is in stark contrast to the six off-the-shelf unaltered devices that performed essentially flawlessly under the same test conditions.

The UV reflectance of the DMD (ignoring losses due to fill factor and diffraction) is less than that of pure aluminum, because the DMD mirror surfaces are made with an aluminum alloy. Regardless on the type of coating, below about 200 nm, the DMD reflectance drops rapidly if they are not protected against the formation of an aluminum oxide layer.  Recoating the mirrors with aluminum is risky due to the possibility of interfering with the underlying electronics, creating short-circuits, or creating stiction if the coating material penetrates between the mirror gaps. In collaboration with GSFC, this operation has extended the use of the DMD to approximately 100 nm, by recoating the DMD with high-purity aluminum and protecting it with a thin film of LiF or AlF{\textsubscript{3}}.

These materials applied after the first pass of Aluminum coating prevent the formation of Aluminum Oxide, disruptive to the FUV mirror's reflectivity [Quijada et al., 2018]. It has been found that recoating the DMDs causes them no damage. The re-coated DMD remained functional with no obvious operational differences between the coated and original regions, except improved reflectance. These initial tests show that DMDs can survive this type of recoating, suggesting that DMDs can be made usable in the 100 - 400 nm range, if the coating is protected with a fluoride film. We plan to continue our collaboration with GSFC coating a number of devices for FUV throughput once the rewindowing process is consolidated. In general, thanks to these advancements in coatings, one can now envision the possibility of building FUV MOS with an all-reflective design.

The overarching goal of the development is to deliver a micromirror array that is mechanically robust and optimized for MOS applications at UV and visible wavelengths. Such a device will enable MOS with unprecedented capabilities, advancing future astronomic exploration. Our target micromirror device is expected to meet the specifications listed in Table 21.

{\footnotesize
\begin{xltabular}{\textwidth}{@{} >{\raggedright\arraybackslash}p{0.24\textwidth} >{\raggedright\arraybackslash}X >{\raggedright\arraybackslash}X @{}}
\caption{DMD TRL 5 demonstration specifications}\label{tab:fig015_mmd}\\
\hline
 & \textbf{MMD -- baseline} & \textbf{MMD -- HWO goal} \\
\hline
\endfirsthead
\caption[]{(continued)}\\
\hline
 & \textbf{MMD -- baseline} & \textbf{MMD -- HWO goal} \\
\hline
\endhead
\hline
\multicolumn{3}{r}{\textit{(continued on next page)}}\\
\hline
\endfoot
\hline
\endlastfoot
Pitch & 30~\textmu m & 100 (\textmu m) \\
Format & 1000 $\times$ 1000 & 2 K $\times$ 2 K \\
Range & Visible-UV & Visible-UV \\
Tilt angle & 15$^\circ$ & 12$^\circ$ to 15$^\circ$ \\
Wavelength & 100 to 1000~nm & 100 to 1000~nm \\
Mirror reflectivity & 96\% & 96\% \\
Contrast ratio & 1:10,000 (blue) \newline 1:5000 (red) & 1:10,000 (blue) \newline 1:5000 (red) \\
Reconfiguration time & $<$ 4 seconds & $<$ 6 seconds \\
scale (6m, f/4) & 0.26$''$/mirror & 0.86$''$/mirror \\
FOV (6m, f/4) & 4.3$' \times$ 4.3$'$ & 29$'$.3$\times$29.3$'$ \\
\end{xltabular}
}
\subsection{UV Gratings \& Filters}
\textbf{Gap Type:} Engineering

\textbf{Est. TRL:} 4 -- 6

\textbf{Importance:} 2 -- Baseline

\textbf{Urgency:} 1 -- Long Term

\subsubsection{Discussion}
\paragraph{UV Gratings}
The dispersive elements needed in any UV spectrograph need to meet performance metrics to offset the higher scatter prevalent in the UV over, say, the visible and NIR bands. Mitigating scatter and improving both reflective throughput and spectral resolution can be achieved by more stringent manufacture methods that focus on increasing the accuracy of machining methods to produce cleaner and better-defined grating structures. This section describes a number of potential ways to do this to build on the current state-of-the-art. Table 22 shows a notional set of requirements for HWO UV Gratings.

\begin{xltabular}{\textwidth}{@{} >{\raggedright\arraybackslash}p{0.22\textwidth} >{\raggedright\arraybackslash}X @{}}
\caption{Notional HWO UV Grating Requirements}\label{tab:16}\\
\hline
\textbf{Parameter} & \textbf{Performance Needed} \\
\hline
\endfirsthead
\caption[]{(continued)}\\
\hline
\textbf{Parameter} & \textbf{Performance Needed} \\
\hline
\endhead
\hline
\multicolumn{2}{r}{\textit{(continued on next page)}}\\
\hline
\endfoot
\hline
\endlastfoot
High Resolution & Echelle gratings capable of achieving R $\geq$ 50,000 with $\geq$ 80\% peak-order groove efficiency and $\lambda$/I{\textsubscript{0}} < 1´10{\textsuperscript{3}} at $\Delta\lambda$ $=$ 1 nm post-coating, with supporting simulations predicting the observed performance. \newline This performance should extend through the far ultraviolet (FUV; 100 - 180 nm) bandpass. \\
Medium/Low Resolution & Aberration correcting (curved grooves) solutions on curved substrates demonstrating groove efficiencies $\geq$ 60\% post-coating and $\lambda$/I{\textsubscript{0}} $<$ 1$\times$10{\textsuperscript{5}} at $\Delta\lambda$ $=$ 1 nm \\
Ultra-low blaze angles & ($\leq$ 2$^{\circ}$) gratings demonstrating $\geq$ 60\% groove efficiency and $\lambda$/I{\textsubscript{0}} $<$ 1$\times$10{\textsuperscript{5}} at $\Delta\lambda$ $=$ 5 nm post-coating \\
UV Coatings on gratings & Demonstrated compatibility with state-of-the-art FUV coating techniques with $<$ 1\% loss in relative diffraction efficiency \\
Grating Characterization & Specialized vacuum characterization facilities for scatter, resolution, ghosting and efficiency measurements \\
\end{xltabular}
\paragraph{UV Filters}
The construction of efficient UV filters has been a relatively recent development (the first Al/MgF2 Fabry-Perot filters, with remarkable efficiency, were developed in the late 1960s-early 1970s\cite{Bates1966, Malherbe1970, Fairchild73} with the construction of Fabry-Perot like cavity interference filters with the appropriate choice and use of metal + dielectric materials of the appropriate thickness and uniformity to produce a well-defined passband that can be used to isolate the densely packed spectral diagnostic lines common in the UV.  The advent of techniques such as ALD and new plasma vapor deposition techniques have now made this a possibility.  Scientifically this allows HWO to perform UV imaging through the UVI (or even the HRI) at the native wavelength as opposed to using a filter technology that upconverts UV photons into visible band photons as was done on the Hubble Space Telescope, with all the losses that came along with that approach.

Table 23 shows a notional set of requirements for HWO UV Gratings.

\begin{xltabular}{\textwidth}{@{} >{\raggedright\arraybackslash}X >{\raggedright\arraybackslash}X @{}}
\caption{Notional HWO UV Filter Requirements}\label{tab:17}\\
\hline
\textbf{Parameter} & \textbf{Performance Needed} \\
\hline
\endfirsthead
\caption[]{(continued)}\\
\hline
\textbf{Parameter} & \textbf{Performance Needed} \\
\hline
\endhead
\hline
\multicolumn{2}{r}{\textit{(continued on next page)}}\\
\hline
\endfoot
\hline
\endlastfoot
Bandpass: & Medium (20 nm) and wide (40 nm) bandpass filter sets covering 100 - 200 nm \\
Compatibility: & Needs to work with shaped optics \\
Throughput: & $>$ 80\% peak throughput over central 15 nm for medium band filters (120 - 200 nm) \newline $>$ 50\% peak throughput over central 15 nm for medium band filters (100 - 120 nm) \newline $>$ 60\% peak throughput over central 30 nm for wide band filters (120 - 200 nm) \newline $<$ 1\% throughput (system) at 121.6 nm and \newline $<$ 0.0001\% to 0.01\% throughput (system) at $>$ 300 nm \\
Stability/Resilience: & Peak throughput decline < 10\% (relative) and out-of-band rejection degradation $<$ 20\% (relative) with exposure to relative humidity $\leq$ 50\% for 12 months. \\
Bandpass Splitting (e.g. Dichroics, edge filters): & Mid-UV dichroic Split: R (FUV) $>$ 0.8, T (NUV) $>$ 0.9 \\
\end{xltabular}
\subsubsection{Current State-of-the-Art}
The current state-of-the-art can be found in Scowen et al. (2026)\cite{Scowen2026}.

\subsubsection{TRL 5 Milestone Definition}
\begin{figure}[htbp]
\centering
\includegraphics[width=0.9\linewidth]{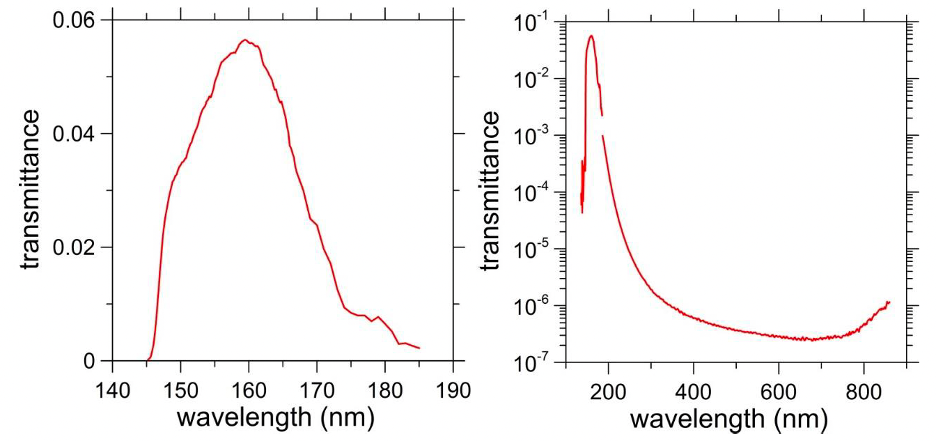}
\caption{Strong rejection of the visible may enable using detectors that are sensitive to the visible. This, alogn with the fact that an imager geometry is simplified with the use of transmittance filters, makes transmittance filters a valuable option.}
\label{fig:31}
\end{figure}
As most grating and filter technologies are relative advanced and high-TRL (or will be upon completion of sub-orbital demonstrations), the remaining development work is aimed at producing engineering development units (EDUs) for HWO-specific designs for TRL 6 and higher demonstrations.

\section{Emerging and Enhancing Technologies}
While the focus of the HWO TMPO is primarily the initial enabling technologies achieving TRL 5 by the Mission Concept Review, it is important to invest in some less mature ``emergent technology'' areas that could provide large benefits. These technologies are most likely targeting second, third and future generations instruments that could be upgrade with servicing.  However, significant progress in certain areas could bump up the possibility of a technology being included in the primary mission. Given that, the investment by the HWO TMPO in this area will be initially targeted and primarily longer term, but the definition of these areas can help justify investments from other technology programs and can help create the potential for one of them to be upgraded to part of the baseline mission.

\subsection{Cryogenic/Superconducting Detectors}
Cryogenic or superconducting detectors include detectors that are photon counting, high quantum efficiency, low (or zero) noise, rad hard, and potential energy resolving.  One possibility for implementation would be to target the visible channel of the coronagraph with either low noise large arrays (e.g., 4K $\times$ 4K format) or with medium size arrays (1K $\times$ 1K) that can energy resolve (R $>$ 100). In addition to the detectors, this topic would include extremely low vibration cooling methods that could achieve $<$ 1K cooling required for many of the superconducting technologies described in Section 4.5.  It would also include ways to implement this cryogenic detector with a warm telescope (e.g., isolating thermal background from the warm telescope using filters or cold windows).

\subsection{Quantum Sensors}
Quantum Sensors - quantum sensors that can achieve super sensitivity (e.g., zero shot noise) or super resolution (e.g., beyond the Rayleigh limit) have been demonstrated at TRL 3 with silicon vacancy centers and Rubidium atom arrays\cite{Mokeev2026}. These detectors may use cryogenic sensors which would benefit from the cooling and IR filtering strategies above. These sensors could help reduce stability requirements or even enhance resolution and signal-to-noise ratios given their lack of shot noise. These detectors are immature but the potential may be large.

\subsection{Metasurfaces}
Metasurfaces are 2-dimensional surfaces made up of sub-wavelength structures that can perform unique optical performance\cite{Rubin2021}. One possible area it helps deal with polarization effects in a coronagraph\cite{Ashcraft2025}.

\subsection{Photonic Integrated Circuits (PICs)}
PIC technology is being developed already for HWO in the areas of nulling for coronagraphs\cite{Sirbu2024} and in the area of laser metrology\cite{Wirth2022}. These technologies can be very mass and volume and cost efficient. However, they often are narrow band.  One possibility is to use PIC nullers in the NIR to achieve better inner working angle\cite{Sirbu2024}. Additional wavelengths will be needed as defined by science needs, and additional performance will be needed for HWO use.

\subsection{Artificial Intelligence \& Machine Learning (AI \& ML)}
Recent advancements in AI/ML provide new opportunities to leverage their capabilities across the HWO TMPO effort. From AI/ML-assisted design, to system-identification as part of wavefront control, and potential applications in post-processing and image analysis\cite{Ansdell2024}, HWO will consider the development and use of AI/ML tools to benefit mission objectives.

\section{Technology funding approach \& synergies with other NASA technology programs}
The funding of HWO technology will undergo a transition during the next few years. Early efforts are funded via a combination of Small Business Innovative Research (SBIR) grants, Strategic Astrophysics Technology (SAT) and Astrophysics Research and Analysis (APRA) investments, and other dedicated Research Opportunities in Space and Earth Sciences (ROSES) programs in segmented ultra-stable telescopes. Investments under internal research and development (IRAD) and internal scientist funding model (ISFM) at NASA centers and in industry were also used. Many of these investments are still active and will remain active over the next few years. An FY24 SAT and APRA call contributed investments to HWO technology development.

Over the next few years, we anticipate that many of these investments will transition over to be managed by HWO TMPO while others finish out under their current programs. Regardless, all investments will be considered as steps along the HWO technology maturation process. The latest ROSES D.19 HWO System Technology Demonstration and Mission Architecture Studies call was released in the Fall of 2024 and was based on roadmaps developed by a NASA-led effort to kickstart technology investments. The award of this call is a major effort managed by the HWO TMPO to advance technologies in the Ultra-stable Telescope Systems track. In addition, testbed work is being done as directed work at the centers to support critical technology evaluations in both the Ultra-stable Telescope Systems and Coronagraph Systems tracks. Longer term, these testbeds will be made available in future calls for external partners to use. Now that all FY24 awards are funded, the HWO TMPO will assess remaining investment gaps and then plan future investments in technologies prioritized based on the risk approach defined in this document. These investments could include both directed efforts and/or competitive calls similar to -- or even directly leveraging -- the ROSES SAT or APRA programs. Each year the technology team will revisit progress and gaps and prioritize investments based on funding available and remaining risks.

\bibliography{report}   % bibliography data in report.bib
\bibliographystyle{spiejour}  

\appendix
\section{Appendix A: Abbreviations and Acronyms}
\begin{xltabular}{\textwidth}{@{} >{\raggedright\arraybackslash}X >{\raggedright\arraybackslash}X @{}}
\label{tab:18}\\
\hline
\textbf{Abbreviation / Acronym} & \textbf{Definition} \\
\hline
\endfirsthead
\caption[]{(continued)}\\
\hline
\textbf{Abbreviation / Acronym} & \textbf{Definition} \\
\hline
\endhead
\hline
\multicolumn{2}{r}{\textit{(continued on next page)}}\\
\hline
\endfoot
\hline
\endlastfoot
ACADIA & ASIC for Control and Digitization of Imagers for Astronomy \\
ACS & Attitude Control System \\
ADC & Analog-to-Digital Converter \\
ADI & Angular Differential Imaging \\
AI & Artificial Intelligence \\
ALD & Atomic Layer Deposition \\
AOX & AOA-Xinetics \\
APRA & Astrophysics Research and Analysis \\
AR & Anti-reflection \\
ARC & Ames Research Center \\
ASIC & Application Specific Integrated Circuit \\
ASSIST & Active Segmented Surrogate for Integrated Systems Tests \\
ATLAST & Advanced Technology Large Aperture Space Telescope \\
BMC & Boston Micromachine \\
BSTA & Backplane Stability Test Article \\
C\&DH & Command \& Data Handling \\
CCD & Charge Coupled Device \\
CGI & (Roman) Coronagraph Instrument \\
CI & Coronagraph Instrument \\
CIC & Clock-Induced Charge \\
CMOS & Complementary Metal Oxide Semiconductor \\
ConOps & Concept of Operations \\
COR & Cosmic Origins \\
COS & Cosmic Origins Spectrograph \\
CTE & Coefficient of Thermal Expansion \\
CTI & Charge Transfer Inefficiency \\
CTR & Coronagraph Technology Roadmap \\
DAC & Digital-to-Analog Converter \\
DLP & Digital Light Projection \\
DM & Deformable Mirror \\
DMD & Digital Micromirror Device \\
DST & Decadal Survey Testbed \\
DZM & Dark Zone Maintenance \\
EAC & Exploratory Analytic Case \\
EAR & Export Administration Regulations \\
EBL & Electron Beam Lithography \\
EDU & Engineering Development Unit \\
EMCCD & Electron Multiplying Charge Coupled Device \\
EPIC-5 & Exoplanet Imaging Coronagraph for TRL 5 \\
EUV & Extreme Ultra-violet \\
ExCam & Exoplanet Camera \\
ExEP & Exoplanet Exploration Program \\
FCM & Focus Control Mechanism \\
FORTIS & Far-UV Off Rowland-circle Telescope for Imaging and Spectroscopy \\
FOV & Field-of-View \\
FPM & Focal Plane Mask \\
FS & Field Stop \\
FSM & Fast Steering Mirror \\
FUV & Far Ultra-violet \\
FWHM & Full-Width Half-Maximum \\
GOMAP & Great Observatory Mission and Technology Maturation Program \\
GSFC & Goddard Space Flight Center \\
H4RG & Hawaii 4k Region-of-Interest Guiding \\
HabEx & Habitable Exoplanet Observatory \\
HCIT & High-Contrast Imaging Testbed \\
HGA & High-Gain Antenna \\
HiCAT & High-contrast imager for Complex Aperture Telescopes \\
HOST & HWO Systems Testbed \\
HOWFS & High-order Wavefront Sensor \\
HRI & High-resolution Instrument \\
HST & Hubble Space Telescope \\
HWO & Habitable Worlds Observatory \\
I\&T & Integration \& Test \\
IBE & Ion-Beam Etching \\
IC & Instrument Carrier \\
IFS & Integral Field Spectrograph \\
IFU & Integral Field Unit \\
IM & Integrated Modeling \\
INFUSE & Integral Field Ultraviolet Spectroscopic Experiment \\
IR & Infrared \\
IRAD & Internal Research and Development \\
IRIS & Image Retrieval In Segments \\
IRU & Inertial Reference Unit \\
ISFM & Internal Scientist Funding Model \\
ITAR & International Traffic in Arms Regulations \\
IWA & Inner Working Angle \\
JHU & Johns Hopkins University \\
JPL & Jet Propulsion Laboratory \\
JWST & James Webb Space Telescope \\
KPM & Key Performance Metric \\
LIDAR & Light Detection and Ranging \\
LM-APD & Linear Mode Avalance Photodiode \\
LoCam & LOWFS Camera \\
LOS & Line-of-Sight \\
LOWFS & Low-order Wavefront Sensor \\
LS & Lyot Stop \\
LUVOIR & Large Ultraviolet / Optical / Infrared Surveyor \\
MAS & Multi-Amplifier Sensing \\
MBA & Main Barrel Assembly \\
MCP & Micro-channel Plate \\
MCR & Mission Concept Review \\
MEMS & Micro-electro-mechanical System \\
MISSE & Materials International Space Station Experiment \\
MKID & Microwave Kinetic Inductance Detector \\
ML & Machine Learning \\
MOS & Multi-object Spectrograph \\
MOU & Memorandum of Understanding \\
MOWFS & Mid-order Wavefront Sensor \\
MSA & Microshutter Array \\
MSE & Mission System Engineer \\
MSFC & Marshall Space Flight Center \\
MUST & Metrology and Ultra-stable Testbed \\
N/A & Not Applicable \\
NASA & National Aeronautics and Space Administration \\
NGMSA & Next Generation Microshutter Array \\
NIR & Near Infrared \\
NIRSpec & Near Infrared Spectrograph \\
NPR & NASA Procedural Requirement \\
NUV & Near Ultra-violet \\
OAP & Off-axis Parabola \\
OTA & Optical Telescope Assembly \\
OTE & Optical Telescope Element \\
OWA & Outer Working Angle \\
PA & Principal Architect \\
PCOS & Physics of the Cosmos \\
PDAA & Pathways to Discovery in Astronomy and Astrophysics \\
PDR & Preliminary Design Review \\
PDT & Project Development Team \\
PI & Principal Investigator \\
PIC & Photonic Integrated Circuit \\
PID & Proportional - Integral - Derivative \\
PISCES & Prototype Imaging Spectrograph for Coronagraphic Exoplanet Studies \\
PM & Primary Mirror \\
PM & Project Manager \\
PSD & Power Spectral Density \\
PSF & Point-spread Function \\
QE & Quantum Efficiency \\
RBCS & Rigid Body Control System \\
RDI & Reference Differential Imaging \\
RH & Relative Humidity \\
RIT & Rochester Institute of Technology \\
RITMOS & RIT Multi-object Spectrograph \\
RMS & Root Mean Square \\
ROIC & Readout Integrated Circuit \\
ROSES & Research Opportunities in Space and Earth Sciences \\
RWA & Reaction Wheel Assembly \\
SAMOS & SOAR Adaptive-Module Optical Spectrograph \\
SAT & Strategic Astrophysics Technology \\
SBIR & Small Business Innovative Research \\
SIG & Science Interest Group \\
SiSeRo & Single-Electron Sensitive Readout \\
SM & Secondary Mirror \\
SME & Subject Matter Expert \\
SMT & Surface Mount Technology \\
SNR & Signal to Noise Ratio \\
SNSPD & Superconducting Nanowire Single Photon Detector \\
SOAR & Southern Astrohphysical Research \\
SPRITE & Supernova Remants and Proxies for Reionization Testbed Experiment \\
ST & Star Tracker \\
STScI & Space Telescope Science Institute \\
SWaP & Size, Weight, and Power \\
TAG & Technical Assessment Group \\
TASTE & Thermally Activated Selective Topography Equilibration \\
TBD & To Be Determined \\
TBR & To Be Revised \\
TCE & Telescope Control Electronics \\
TEC & Thermal Electric Cooling \\
TES & Transition Edge Sensor \\
TI & Texas Instruments \\
TMPO & Technology Maturation Project Office \\
TPF & Terrestrial Planet Finder \\
TPX & TimePix \\
TRL & Technology Readiness Level \\
ULE & Ultra-Low Expansion \\
UQ & Uncertainty Quantification \\
USORT & Ultra-stable Optics Roadmap Team \\
USSL & Ultra-stable Structure Laboratory \\
UV & Ultra-violet \\
UVI & Ultra-violet Instrument \\
VIS & Vibration Isolation System \\
WFE & Wavefront Error \\
WFI & (Roman) Wide-field Instrument \\
WFS & Wavefront Sensor \\
WFS\&C & Wavefront Sensing \& Control \\
XS & Cross-strip \\
\end{xltabular}

\section{Appendix B: Visible Light Coronagraph Detector Technologies and References}
Table courtesy of Nathan Bush (JPL).
\begin{landscape}
{\footnotesize
\begin{xltabular}{\linewidth}{@{} >{\raggedright\arraybackslash}p{0.10\linewidth} >{\raggedright\arraybackslash}p{0.28\linewidth} >{\raggedright\arraybackslash}p{0.14\linewidth} >{\raggedright\arraybackslash}X @{}}
\caption{Detector technology reference summary}\label{tab:detector_refs}\\
\hline
\textbf{Technology} & \textbf{Title} & \textbf{Reference} & \textbf{Notes} \\
\hline
\endfirsthead
\caption[]{(continued)}\\
\hline
\textbf{Technology} & \textbf{Title} & \textbf{Reference} & \textbf{Notes} \\
\hline
\endhead
\hline
\multicolumn{4}{r}{\textit{(continued on next page)}}\\
\hline
\endfoot
\hline
\endlastfoot
\multirow{5}{*}{MKIDs} & KID Detector Readout Electronics Development for Habitable Worlds Observatory & Bryan et al. 2026~\cite{bryan2026} & Prototype readout architecture with 100,000 pixel readout with $<$ 1 kW total power consumption. \\ \\
 & Characterizing the dark count rate of a large-format MKID array & Swimmer et al. 2023~\cite{swimmer2023} & Dark count rate of (9.3$\pm$0.9)$\times$10$^{-4}$ measured on an MKID pixel across 1310-808 nm. \\ \\
 & High-Resolution Arrayed-Waveguide-Gratings in Astronomy: Design and Fabrication Challenges & Stoll et al. 2017~\cite{stoll2017} & Describes a path to improving intrinsic resolution of R$\sim$40 to R $>$ 103 using photonic waveguides integrated with MKIDs. \\ \\
 & Die Separation for Mitigation of Phonon Bursts in Superconducting Circuits & Moshel et al. 2025~\cite{moshel2025} & Demonstration of resilience to cosmic rays through manufacturing process that isolate die. \\ \\
 & Model and Measurements of an Optical Stack for Broadband Visible to Near-Infrared Absorption in TiN MKIDs & Kouwenhoven et al. 2022~\cite{kouwenhoven2022} & Internal absorption of $>$ 80\% across 400-1550 nm, corresponding to an estimated system detection efficiency of circa 60-70\%. \\ \hline\hline \\
\multirow{4}{*}{EMCCDs} & Assembly, Optimization and Calibration of the Roman Coronagraph Camera Systems: The Exoplanetary Systems Camera (EXCAM) and the Low Order Wavefront Sensing Camera (LOCAM) & Bush et al. 2025~\cite{Bush2025} & Roman CGIs TRL-8 EMCCD camera system. The current ``state-of-the-art'' for EMCCDs for space applications. Provides pathways for improvements for HWO. Dark count rate of 1.5$\times$10$^{-5}$ e-/pix/s demonstrated. \\ \\
 & Proton Induced traps in Electron Multiplying CCDs & Bush et al. 2021~\cite{bush2021} & Most detailed characterization of silicon defects in EMCCDs currently available; vital for lifetime improvements \\ \\
 & Measurement and optimization of clock induced charge in Electron-Multiplying Charge Coupled Devices & Bush et al. 2021~\cite{bush20212} & Lowest demonstration of CIC at 6.9$\times$10$^{-4}$ e-/pix/frame, with path to improved performance. \\ \\
 & Flight photon counting electron multiplying charge coupled device development for the Roman Space Telescope coronagraph instrument & Morrissey et al. 2023~\cite{Morrissey2023} & Radiation damage study. Technology advancement of the CCD201 into the CCD311 – flight device for Roman CGI. \\ \hline \hline \\
\multirow{3}{*}{sCMOS (Fairchild)} & The Lazuli Space Observatory: Architecture \& Capabilities & Roy et al. 2026~\cite{roy2026} & Example of architecture also considering HWK4123 for space applications \\ \\
 & QCMOS Characterization Summary & JPL internal~\cite{detinternal10} & Verification of 0.24e- RMS read noise, dark current and low flux imaging performance. \\ \\
 & Radiation Tolerance of HWK4123 for Space Applications & Hu et al.\cite{HuFairchild} & Proton, gamma and heavy ion testing of HWK4123 sensor and camera system. Manageable dark current increase and no destructive latchup reported.\\ \\
 & Characterizing radiation-tolerant single photon resolving CMOS detectors & Gallagher et al. 2024~\cite{gallagher2024} & HWK4123 independent verification of 0.24 e- RMS read noise. Discussion of radiation testing. \\ \hline \hline \\
\multirow{4}{*}{sCMOS (Teledyne)} & Characterization of the Teledyne COSMOS Camera: A Large Format CMOS Image Sensor for Astronomy & Layden et al. 2025~\cite{layden2025} & COSMOS sensor – not the same as the CIS300 series, but similar architecture. 0.7 e- RMS measured. Charge transfer issues reported. \\ \\
 & New developments for large-area high-performance visible detectors from VUV to NIR & Jordan et al. 2025~\cite{jordan2025} & 1.7 e- noise measured on the CIS300 \\ \\
 & SAT21-0039 Annual Report: High Performance FUV, NUV, and UV/Optical CMOS Imagers & JPL internal~\cite{detinternal14} & Dark Current/Glow Measurements in the CIS120. Low signal charge transfer performance for Teledyne’s CIS architecture. Demonstration of the need for cryogenic irradiations. SAT still ongoing. \\ \\
 & Detector developments for UV Space missions & Skottfelt et al. 2025~\cite{skottfelt2025} & Selection of CIS300 architecture for CASTOR baseline. \\ \hline \hline \\
\multirow{5}{*}{Skipper CCDs} & Optimizing Charge-coupled Device Readout Enabled by the Floating-Gate Amplifier & Lin et al. 2025~\cite{lin2025} & Alternative architecture for faster skipper readout. 0.9 e- RMS demonstrated with the equivalent of a 17.5 s readout time for a 1MPix array. \\ \\
 & Characterization of proton-induced damage in thick, p-channel skipper-CCDs & Cervantes-Vergara et al. 2025~\cite{cervantesvergara2025} & 0.134 traps/pixel reported. CTE not reported, but 1 trap per 10 pixels implies a performance concern to be investigated further. \\ \\
 & SENSEI: Characterization of Single-Electron Events Using a Skipper Charge-Coupled Device & Barak et al. 2022~\cite{barak2022} & Demonstration of 1.52$\times$10$^{-4}$ e-/pix/frame of CIC (with full well penalty and ultra slow readout - 5 hours) \\ 
 &  &  & Demonstration of ultra-low dark current 10$^{-4}$ e-/pix/day \\ \\
 & Developing a skipper-CCD instrument to search for Dark Matter from Low Earth Orbit & Alpine et al. 2024~\cite{alpine2024} & 6U CubeSat with 4 skipper CCDs \\ \hline \hline \\
\multirow{2}{*}{Skipper CMOS} & Skipper-in-CMOS: Nondestructive Readout With Sub-electron Noise Performance for Pixel Detectors & Lapi et al. 2024~\cite{lapi2024} & Demonstration of concept on test device. 0.15 e- RMS demonstrated. \\ \\
 & Simulations and Design of a Single-Photon CMOS Imaging Pixel Using Multiple Non-Destructive Signal Sampling & Stefanov et al. 2020~\cite{stefanov2020} & Study on proof of concept for skipper CCD in CMOS. \\ \hline \hline \\
\multirow{4}{*}{SNSPDs} & A superconducting nanowire single-photon camera with 400,000 pixels & Oripov et al. 2023~\cite{oripov2023} & Dark count rate of 1.0$\times$10$^{-4}$ cps demonstrated. \\ \\
 & Detecting single infrared photons with 99.5\% system detection efficiency & Hu et al. 2020~\cite{hu2020} & System Detection efficiency of 99.5 \% for single pixel SNSPD through use of optical cavities – represents demonstration of high detection. \\ \\
 & High-efficiency, high-count-rate 2D superconducting nanowire single-photon detector array & Fleming et al. 2025~\cite{fleming2025} & 2D array (8 $\times$ 8), more realistic system detection efficiency of 65 \% measured at 1550 nm. Note, may be different in visible spectrum. \\ \\
 & A Four-Stage Continuous ADR for Space Missions & Shinozaki et al. 2010~\cite{shinozaki2010} & State of the art cooling targeted at vibration isolation. \\ \hline \hline \\
\multirow{4}{*}{SPADs} & A 3.2 Megapixel 3D-Stacked Charge-Focusing SPAD for Low-Light Imaging & Morimoto et al. 2021~\cite{morimoto2021} & Demonstration of large format, backside illuminated SPAD with near 100\% fill factor. 2 cps dark signal at room temperature. \\ \\
 & Radiation hardness properties and DCR reduction via laser annealing of InGaAs/InP SPADs & Finazzi et al. 2025~\cite{finazzi2025} & Radiation damage study and use of annealing via lasers as a mitigation strategy \\ \\
 & Comparison of Proton and Gamma Irradiation on Single-Photon Avalanche Diodes & Xun et al. 2024~\cite{xun2024} & Comparison of proton/gamma effects. Displacement damage effects dominate the degradation. \\ \\
 & Operation of silicon single photon avalanche diodes at cryogenic temperature & Rech et al. 2007~\cite{rech2007} & Cryogenic cooling in an attempt to lower the count rate – 1 cps at 120 K. \\ \hline \hline \\
\multirow{3}{*}{QIS} & A 0.19e- rms Read Noise 16.7Mpixel Stacked Quanta Image Sensor With 1.1~\textmu m-Pitch Backside Illuminated Pixels & Ma et al. 2021~\cite{ma2021} & 0.17 e- RMS reported at 253 K. 4K $\times$ 4K array. \\ \\
 & High Dynamic Range Imaging using Quanta Image Sensors & Chi et al. 2020~\cite{chi2020} & Multi-bit readout with resets per integration enabled dynamic range up to 120 dB. \\ \\
 & Radiation tolerance of a single-photon counting complementary metal-oxide semiconductor image sensor & Gallagher et al. 2024~\cite{gallagher20242} & Radiation effects in QIS are similar to standard CMOS image sensors. Introduction of RTS pixels contributes to degradation. \\ \hline \hline \\
\multirow{3}{*}{LmAPDs} & Four megapixel sensor for ultra-low-background shortwave infrared astronomy & Claveau et al. 2022~\cite{claveau2022} & Describes latest development effort via a funded SAT program. \\ \\
 & Progress towards a megapixel linear-mode avalanche photodiode array for ultra-low background shortwave infrared astronomy & Claveau et al. 2022~\cite{claveau20222} & 1K $\times$ 1K device, 0.3 e- read noise demonstrated via non destructive sampling in combination with a lower gain mode (for low dark current). \\ \\
 & Effects of proton irradiation on a SAPHIRA HgCdTe avalanche photodiode array & Sun et al. 2022~\cite{sun2022} & Annealing (343 K) used to bring dark current down to pre-irradiation levels. \\ \hline \hline
\end{xltabular}
}
\end{landscape}

\end{document}